\documentclass[acmsmall]{acmart}

\usepackage{xcolor}
\usepackage{colortbl}

\usepackage{graphicx}
\usepackage{romannum}
\usepackage{color}
\usepackage{multirow}
\usepackage{tikz}
\usepackage{wasysym}
\usepackage{longtable}
\usepackage{tabularx}
\usepackage{pdflscape}
\usepackage{array}
\usepackage{boldline}
\usepackage{csquotes}
\usepackage{romannum}

\usepackage[margin=20pt]{subcaption}

\newcommand{\etal}{\emph{et al.}}
\newcommand{\eg}{\emph{e.g.}}
\newcommand{\ie}{\emph{i.e.}}
\newcommand{\etc}{\emph{etc}}

\newcommand*\rot{\rotatebox{90}}

\definecolor{Blue}{RGB}{0,0,139}

\acmJournal{CSUR}

\begin{document}

\title{Deep Learning-Based Anomaly Detection in Cyber-Physical Systems: Progress and Opportunities}

\author{Yuan Luo}
\affiliation{%
  \streetaddress{School of Cyber Science and Engineering}
  \institution{Wuhan University}
  \state{Hubei, China}
 \postcode{430072}
 }
\email{leonnewton@whu.edu.cn}

\author{Ya Xiao}
\affiliation{%
 \streetaddress{Department of Computer Science}
  \institution{Virginia Tech}
  \city{Blacksburg}
  \state{VA}
 \postcode{24060}
 }
\email{yax99@vt.edu}

\author{Long Cheng}
\affiliation{%
 \streetaddress{School of Computing}
  \institution{Clemson University}
  \city{Clemson}
  \state{SC}
 \postcode{29634}
 }
\email{lcheng2@clemson.edu}

\author{Guojun Peng}
\authornote{Corresponding authors}
\affiliation{%
  \streetaddress{School of Cyber Science and Engineering}
  \institution{Wuhan University}
  \state{Hubei, China}
 \postcode{430072}
 }
\email{guojpeng@whu.edu.cn}

\author{Danfeng (Daphne) Yao}
\authornotemark[1]
\affiliation{%
 \streetaddress{Department of Computer Science}
  \institution{Virginia Tech}
  \city{Blacksburg}
  \state{VA}
 \postcode{24060}
 }
\email{danfeng@vt.edu}

\renewcommand{\shortauthors}{Yuan Luo, Ya Xiao, et al.}

\pagenumbering{arabic}





\begin{abstract}
\textcolor{black}{Anomaly detection is crucial to ensure the security of cyber-physical systems (CPS). However, due to the increasing complexity of CPSs and more sophisticated attacks, conventional anomaly detection methods, which face the growing volume of data and need domain-specific knowledge, cannot be directly applied to address these challenges. To this end, deep learning-based anomaly detection (DLAD) methods have been proposed. In this paper, we review state-of-the-art DLAD methods in CPSs. We propose a taxonomy in terms of the type of anomalies, strategies, implementation, and evaluation metrics to understand the essential properties of current methods. Further, we utilize this taxonomy to identify and highlight new characteristics and designs in each CPS domain. Also, we discuss the limitations and open problems of these methods. Moreover, to give users insights into choosing proper DLAD methods in practice, we experimentally explore the characteristics of typical neural models, the workflow of DLAD methods, and the running performance of DL models. Finally, we discuss the deficiencies of DL approaches, our findings, and possible directions to improve DLAD methods and motivate future research.}
\end{abstract}



\maketitle

\section{Introduction}
\label{intro}

Cyber-physical systems (CPS) are increasingly being deployed in critical infrastructures. The CPS market is expected to expand by 9.7\% each year, which will reach \$9563 million by 2025~\cite{cps_market}. Prominent applications of CPS include industrial control systems (ICS), smart grid, intelligent transportation systems (ITS), and aerial systems. CPSs have evolved to be complex, heterogeneous, and integrated to provide rich functionalities. However, such characteristics also expose CPSs to broader threats. In H1 2019, 41.6\% of ICS computers that installed Kaspersky products detected attacks~\cite{Kaspersky_threat_2019}. According to FireEye's report, insiders, ransomware, market manipulation, \etc ~are among the top attack types in ICS~\cite{fireeye_top20}. Recent incidents (\eg, Stuxnet~\cite{stuxnet_2020}, Ukraine power grid outage~\cite{Ukraine_december_2019}, auto-driving crashes~\cite{accident_list_2019}, robot malfunction~\cite{assemblylinekill}) have shown that sophisticated and stealthy attacks (and faults) can result in catastrophic consequences to the economy, environment, and even human lives. Thus, it is paramount important to ensure the security of CPSs.

To detect attacks and unexpected errors in CPSs, anomaly detection methods are proposed to mitigate these threats. For example, rule, state estimation (\eg, Kalman filter), statistical model (\eg, Gaussian model, histogram-based model) based methods are utilized to learn normal status of CPSs~\cite{lun2019state}. However, these methods usually require expert knowledge (\eg, operators manually extract certain rules), or need to know the underlying distribution of normal data. Machine learning approaches do not rely on domain-specific knowledge~\cite{chandola2009anomaly}. But they usually require a large quantity of labeled data (\eg, classification-based methods). Also, they cannot capture the unique attributes of CPSs (\eg, spatial-temporal correlation)~\cite{schneider2018high}. Intrusion detection methods are dedicated to ensuring network communication security~\cite{mitchell2014survey, zhang2016causality}. Physical properties are captured to depict the immutable nature of CPSs~\cite{giraldo2018survey}. Program execution semantics are characterized to protect control systems~\cite{cheng2017orpheus, xu2016sharper, shu2015unearthing}. However, as CPSs become more complicated and attacks are more stealthy (\eg, APT attacks), these methods are hard to ensure the overall status of CPSs (\eg, protect multivariate physical measurement) and need more domain knowledge (\eg, more components and correlation). Anomaly detection systems need to adapt to capture new characteristics of CPSs.

To this end, deep learning-based anomaly detection (DLAD) methods have been proposed to identify anomalies in CPS. Current studies have explored different neural network architectures (\eg, ConvLSTM) to mitigate various threats (\eg, false data injection attacks) in different CPS domains (\eg, smart grid). However, since these studies are not introduced in a unified way, a systematic survey is needed to review existing methods and provide guidance for future solutions. \textcolor{black}{Specifically, we need to answer the following four research questions}:

\begin{itemize}
\item \textcolor{black}{What are the characteristics of existing approaches? How existing DLAD methods can be categorized in terms of the threat model, detection strategies, implementation, and evaluation metrics?}
\item \textcolor{black}{How a DL model can be applied to solve a problem? For example, what are the characteristics of each neural model and how to use a neural model to build a DLAD method (\ie, the workflow)?}
\item \textcolor{black}{What are the limitations and deficiencies of DL approaches when being applied to the anomaly detection task in CPS?} 
\item \textcolor{black}{How can researchers address the limitations and improve DLAD methods?}
\end{itemize}

Answering these questions helps to understand the fundamentals of DLAD methods, evaluate proposed DLAD models, and explore new solutions.
This motivates our work to summarize and identify progress, challenges, and future research directions of DLAD methods. Our contributions are as follows.

\begin{itemize}
\item We systematically review existing deep learning-based anomaly detection methods that target at detecting faults and attacks in CPS. To this end, we propose a new taxonomy that is based on \romannum{1}) type of anomalies (\ie, threat model), \romannum{2}) detection strategies (\ie, input data, neural network designs, anomaly scores), and \romannum{3}) implementation and evaluation metrics. Further, we explore and categorize peer-reviewed research papers from conferences and journals under the setting of this taxonomy.

\item \textcolor{black}{We identify and highlight characteristics that are essential to building a DLAD method. First, we discuss existing methods in representative CPS domains (\ie, ICSs, smart grid, ITSs, and aerial systems). Then, we report unique designs and trends in each domain. All these findings are summarized according to our taxonomy. Meanwhile, we summarize and discuss the limitations and open problems of current methods.}

\item \textcolor{black}{We experimentally explore typical neural models to capture different characteristics of CPSs. We show the workflow to build a DLAD method and present the running performance of neural models.}

\item \textcolor{black}{We identify the limitations and deficiencies of deep learning approaches when being applied to the anomaly detection task in CPS. We present our findings and takeaways to improve the design and evaluation of DLAD methods. Also, we discuss several promising research directions and open problems that motivate future research efforts.}

\end{itemize}

\section{Background}
In this section, we introduce a generic architecture of cyber-physical systems and threats that are typically studied in existing DLAD methods (Section~\ref{cpsandthreat}), the workflow of DLAD methods (Section~\ref{dadmethod}). We discuss the key differences between our work and the existing survey papers in CPS (Section~\ref{relatedworksection}). 

\subsection{\textsc{Cyber-physical systems and threats}}\label{cpsandthreat}

\begin{figure}[!h]
	\centering
	\includegraphics[width=0.7\textwidth]{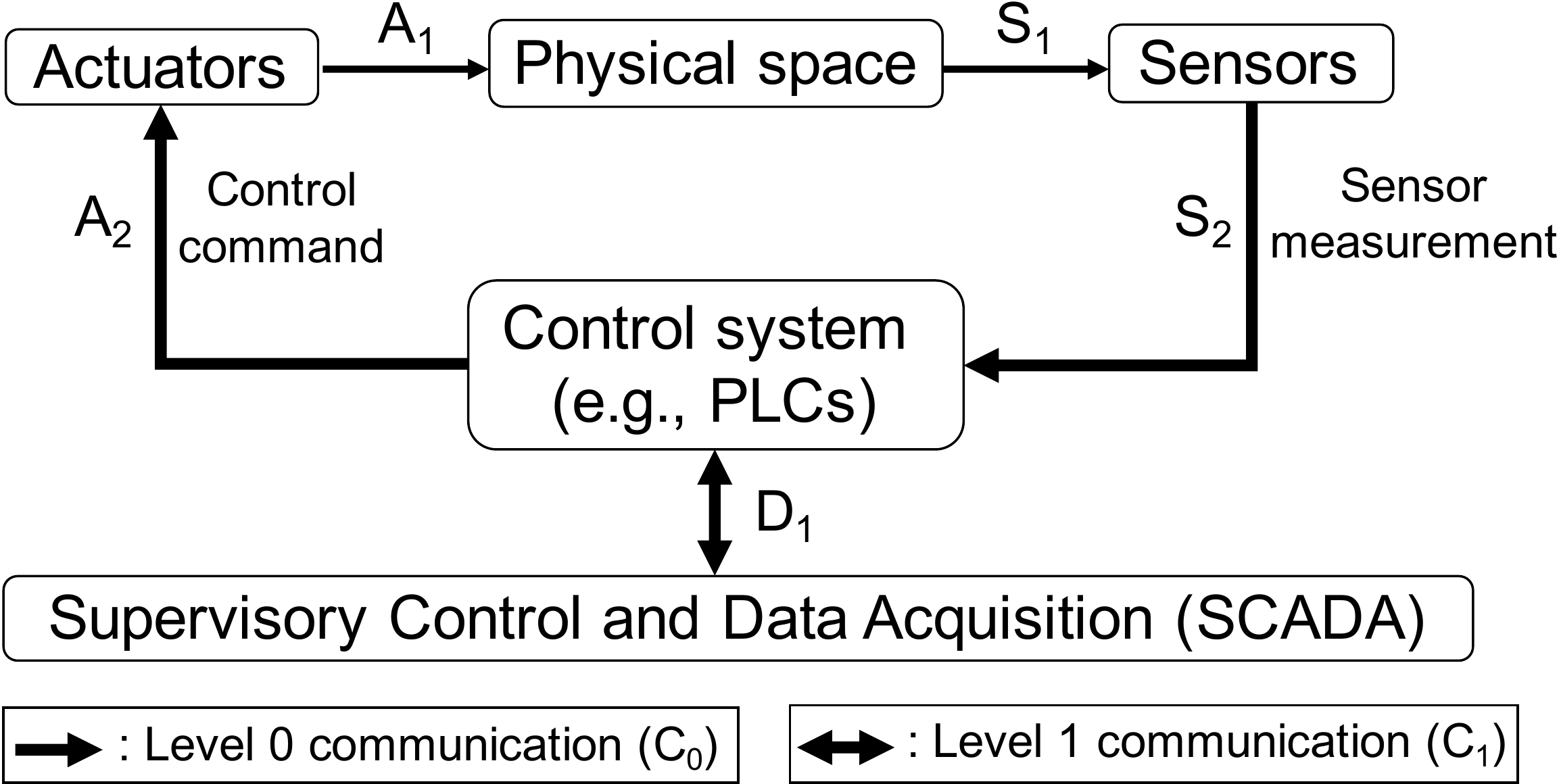} %
	\vspace{-3pt}
	\caption{A generic CPS architecture. Deep anomaly detection methods mainly aim to protect sensors, actuators, level 0 and level 1 communication, and control systems.}
	\label{cpsarchitecture} 
\end{figure}

\textbf{The generic definition of CPS}. As illustrated in Figure~\ref{cpsarchitecture}, CPSs typically consist of five components: \textit{The physical space} contains physical components of CPSs, \eg, engines, tanks, wheels. \textit{Actuators} receive control commands (denoted as $A_2$) from control systems and change the running parameters of physical devices ($A_1$). \textit{Sensors} measure the running status of devices ($S_1$) and report to the control systems ($S_2$). \textit{Control systems} obtain sensor measurement ($S_2$) and send control commands to actuators ($A_2$), which follows the predefined control logic. \textit{Supervisory control and data acquisition (SCADA)} systems are used to gather data from control systems ($D_1$) and monitor the running status of CPSs for users.

We define communication between sensors (actuators) and control systems as level 0 communication (denoted as $C_0$). The content of $C_0$ communication traffic is sensor measurement ($S_2$) and control commands ($A_2$). Similarly, communication between control systems and SCADA is defined as level 1 communication ($C_1$). The content of $C_1$ is $D_1$. Specifically, our work focuses on four representative types of CPSs, \ie, Industrial Control Systems (ICSs), smart grid, Intelligent Transportation Systems (ITSs) and aerial systems. Actual devices may vary in these four CPSs (\eg, actuators can be pumps in ICS and brakes in ITS) but they share the same generic architecture.

\textbf{Threat model}. We then present threats that are studied by DLAD methods in our work. Threats can be classified as attacks and faults. We observe that most existing studies usually do not obtain data directly from physical space. Namely, these two data sources are not adopted: \romannum{1}) running status data of physical components from physical devices to sensors ($S_1$), \romannum{2}) control parameters from actuators to physical devices ($A_1$). Instead, $S_2$ (values sent to control systems) and $A_2$ (commands sent to actuators) are commonly utilized by existing work. We focus our investigation on:

\begin{itemize}

\item Sensor and actuator anomalous values. Sensors and actuators either can be compromised under attacks or failed due to various reasons (\eg, lack of maintenance). Attackers may physically tamper with field sensors and actuators under this scenario. In Figure~\ref{cpsarchitecture}, $S_2$ and $A_2$ are affected under this threat model ($S_2 \neq S_1$, $A_2 \neq A_1$).

\item Manipulated level 0 and level 1 communication traffic. Attackers can manipulate two types of communication signals: \romannum{1}) network traffic between sensors (actuators) and control systems ($C_0$), \romannum{2}) traffic data between control systems and SCADA ($C_1$).

\item Compromised control systems. Control systems are connected to field devices and central operating centers, which makes it prone to remote attacks. For example, attackers can plant malware and send false control signals. Also, internal faults (\eg, logic errors) can cause wrong control commands. $A_2$ and $D_1$ are affected in this scenario.

\end{itemize}

\subsection{\textsc{The workflow of typical DLAD methods}}\label{dadmethod}

\begin{figure}[!h]
	\centering
	\includegraphics[width=\textwidth]{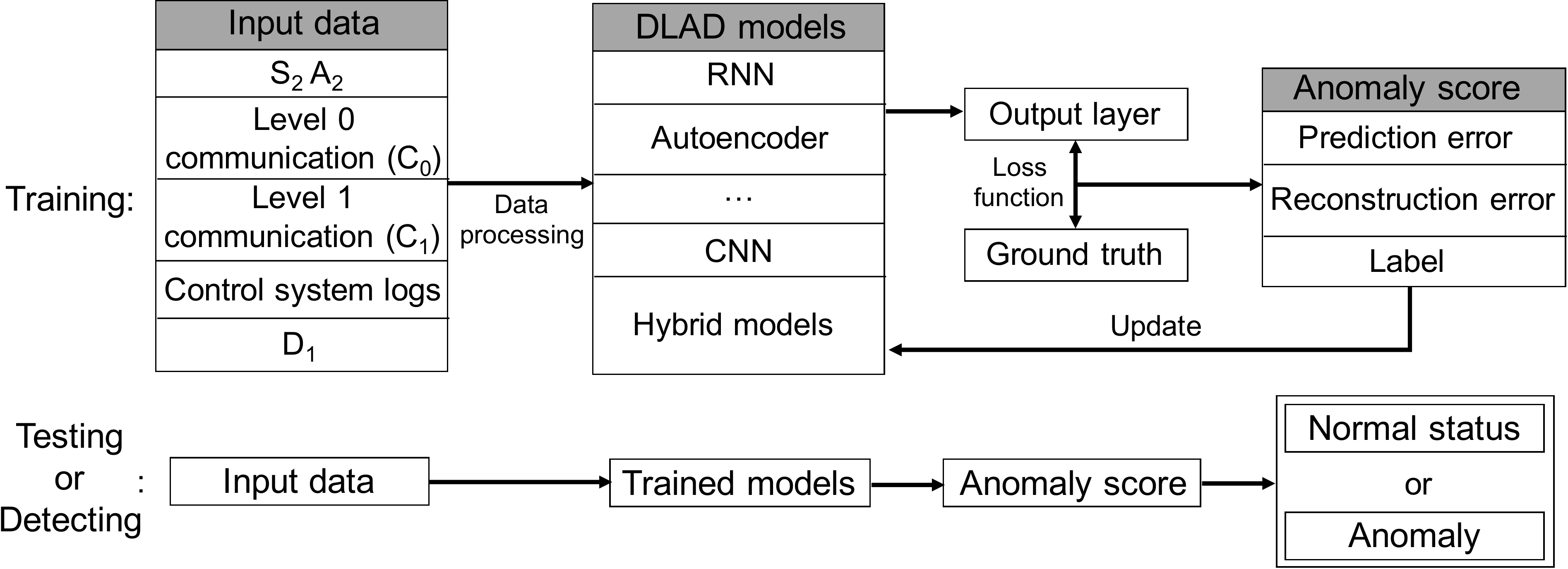} %
	\vspace{-10pt}
	\caption{The workflow of a typical DLAD method. The input data is used to train or test DLAD models. The anomaly score is used to optimize DLAD models. Trained DLAD models are applied to decide whether the input data is an anomaly at the online detection phase.}

	\label{DADarchitecture} 
\end{figure}

Anomaly detection has developed for many different applications~\cite{yao2017anomaly, chandola2009anomaly}, \eg, intrusion detection, fraud detection. In this work, we focus on new research efforts that detect anomalies in CPS with the help of emerging deep learning methods. As illustrated in Figure~\ref{DADarchitecture}, we characterize the generic workflow of DLAD methods. Typically, DLAD methods consist of training and testing phases. At the training phase, a large quantity of input data is first collected. Sensor and actuator data, level 0 and level 1 communication traffic, and control system logs are commonly used data sources. Various customized data processing approaches are applied to the input data, which is then fed to neural network models. Then, the main contribution of new methods lies in different DLAD models (\eg, RNN, autoencoders, CNN, and customized models) in different application scenarios. Further, DLAD models utilize loss functions to compute differences between output data from the output layer and ground truth data. We denote these differences as anomaly scores.  There are three types of anomaly scores: (1) Prediction error (2) Reconstruction error, and (3) Predicted labels (details in Section~\ref{detectionstrategy}). Anomaly scores are used to optimize and update DLAD models. At the testing phase, collected or real-time input data is fed to trained models and determine whether the input is an anomaly.

\begin{table}[h]
\caption{Summary of techniques, applications, and scope covered by related surveys. \Circle ~and \CIRCLE ~means NO and YES respectively. \LEFTcircle ~means related but not fully covered. \enquote{DL} and \enquote{AD} denotes deep learning and anomaly detection respectively.}

\resizebox{14cm}{!}{
\begin{tabular}{|c|c|c|c|c|c|c|}
\hline
 {\cellcolor{lightgray}   \begin{tabular}[c]{@{}c@{}} \textbf{Related} \\ \textbf{work} \end{tabular} } & \cellcolor{lightgray}{\textbf{Techniques}}                                                                                                               & \cellcolor{lightgray}{\textbf{DL?}} & \cellcolor{lightgray}{\textbf{Application}}                                                                                                     & \cellcolor{lightgray}{\textbf{CPS?}} & \cellcolor{lightgray}{\textbf{Scope}   }                                                                         & \cellcolor{lightgray}{ \textbf{AD?} }\\ \hline \hline
Chandola~\etal~\cite{chandola2009anomaly}   & \multicolumn{1}{l|}{\begin{tabular}[c]{@{}l@{}}Classiflcation-Based, Clustering-Based, \\ Statistical, \etc anomaly detection methods\end{tabular}} & \Circle   & \multicolumn{1}{l|}{\begin{tabular}[c]{@{}l@{}}Cyber Intrusion Detection,\\ Fraud Detection, \etc\end{tabular}} & \Circle     & \multicolumn{1}{c|}{\begin{tabular}[c]{@{}l@{}}Anomaly detection\end{tabular}} & \CIRCLE   \\ \hline
  Celik~\etal~\cite{celik2019program}         &   Program Analysis       &   \Circle   &   Commodity IoT    &  \LEFTcircle    &  App security and privacy   &  \Circle    \\ \hline
   Giraldo~\etal~\cite{giraldo2018survey}         &   Physical properties      &   \Circle   &   CPS    &   \CIRCLE   &  Anomaly detection   &   \CIRCLE    \\ \hline
     Chalapathy~\etal~\cite{chalapathy2019deep}       &   Deep learning      &   \CIRCLE    &  \multicolumn{1}{l|}{\begin{tabular}[c]{@{}l@{}}Cyber Intrusion Detection,\\ Fraud Detection, \etc \end{tabular}}   &   \Circle   &  Anomaly detection   &   \CIRCLE    \\ \hline
 Cherdantseva~\etal~\cite{cherdantseva2016review} &   Attack tree, model-based     &  \Circle    &  \multicolumn{1}{c|}{\begin{tabular}[c]{@{}l@{}} CPS (focus on SCADA) \end{tabular}}   &  \CIRCLE   & Cyber risk assessment  &    \LEFTcircle  \\ \hline
 Veith~\etal~\cite{veith2019analyzing}   &   Deep learning     &  \CIRCLE   &  \multicolumn{1}{c|}{\begin{tabular}[c]{@{}l@{}} CPS \end{tabular}}   &  \CIRCLE   &  \multicolumn{1}{c|}{\begin{tabular}[c]{@{}l@{}}Analyzing applications \\ of DL in CPS  \end{tabular}} &     \Circle  \\ \hline
 Mitchell~\etal~\cite{mitchell2014survey}   &  \multicolumn{1}{c|}{\begin{tabular}[c]{@{}l@{}}  Knowledge-Based, Behavior-Based \\ Intrusion Detection system  \end{tabular}}  &   \Circle   &  \multicolumn{1}{c|}{\begin{tabular}[c]{@{}l@{}} CPS \end{tabular}}   &  \CIRCLE   &  \multicolumn{1}{c|}{\begin{tabular}[c]{@{}l@{}} Anomaly detection \end{tabular}} &    \CIRCLE  \\ \hline
 Nazir~\etal~\cite{nazir2017assessing} &  \multicolumn{1}{c|}{\begin{tabular}[c]{@{}l@{}}  Intrusion Detection system, \\ machine learning, honey pots  \end{tabular}}  &   \Circle   &  \multicolumn{1}{c|}{\begin{tabular}[c]{@{}l@{}} CPS \end{tabular}}   &  \CIRCLE   &  \multicolumn{1}{c|}{\begin{tabular}[c]{@{}l@{}} Cyber security \end{tabular}} &   \LEFTcircle  \\ \hline
  Heartfield~\etal~\cite{heartfield2018taxonomy} &  \multicolumn{1}{c|}{\begin{tabular}[c]{@{}l@{}}  -  \end{tabular}}  &   -  &  \multicolumn{1}{c|}{\begin{tabular}[c]{@{}l@{}} Smart home IoT \end{tabular}}   &  \LEFTcircle   &  \multicolumn{1}{c|}{\begin{tabular}[c]{@{}l@{}} Taxonomy of threats,\\ not detection methods \end{tabular}} &   \Circle\\ \hline
   Lun~\etal~\cite{lun2019state} &  \multicolumn{1}{c|}{\begin{tabular}[c]{@{}l@{}} Plant models, noise-based detection, \\ state estimation, \etc \end{tabular}}  &  \Circle   &  \multicolumn{1}{c|}{\begin{tabular}[c]{@{}l@{}} CPS \end{tabular}}   &   \CIRCLE   &  \multicolumn{1}{c|}{\begin{tabular}[c]{@{}l@{}} Anomaly detection \end{tabular}} &   \CIRCLE \\ \hline    
Mohammadi~\etal~\cite{mohammadi2018deep} &  \multicolumn{1}{c|}{\begin{tabular}[c]{@{}l@{}} Deep learning \end{tabular}}  &   \CIRCLE   &  \multicolumn{1}{c|}{\begin{tabular}[c]{@{}l@{}} IoT \end{tabular}}   &   \LEFTcircle  &  \multicolumn{1}{c|}{\begin{tabular}[c]{@{}l@{}} Data analytics \end{tabular}} &   \Circle \\ \hline    
Ours &  \multicolumn{1}{c|}{\begin{tabular}[c]{@{}l@{}} Deep learning \end{tabular}}  &   \CIRCLE   &  \multicolumn{1}{c|}{\begin{tabular}[c]{@{}l@{}} CPS \end{tabular}}   &   \CIRCLE  &  \multicolumn{1}{c|}{\begin{tabular}[c]{@{}l@{}} Anomaly detection \end{tabular}} &   \CIRCLE\\ \hline    
     
\end{tabular}
}

\label{relatedwork}
\end{table}
        
\subsection{\textsc{Related survey}}\label{relatedworksection}
There are a number of recent related surveys, which are different in focus and domain from our work. As illustrated in Table~\ref{relatedwork}, we summarize these papers in terms of techniques, applications, and scope. Chandola~\etal~provided a comprehensive overview of anomaly detection methods~\cite{chandola2009anomaly}. As an early effort to review anomaly detection methods, they did not consider deep learning-based methods and did not include CPS. Commodity IoT systems have transformed the way people live. For example, emerging smart home applications allow users to interact with home appliances automatically. Program analysis methods are proposed to protect the privacy and discover vulnerabilities in these applications~\cite{celik2019program}. Meanwhile, Giraldo~\etal ~reviewed anomaly detection methods that utilize the physical properties of CPSs (\eg, \textcolor{black}{the evolution of the physical system under control})~\cite{giraldo2018survey}. Studies in terms of network security of SCADA systems are summarized with a focus on risk assessment techniques~\cite{cherdantseva2016review}. Mitchell~\etal~\cite{mitchell2014survey}, Nazir~\etal~\cite{nazir2017assessing}, Lun~\etal~\cite{lun2019state} provided a review of anomaly detection approaches in CPS. But the techniques did not include deep learning methods and are conventional, \eg, state estimation, intrusion detection based methods. There is work that studied deep learning-based anomaly detection methods but did not focus on CPS~\cite{chalapathy2019deep}. While Veith~\etal ~investigated applications of deep learning methods in CPS, it did not cover anomaly detection~\cite{veith2019analyzing}. Heartfield~\etal ~examined the taxonomy of threats in smart home IoT, which did not consider anomaly detection methods~\cite{heartfield2018taxonomy}. Finally, Mohammadi~\etal ~studied data analysis approaches that use deep learning methods in IoT~\cite{mohammadi2018deep}. To the best of our knowledge, our work is the first work that studies deep learning-based anomaly detection methods in CPS, which differs from the above existing surveys.

\section{Taxonomy}

\begin{figure}[!h]
	\centering
	\includegraphics[width=0.9\textwidth]{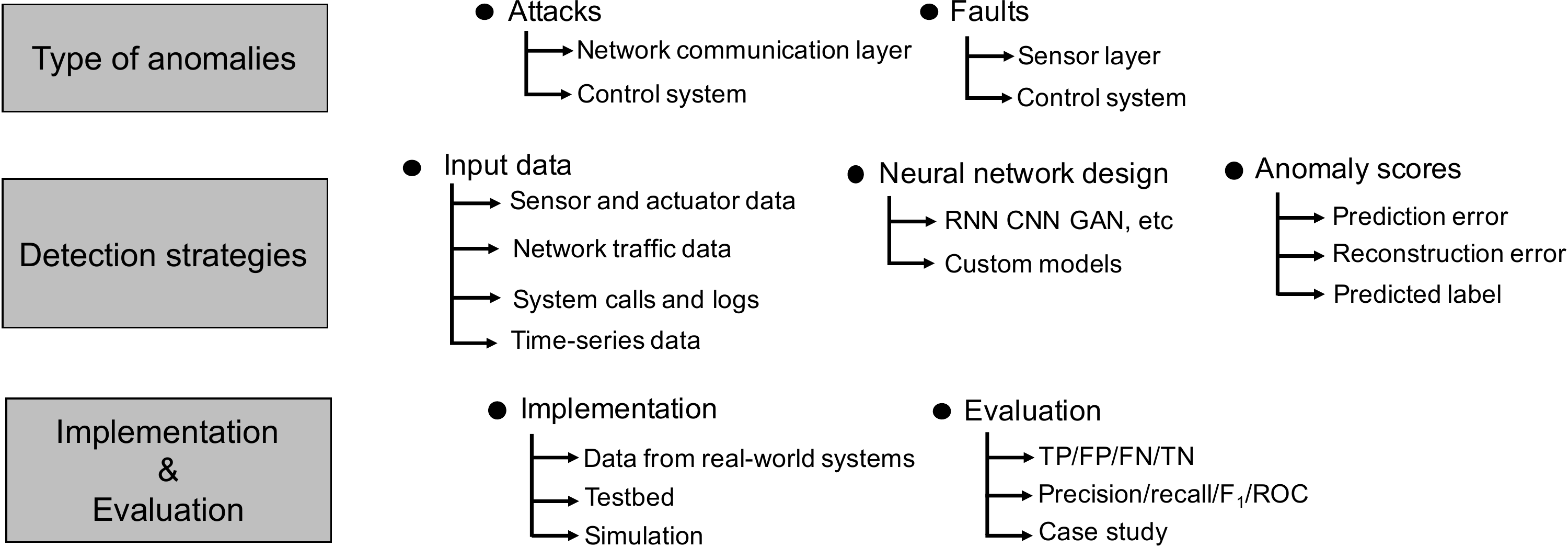} %
	\vspace{-3pt}
	\caption{Taxonomy of deep learning-based anomaly detection methods in cyber-physical systems.}
	\label{taxonomyfigure} 
\end{figure}

In this section, we present our taxonomy to classify existing work. In particular, our taxonomy consists of three aspects: (1) Type of anomalies. DLAD methods first need to decide what type of anomalies they intend to detect. (2) Detection strategies. Based on different anomalies, different strategies (\eg, neural network design) are adopted. (3) Implementation and evaluation metrics. Once a strategy is decided, appropriate implementation and evaluation metrics are selected to assess the performance of methods. Our taxonomy is depicted in Figure~\ref{taxonomyfigure} and we elaborate the details as follows.

\subsection{\textsc{Type of anomalies}}

We elaborate anomalies described in Section~\ref{cpsandthreat}. Anomalies can be broadly categorized as: (1) attacks; (2) faults.

\noindent \textbf{Attacks.} Since CPSs usually manage critical infrastructure (\eg, ICS, medical devices, and power grid), they are always under the threat of various attacks. An attacker who has the motive (\eg, financial interest, privacy theft, and state operations) can conduct attacks. These attacks can target different parts of CPSs:

\begin{enumerate}

\item \textit{Network communication layer.} Field devices (\eg, sensors and actuators) rely on communication networks to cooperate with each other. Also, sensor values, device status are reported to data centers and control commands are sent by control systems through the network. In this case, level 0 communication ($C_0$) and level 1 communication ($C_1$) can both be targeted. Note that $S_2$, $A_2$, $D_1$ (contained in $C_0$ and $C_1$ traffic) can also be manipulated under these attacks. We identify three types of attacks:

\begin{itemize}

\item \textit{Denial-of-service (DoS) attacks.} DoS attacks bring a significant threat to the functionalities of real-time applications in CPSs. For example, it would cause a collision of aircraft or low traffic utilization if the ADS-B system is out of service. Meanwhile, the broadcast feature in some CPS communication protocols (\eg, the CAN protocol in smart car systems) makes the network prone to DoS attacks. 

\item \textit{Man-in-the-middle (MITM) attacks.} CPSs adopt many newly designed protocols, which may lack a well-designed authentication mechanism. Also, Ethernet used in CPS can be exploited to conduct MITM attacks. Packet content may be manipulated and sensitive information can be leaked through MITM attacks~\cite{feng2017multi}.

\item \textit{Packet injection.} If attackers gain access to the network, they are able to inject an arbitrary packet to send control command into the system. False control commands can cause severe damage to running devices and even place human lives under risk. For example, a false engine and brake control command could possibly induce a car crash~\cite{habler2018using}.

\end{itemize}

\item \textit{Control system.} As the core of one CPS, control systems take sensor values as input and give control signals to actuators or field devices. Due to harsh working environments or limited hardware resources, the protection mechanism may not well-established in control systems. Once control systems are compromised, data sent to SCADA systems ($D_1$) and commands sent to actuators ($A_2$) can be altered. We find two types of attacks that target control systems:

\begin{itemize}

\item \textit{Malware.} For the long-term monitoring and information leakage, attackers would place malware in the control system. Moreover, malware can be used to launch a stealthy attack (\eg, APT attack) at a certain critical moment. Sensor readings can be manipulated by malware. Under certain circumstances, malware may also cause physical damage to devices~\cite{stuxnet_2020}. 

\item \textit{False control signals.} Devices operate deviating from regular working status when receiving false control signals. Wrong operations shorten the working life of devices and can even damage devices directly. Attackers usually conceal their unauthorized access to the system and send false control commands at a critical time point.

\end{itemize}

\end{enumerate}

\noindent \textbf{Faults.} The complexity of systems and heterogeneity of devices lead CPSs to generate unexpected faults. For example, industrial control systems typically consist of multiple stages and a lot of components in each stage. Many devices operate in a harsh environment (\eg, high humidity or temperature). Also, mechanical parts are vulnerable to abrasion and vibration. $S_2$, $A_2$, and $D_1$ can all be anomalous due to faults. We find that faults typically happen in two layers:

\begin{enumerate}

\item \textit{Sensor layer.} False sensor value is a common fault in the sensor layer.  First, physical damage or flaw lead sensors to report inaccurate and even wrong sensor values. Also, previously unseen circumstances may cause sensors to work beyond their abilities. For example, sensors on spacecraft may come across unexpected conditions~\cite{hundman2018detecting, tariq2019detecting}.

\item \textit{Control system.} CPSs typically hold the dynamic running characteristic, which means there are always situations that may not be covered during the system design stage. For example, different orders and timings of events in the PLC code can cause object collisions of an assembly line in industrial plants~\cite{zhang2019towards}.

\end{enumerate}

\subsection{\textsc{Detection strategies}}\label{detectionstrategy}

DLAD methods choose their detection strategies from three aspects:

\noindent \textbf{Input data}. 
DLAD methods first need to decide what type of data to take as input, which depends on specific anomalies they tend to detect. Based on the layer and source where data is collected, we conclude four types of input data: (1)  Sensor and actuator data. (2) Network traffic data. (3) System calls and logs. (4) Time-series data, which is preprocessed sensor, network, and log data in numeric time-series form. DLAD methods adopt semi-supervised and unsupervised learning to resolve the lack of labeled data (especially anomalous data).

\noindent \textbf{Neural network design}. DLAD methods adopt different neural network designs based on input data and tasks. The deep network can be stacked models (\eg, LSTMs) or hybrid combinations of models (\eg, the combination of LSTM and CNN). Although neural network designs can be in various forms, we found several basic models used to build the neural network. (1) RNN. LSTM models (one type of RNN) are often used to capture characteristics of time-series data~\cite{hundman2018detecting}. (2) Autoencoder. Autoencoders are applied to handle imbalanced data and achieve unsupervised learning~\cite{schneider2018high}. (3) CNN. CNN models can capture correlations and context information of multivariate measurement data~\cite{canizo2019multi}.

\noindent \textbf{Anomaly scores}. There exist three metrics to calculate the detection error: (1) \textit{Prediction error.} DLAD methods take past data as input to predict future sensor or actuator values. Then, the error between predicted and real values is measured. Anomalous data usually deviate from predicted values. (2) \textit{Reconstruction error.} Input data is fed to the model and compressed to hidden layers, which represents low dimensional space. The data is then reconstructed to the size of the original dimension. Similarly, the error between reconstructed and origin values is calculated. A threshold of error is usually selected to identify anomalous data. (3) \textit{Predicted label or class}. If labeled data is relatively sufficient in some domain (\eg, SWaT~\cite{itrust} testbed in ICS), DLAD models can be trained to predict labels of input data. The assumption is that latent features learned from neural networks can be used to identify anomalies. We observe very few methods to adopt this design since a large quantity of labeled data needs profound manual effort.

\vspace{-5pt}
\subsection{\textsc{Implementation and evaluation metrics}}
We summarize the implementation of existing work with a focus on platforms where data is collected. Then, metrics that are used to evaluate the effectiveness and performance of DLAD methods are identified. 

\noindent \textbf{Implementation}. As data-driven techniques, DLAD methods consume a large quantity of data to train and test models. We summarize three types of environments where data is collected: (1) Data from real-world systems. (2) Testbed. Researchers build scaled-down yet entirely functional testbeds, where experiments can be done without the risk of damaging real CPSs. (3) Simulation. The advantage of data from real-world systems is that it reflects the intrinsic principle of real systems, although the data is hard to harvest and the number of systems is limited. Simulation is easy to operate but can not capture problems that only exist in real systems. A scaled-down testbed could balance the data distortion and operability. 

Similarly, anomalous data can be collected from real-world systems and manually created. There can be insufficient real-world anomalous data since anomalies are hard to harvest. For example, in smart cars and medical domain, anomalies in real devices may place human lives at risk. So existing studies tackle this problem by manually creating three kinds of anomalies: (1) Point anomaly. Through investigating anomalies that can possibly happen, several independent abnormal cases can be injected into the normal data series. For instance, Taylor \etal~\cite{taylor2016anomaly} and Russo \etal~\cite{ russo2018anomaly} injected several attack cases into the sequence of CAN bus packets. (2) Statistical anomaly. Anomalies that follow certain statistical patterns are injected into normal data as an abnormal period~\cite{zhang2019deep}. (3) Simulated attacks. Various attacks are simulated in the testbed, where sensor values and system logs can be easily collected. Zhang \etal~\cite{zhang2019cyber} created cyber attacks in transactive energy systems.

\noindent \textbf{Evaluation metrics}. Metrics are proposed to measure the effectiveness of DLAD methods. We conclude that the most commonly used metrics are precision, recall, and $F_1$ score. Given imbalanced datasets, these metrics consider false positives and false negatives, which are better than metrics such as accuracy. The precision is defined as $TP/(TP+FP)$, where $TP$ stands for True Positives and $FP$ means False Positives. The recall is defined as $TP/(TP+FN)$, where $FN$ denotes False Negatives. $F_1$ is defined as 2*Precision*Recall/(Precision+Recall). Also, the Receiver Operating Characteristic (ROC) curve is used to manage tradeoffs between $FP$ and $TP$. Meanwhile, methods are often compared with baseline methods to examine the improvement. Some error-based metrics are also applied to measure the prediction and reconstruction performance such as Mean Absolute Error (MAE) and Relative Errors (ReErr)~\cite{zohrevand2017deep, Salmanccs20}.




\vspace{-5pt}
\section{Review of deep learning-based anomaly detection methods}\label{summary}
In this section, we present novel ideas and our findings in each domain of CPSs. We identified that current research efforts mainly focus on four types of systems: (1) industrial control systems (ICSs); (2) smart grid; (3) intelligent transportation systems (ITSs); and (4) aerial systems. Also, we investigate general-purpose methods that analyze time-series data. We have summarized existing work under our taxonomy in Table~\ref{summarytable}. The metrics of the taxonomy are listed in the column while current methods that target different CPSs are presented in each row. \textcolor{black}{We also provide a list of public datasets used in DLAD methods\footnote{https://github.com/leonnewton/DLAD-Survey}.}


\begin{table}[]
\caption{Summary of existing work on deep learning-based anomaly detection in Cyber Physical Systems.  \enquote{\CIRCLE}, \enquote{-}, \enquote{\LEFTcircle} means \enquote{Yes}, \enquote{No}, \enquote{Does not clear but inferred to be Yes} respectively. 1: GAN 2: DBN}  
\label{summarytable}
\vspace{5pt}
\resizebox{14cm}{!}{
 \begin{tabular}{l|c|c|c|c|c|c|c|c|c|cV{5}c|c|c|c|c|c|c|c|c|c|c|cV{5}c|c|c|c|c|c}
\hline
                                                     &    &      \multicolumn{9}{cV{5}}{\textbf{Type of anomalies}} & \multicolumn{12}{cV{5}}{\textbf{Detection strategies}}   & \multicolumn{6}{c}{\textbf{Implementation} \& \textbf{Metrics}}       \\ \cline{3-29}
                                 \multirow{3}{*}{\rot{\textbf{CPS Systems}} }                              &       \multirow{3}{*}{\rot{\textbf{Existing work}} }                      &     \multicolumn{5}{c|}{\textbf{Attacks}}    & \multicolumn{3}{c|}{\textbf{Faults}}     &      & \multicolumn{4}{c|}{\textbf{Input data}}   & \multicolumn{5}{c|}{ \begin{tabular}[c]{@{}c@{}} \textbf{Neural network} \\\textbf{design} \end{tabular} } & \multicolumn{3}{cV{5}}{ \begin{tabular}[c]{@{}c@{}}  \textbf{Anomaly} \\\textbf{ score}   \end{tabular} }  &  \multicolumn{3}{c|}{\begin{tabular}[c]{@{}l@{}} \textbf{Implemen}\\\textbf{-tation}   \end{tabular}    }        & \multicolumn{3}{c}{\begin{tabular}[c]{@{}c@{}} \textbf{Evaluation}\\ \textbf{metric}\end{tabular}}  \\ \cline{3-29}
                                            &    & \rot{\textbf{DoS}} & \rot{\textbf{MITM}} & \rot{\begin{tabular}[c]{@{}l@{}}\textbf{Packet injection}\end{tabular}} & \rot{\textbf{Malware}} & \rot{\begin{tabular}[c]{@{}l@{}} \textbf{False control signals}\end{tabular}} & \rot{\begin{tabular}[c]{@{}l@{}} \textbf{Sensor layer}\end{tabular}} &    \rot{\textbf{Network layer}}         & \rot{\begin{tabular}[c]{@{}l@{}} \textbf{Control system}\end{tabular} }&      \rot{\textbf{Manually created}}                                                  & \rot{\textbf{Sensor data}} & \rot{\textbf{Network traffic}} & \rot{\begin{tabular}[c]{@{}l@{}}\textbf{System calls/logs}\end{tabular} } & \rot{\textbf{Time-series}} & \rot{\textbf{DNN}} & \rot{\textbf{RNN} }      & \rot{\textbf{Autoencoder}}       & \rot{\textbf{CNN}}  &  \rot{\textbf{RNN+Autoencoder}}  & \rot{\begin{tabular}[c]{@{}l@{}}\textbf{Prediction error}\end{tabular}} & \rot{\begin{tabular}[c]{@{}l@{}}\textbf{Reconstruction error}\end{tabular} } &  \rot{\begin{tabular}[c]{@{}l@{}}\textbf{Predicted label}\end{tabular}}& \rot{\textbf{Real-world}}     & \rot{\textbf{Testbed}} & \rot{\textbf{Simulation}} & \rot{\textbf{TP/FP/FN/TN}} & \rot{\begin{tabular}[c]{@{}l@{}}\textbf{Precision/recall/F1/ROC}\end{tabular} }    &  \rot{\textbf{Case study}}   \\ \hline \hline
\multirow{18}{*}{\rot{Industrial control systems}} & Schneider~\etal~\cite{schneider2018high}  & \CIRCLE   &   \CIRCLE    &        \LEFTcircle           &     -    &       -     &    -  & -  &      -     &       -     &    -         &    \CIRCLE              &                   -                                              &       -    & -  &     -      &       \CIRCLE            &     -   & - &      -      &    \CIRCLE      &    -     &    -      &     \CIRCLE      &    \CIRCLE          &     -        &              \CIRCLE                  &  - \\
                                             & Kravchik~\etal~\cite{kravchik2018detecting}  &  -   &   \CIRCLE    &   -    &    -     &  \CIRCLE           &       -  &  -  &      -       &        -           &     \CIRCLE          &          -       &                            -                                     &       -& -      &   -   &          -         &     \CIRCLE   & -   &              \CIRCLE                                                &                          -                                      &                                -                           &        -        &   \CIRCLE        &      -      &        -     &          \CIRCLE                                                      &  -    \\
                                             & Zohrevand~\etal~\cite{zohrevand2017deep}  &  -   &  -  & - &    -     &    -     &      -  & -  &      -       &     \LEFTcircle   &                -                  &      -       &        -         &      \CIRCLE  & -     &      -       &      -     &      -             &    \CIRCLE       &          -          &          -      &          -           &         \CIRCLE         &     -    &      -      &       -      &              -          &   -  \\
                                             & Su~\etal~\cite{su2019robust}  &  -  &  - &   -   &     -       &     -    &   \CIRCLE     & -    &       -    &                                 -                         &      \CIRCLE      &      \CIRCLE         &        \CIRCLE           &                                  -           &            -        &     \CIRCLE           &      -     &      -             &      -    &          -              &             \CIRCLE        &        -          &        \CIRCLE         &    -     &   -         &       -      &            \CIRCLE                                      &        -          \\
                                             & Eiteneuer~\etal~\cite{eiteneuerdimensionality}  & - &  - &   -   &         -        &    -     &     \CIRCLE   & -  &         -            &       \CIRCLE     &      \CIRCLE          &      -       &         -        &                                           -        &   -           &      -       &    \CIRCLE         &        -           &     -     &         -   &          \CIRCLE            &           -                                                &      \CIRCLE             &    \CIRCLE        &     -       &        -     &             \CIRCLE                                 &           -              \\
                                             & Goh~\etal~\cite{goh2017anomaly} &  -  & \CIRCLE  &  -    &      -           &   \CIRCLE       &         -       &  -      &       -        &           -        &        \CIRCLE                &     -        &            -     &              -     &     - &     \CIRCLE         &       -    &          -         &     -     &         \CIRCLE            &     -      &            -       &       -         &  \CIRCLE        &           - &      -       &                     -                                   &       \CIRCLE        \\
                                             & Feng~\etal~\cite{feng2017multi} & \CIRCLE    & - &  \CIRCLE      &       -        &    \CIRCLE       &          -      &     -    &            -       &               -         &     -                                                       &     \CIRCLE          &         -        &         -    & -  &     \CIRCLE          &      -     &           -        &     -     &      \CIRCLE                   &      -                                                          &             -          &        -        &   \CIRCLE       &     -       &       -      &        \CIRCLE               &          -        \\
                                             & Inoue~\etal~\cite{inoue2017anomaly} & - &  \CIRCLE    &   -   &              -                         &           \CIRCLE             &    -  & -  &               -       &        -                                                &        \CIRCLE                &          -   &      -       &         -        &             \CIRCLE           &       -      &      -     &         -          &     -     &                                      \CIRCLE                         &           -        &              -                &        -        &     \CIRCLE       &       -     &       -      &           \CIRCLE                                                  &     -     \\
                                             & Ferrari~\etal~\cite{ferrari2019performance} & -  & - &   -   &      -         &         -           &           -              &    \CIRCLE      &     -        &       -         &        -                                                  &          \CIRCLE        &      -       &      -           &               -        &    \CIRCLE          &     -      &        -           &    -      &      \CIRCLE                                                       &    -          &          -         &        -        &   \CIRCLE       &     -       &      -       &        -            &      \CIRCLE                  \\
                                             & Legrand~\etal~\cite{legrand2018study} &  - & - &  -    &     -           &      -          &        \CIRCLE        &    -     &       -      &          -        &    \CIRCLE                                                        &       -    &       -      &        -         &            -            &      -       &    \CIRCLE         &       -            &    -      &        -                                                    &        \CIRCLE          &             -              &       -         &   \CIRCLE        &    -        &     -        &     \CIRCLE         &      -                                  \\
                                             & Wu~\etal~\cite{wu2018weighted} & -  &-  &   -   &      -        &    -        &          \CIRCLE             &    -     &            -               &          -          &      \CIRCLE                                                     &         -       &       -      &        -        &          -      &      \CIRCLE        &     -      &      \CIRCLE              &    -      &     -                                                        &        -     &        \CIRCLE         &      \CIRCLE           &    -     &        -    &      -       &           \CIRCLE                                 &        -                 \\
                                             & Li~\etal~\cite{li2019deep} & - &  - &   -   &    -     &     -          &        \CIRCLE           &    -     &         -      &        \CIRCLE           &      \CIRCLE        &        -                                                    &      -       &        -         &           -            &      -       &       -    &          -         &     \CIRCLE       &         \CIRCLE                 &       -                                                         &          -            &        -        &     \CIRCLE      &   -         &      -      &       \CIRCLE                                      &       -                 \\
                                             & Lindemann~\etal~\cite{lindemann2019anomaly} & -  & - &   -   &         -      &   -        &     \CIRCLE     &      -   &        -             &       \CIRCLE          &       \CIRCLE                                                   &    -     &     -        &      -           &          -       &      -       &     -      &         -          &      \CIRCLE     &      -             &       \CIRCLE                                                          &         -             &   \CIRCLE              &    -     &       -     &     -        &      -            &   \CIRCLE                            \\
                                             & Canizo~\etal~\cite{canizo2019multi} & - &  -  &  -    &     -      &      -         &           \CIRCLE       &     -    &            -           &            \CIRCLE                &        \CIRCLE                                                      &             -          &       -      &        -         &             -            &       \CIRCLE          &     -      &      \CIRCLE                 &    -      &                                       -           &       -        &        \CIRCLE            &        \CIRCLE            &    -     &     -       &     -        &    \CIRCLE         &    -              \\
                                             & Khan~\etal~\cite{khan2019malware} & \CIRCLE & -  &  -    &   \CIRCLE      &    \LEFTcircle             &        -           &     -    &        -                &  \CIRCLE                                                      &         \CIRCLE          &           -        &      -       &        -         &            \CIRCLE                  &       -      &     -      &       -           &     -     &         \CIRCLE                                                        &        -          &     -          &       -         &     \CIRCLE         &     -       &       -      &         \CIRCLE                &    -                      \\
                                             & Xiao~\etal~\cite{xiao2017nipad} & - & -  &  -    &    \CIRCLE              &     -       &       -        &    -     &       -              &        \LEFTcircle          &         \CIRCLE                                                  &        -         &     -       &      -           &         -          &     \CIRCLE         &    -       &          -         &      -    &       \CIRCLE                                                      &              -           &           -          &       -        &     \CIRCLE     &      -      &      \CIRCLE          &         -            &          -            \\
                                             & Li~\etal~\cite{li2019mad} & -  &    \CIRCLE    &  -    &    -       &       \CIRCLE              &        -    &     -    &          -          &       -         &    \CIRCLE                                                &       -              &     -        &        -         &              1            &      -      &    -       &        -           &  -        &     \CIRCLE                                                         &            \CIRCLE       &           -           &        -        &      \CIRCLE     &        -    &     -       &  \CIRCLE     &    -   \\ \hline 
                           \multirow{9}{*}{\rot{Smart grid}}  & Tasfi~\etal~\cite{tasfi2017deep} &  -  & -  &   -   &     -      &     -     &      -      &  -       &     -           &  \CIRCLE    &    \CIRCLE       &       -                                                     &       -      &        -         &       -        &      -       &     \CIRCLE      &         -          &     -     &     -          &       \CIRCLE               &      \CIRCLE                                                     &       \CIRCLE         &    -     &     -       &   -          & -  & \CIRCLE \\ 
                             & Zhang~\etal~\cite{zhang2019cyber} &  \CIRCLE    &  \LEFTcircle   &   \CIRCLE     &      \CIRCLE       &       \CIRCLE               &     -      &     -    &       -           &        \CIRCLE        &            \CIRCLE             &    -                                                        &     -        &        -         &       -         &    -         &       \CIRCLE       &        -           &    -      &        -           &          \CIRCLE                                                         &        -    &        -        &   -      &     \CIRCLE          &       \LEFTcircle        & -  &  -\\ 
                              & Wang~\etal~\cite{wang2018distributed} &  -  & \CIRCLE  &     - &     -      &          -          &      -        &    -     &         -              &   \CIRCLE    &       \CIRCLE          &    -         &    -         &       -          &      -       &      -       &      \CIRCLE       &         -          &    -      &    -        &       \CIRCLE         &           -       &     -           &   -      &     \CIRCLE         &     -        &  \CIRCLE   & - \\ 
                               & Deng~\etal~\cite{deng2018false} &  -  &  \CIRCLE  &   -   &      -     &     -       &   -   &     -    &       -      &  \CIRCLE      &     \CIRCLE              &       -                                                     &     -        &         -        &        -         &     \CIRCLE         &     -      &         -          &    -      &   \CIRCLE                                                          &       -           &             -          &       -         &     -    &     \CIRCLE        &       -      &  \CIRCLE  & - \\ 
                                & Niu~\etal~\cite{niu2019dynamic} &  -  &  \CIRCLE &  -    &     -      &          -          &     -   &     -    &    -         &     \CIRCLE      &         \CIRCLE           &     \CIRCLE                                                        &     -        &      -           &       -        &        \CIRCLE    &     -      &      \CIRCLE    &     -      &     \CIRCLE                                                      &   -      &     -        &        -        &    -     &       \CIRCLE       &    -         &   \CIRCLE  & - \\ 
                                 & Wang~\etal~\cite{wang2019deep} &   - & \CIRCLE &   -   &      -     &           -         &    -        &   -      &     -        &     \CIRCLE     &  \CIRCLE                                                        &     -   &     -        &        -         &       2        &       -      &      -     &     -  &     -     &    \CIRCLE          &     -                                                           &        -      &      -          &   -      &    \CIRCLE        &     -        & -  &  \CIRCLE \\ 
                                  & Basumallik~\etal~\cite{basumallik2019packet} & -   &  \CIRCLE     &   -   &     -      &        -            &      -      &   -      &       -          &    \CIRCLE                                                  &      \CIRCLE        &     -             &     -        &       -          &        -       &     -        &   -        &       \CIRCLE               &     -     &   -                                                         &      -         &       \CIRCLE         &         -       &    -     &     \CIRCLE          &    -         &  \CIRCLE    & - \\ 
                                   & Fan~\etal~\cite{fan2018analytical} & -   & -   &  -    &    -      &    -   &       \CIRCLE         &      -   &        -         &       -         &        \CIRCLE                                                      &       -   &      -       &        -         &       -      &     -        &      \CIRCLE         &      -     &     -     &     -                                                       &       \CIRCLE       &       -         &         \CIRCLE           &   -      &      -      &      -       &   \CIRCLE    & - \\ 
                                    & Wang~\etal~\cite{wang2019wide} &  -  & \CIRCLE   &   -   &       -    &     -    &     -    &   -      &      -        &      \CIRCLE         &     \CIRCLE      &     -                                                       &      -       &         -        &      -               &   \CIRCLE          &    -       &        -           &   -       &      -                                                      &         -         &     \CIRCLE         &      -          &     -    &    \CIRCLE        &   \CIRCLE          &  - &  -\\ \hline
                             \multirow{7}{*}{\rot{ITS}}   & Khanapuri~\etal~\cite{khanapuri2019learning} &  -  & -  &   -   &     -     &    -      &      -      &     -    &   -      &  -     &    \CIRCLE                                                         &  -     &    -         &       -          &     -        &     -        &     -      &        \CIRCLE              &   -       &        -                                                    &      -         &      \CIRCLE        &       -         &    -     &      \CIRCLE         &     -        &  \CIRCLE     &   -          \\ 
                               & Wyk~\etal~\cite{van2019real} & -   &  \CIRCLE     &  -     &     -      &    -       &     \CIRCLE       &   -      &   -            &   \CIRCLE       &     \CIRCLE                                                       &     -      &       -      &       -          &      -         &      -       &      -     &     \CIRCLE                &    -      &      -                                                      &     -      &    \CIRCLE      &      \CIRCLE            &    -     &      -      &    -         & \CIRCLE     &  -     \\
                                & Taylor~\etal~\cite{taylor2016anomaly} &  -  & -  &   \CIRCLE     &      -         &        \CIRCLE                 &        -              &     -      &       -                                                          &      \CIRCLE      &      -      &      \CIRCLE       &     -        &       -          &      -     &       \CIRCLE        &    -       &       -            &   -       &                                     \CIRCLE                         &        -           &          -        &        \CIRCLE          &  -       &       -     &    -         &   \CIRCLE   &  -    \\
                                 & Russo~\etal~\cite{russo2018anomaly} &  \CIRCLE   &  -  &   \CIRCLE    &       -        &       -       &  -     &     -    &   -     &    \CIRCLE    &   -     &       \CIRCLE                                                      &      -       &         -        &   -        &      \CIRCLE        &     -      &      -             &    -      &    \CIRCLE                                                         &     -      &      -       &       \CIRCLE          &    -     &      -      &     \CIRCLE         & -   &   -           \\
                                  & Kieu~\etal~\cite{kieu2018outlier} & -   &   -  &   -   &       -        &       -        &      \LEFTcircle         &    -     &     -           &       -                                                 &      \CIRCLE     &       -     &      -       &        -         &   -   &      -       &    \CIRCLE         &       \CIRCLE              &    -      &  -                                                          &     \CIRCLE      &    -     &        \CIRCLE          &  -       &     -       &      -       &  \CIRCLE    &    -          \\
                                   & Zhu~\etal~\cite{zhu2019mobile} &   \CIRCLE    &   \CIRCLE     &    \CIRCLE     &     -          &        \CIRCLE    &          -            &    -     &     -                                                           &     \CIRCLE     &   -    &      \CIRCLE    &     -        &    -             &  -       &       \CIRCLE         &   -        &      -  &    -      &   -                                                         &     -     &           \CIRCLE               &       \CIRCLE            &   -      &    -        &    -         &    \CIRCLE   &   -  \\
                                    & Jichici~\etal~\cite{jichici2018examining} &  -  &    \CIRCLE     &    \CIRCLE      &      -         &       -      &      -  &   -      &     -                                                            &   \CIRCLE &       -        &     \CIRCLE        &      -       &       -          &       \CIRCLE        &     -        &     -      &     -       &    -      &    -                                                        &     -       &      \CIRCLE         &      -          &   -      &        \CIRCLE        &     \CIRCLE            &  -  &  -          \\ \hline
                          \multirow{7}{*}{\rot{Aerial systems} } & Hundman~\etal~\cite{hundman2018detecting} &  -  & -   &  -    &    -       &  -     &    \CIRCLE     &    -     &   -                                                              &   -  &    -    &    \CIRCLE      &  -  &   -     &    -    &      \CIRCLE          &      -     &      -     &     -     &   \CIRCLE      &    -                                                            &     -     &     \CIRCLE    &   -      &     -       &      -       &  \CIRCLE   &     -           \\ 
                                     & Tariq~\etal~\cite{tariq2019detecting} &  -  & - &   -   &  -   &   -    &    \CIRCLE    &   -      &  -     &  -     &     \CIRCLE       &   -     &   -  &  -   & -      &      \CIRCLE         &     -      &     -   &  -  &   \CIRCLE        &   \CIRCLE   & -     &      \CIRCLE            &    -     &      -      &      -       &   \CIRCLE   &   -       \\
         & Ezeme~\etal~\cite{ezeme2019dream} &  -  &  -     &   -   &   \LEFTcircle   &   \CIRCLE      &   -     &    -     &   -     &  -    &   -      &     -      &    \CIRCLE        &   -  &   -                                                              &    \CIRCLE          &      -     &    -    &    -      &  \CIRCLE         &     -      &     -    &      \CIRCLE           &   -      &      -      &     \CIRCLE         & -   &   - \\
         & Gunn~\etal~\cite{gunn2018anomaly} & -   & -  &   -   &  -   &  -  & -   &   \CIRCLE       &    -     &    -       &   -   &  \CIRCLE      &  -   &    - &  -   &  \CIRCLE     &  -  &  -   &  -  &   \CIRCLE                                                          &    -     &   -     &   \CIRCLE     &    -     &     -       &    -   &  -  &    -          \\
          & Nanduri~\etal~\cite{nanduri2016anomaly} &  -  & -   &   -   &  -    &   -    &    -    &    -     &    \CIRCLE     &    \CIRCLE   &  -   &   -  &   \CIRCLE  & -  & - &   \CIRCLE  & -  & - &  -  &  \CIRCLE                                                          &    - &  -  &    -   &   -  &    \CIRCLE        &     -        &  \CIRCLE  &  -        \\
         & Habler~\etal~\cite{habler2018using} &  -  &  -  &  -  &  -  &   -  &  -  & - &  -  &  \CIRCLE & -  &  \CIRCLE  &  -   &   -   & -   &  - &  - &    -  &     \CIRCLE      &  -                                                          &     \CIRCLE     &    -  &    \CIRCLE  &    -     &     -       &        \CIRCLE      &  -  &   -  \\
          & Ezeme~\etal~\cite{ezeme2019deepanom} &  -  & -   &   -   &   -   &    -    &    -   & -    &  -    &   \CIRCLE  &  -   &  -    &    \CIRCLE  &   -    &  - &    \CIRCLE    &  -  &   -  &  -   &     \CIRCLE                                                         &   -   &  -  &   -  &    -     &     \CIRCLE         &     -        &  -  & -    \\ \hline
                          
\end{tabular}
}
\end{table}


\subsection{\textsc{DLAD methods in ICSs}}\label{ics}

\begin{figure}[!h]
	\centering
	\includegraphics[width=0.6\textwidth]{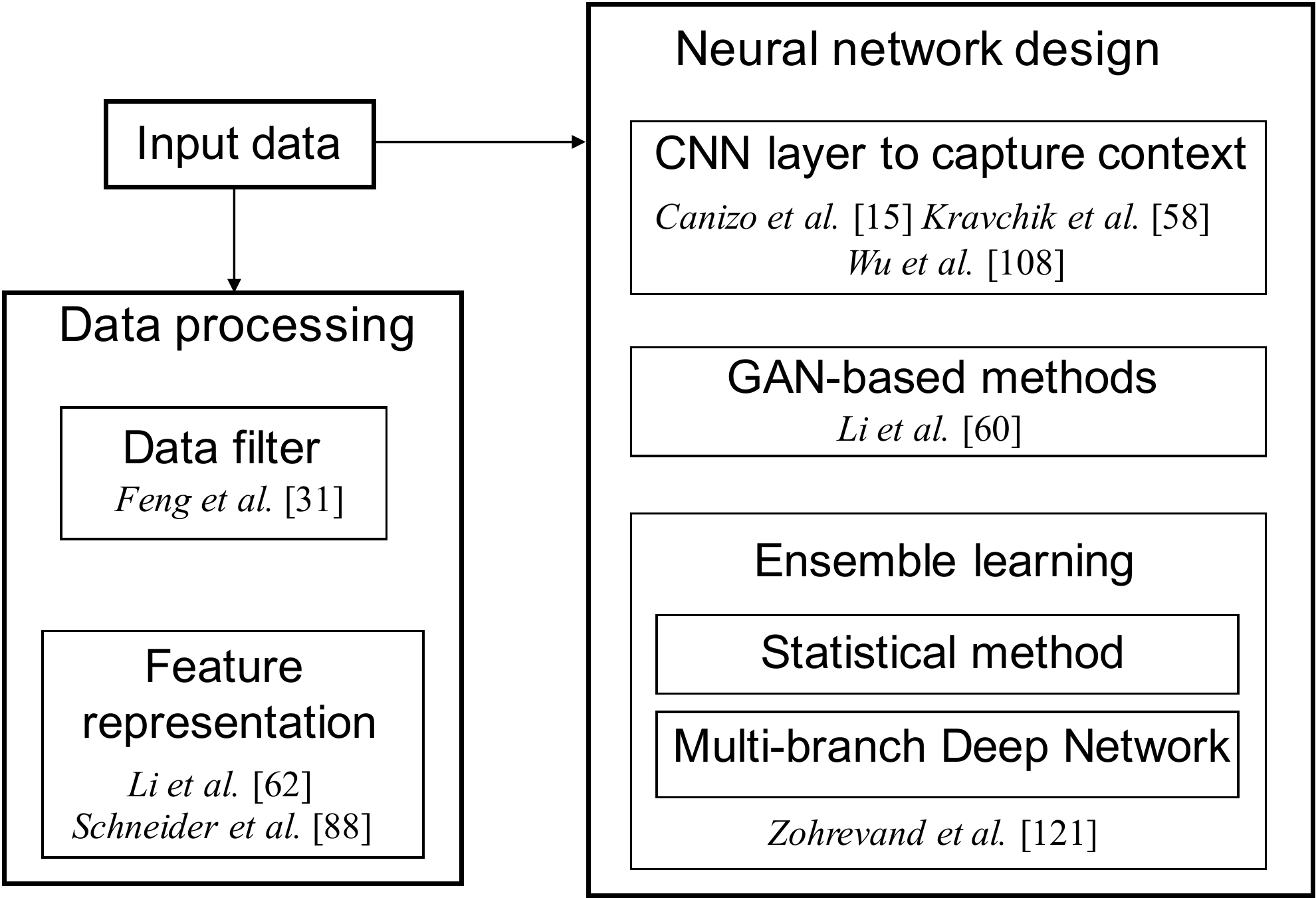} %
	\vspace{-3pt}
	\caption{An illustration of representative new techniques in ICS. These methods can also be applied to other domains.}
	\label{dladinics} 
\end{figure}

\textbf{Characteristics of DLAD methods in ICSs}. DLAD methods in ICS detect both attacks~\cite{schneider2018high, kravchik2018detecting, goh2017anomaly, feng2017multi, inoue2017anomaly, khan2019malware, xiao2017nipad, li2019mad} and faults~\cite{su2019robust, eiteneuerdimensionality, ferrari2019performance, legrand2018study, wu2018weighted, li2019deep, lindemann2019anomaly, canizo2019multi}. The attack types include injecting false control commands, altering communicating traffic packets, and spoofing sensor values. On the other hand, much of the research effort in ICS is on detecting faults, which have been less studied in other applications of CPSs. The complexity of infrastructures and the harsh working conditions of field devices can cause unexpected faults. The majority of existing work detects anomalies from sensor and actuator values, which are easy to be obtained. Only several studies handle network traffic data since there are inadequate real-world traffic data and proprietary communication protocols. Very few studies target control systems (\eg, system logs) and thus we did not find such a dataset in ICS. For neural network architectures, LSTMs and autoencoders (and their variations) are the most commonly used. Typically, LSTMs are used to capture the temporal relation of sensor values and unsupervised learning is achieved through autoencoders. Most solutions adopt the prediction error to measure the deviation of an anomaly. Testbeds are usually used to evaluate proposed methods and the SWaT testbed~\cite{goh2016dataset} is a popular platform to conduct the evaluation. Precision, recall, $F_1$, and ROC are de facto evaluation metrics.
In addition to the above characteristics, we also find some new techniques and explorations used by DLAD methods in ICS. As illustrated in Figure~\ref{dladinics}, in what follows, we discuss representative new techniques in ICS. Note that these methods can also be applied to other domains.

\subsubsection{Representative new techniques}

\noindent \textbf{Applying filters before DLAD methods to improve efficiency}. Applying DLAD methods in ICS, where running environments are usually resource-constrained, must consider the efficiency factor. A lightweight and efficient conventional detecting method could be utilized before DLAD methods to decrease data to be checked significantly. Feng \etal~\cite{feng2017multi} proposed a combined anomaly detection framework. The main idea is to first apply a Bloom filter to traffic data and then pick suspicious packets to the follow-up LSTM-based detector. The fast and lightweight filter reduces the burden of the LSTM detector, which enhances the detection efficiency. This method aims to identify cyber attacks in the communication layer of a SCADA system. The attack types include injecting malicious commands (\eg, state, parameter, and function code) and DoS attacks. Also, the LSTM detector stacks two LSTM layers using signatures of previous packets to predict the signature of the next packet. Then, the predicted signature is checked to examine whether it is in the normal signature database. The method is evaluated on a gas pipeline system in a laboratory environment, which outperforms baseline methods (\eg, Bayesian Network, Isolation Forest) in the recall, accuracy, and $F_1$ score. 

\textcolor{black}{ \textit{Open problems}. Computing resources of CPS devices are limited. Utilizing filters to boost detecting efficiency makes deploying DLAD methods in CPS more practical. However, there are two open questions in terms of how to design the filter. First, the interrelationship between the filter and DL models is not studied. Namely, the methodology to find a proper filter that works best with DLAD methods can be further investigated. Second, the authors found that if the filter can remove noise, the detection performance will be improved. Hence how to design a denoising filter can be studied.}

\noindent \textbf{Deep learning-based feature representation}. We identify three types of feature representation in DLAD methods: (1) raw data (directly fed to models) (2) data processing (\eg, inner products of two sensor time series) (3) deep learning-based embedding. Data processing helps to identify discriminative characteristics of data, which is also used in conventional detection methods. We find that deep learning methods are utilized to integrate features and reduce dimensions of feature space. For example, Li \etal~\cite{li2019deep} and Schneider \etal~\cite{schneider2018high} proposed deep autoencoders to automatically compress raw input to lower-dimension hidden layer representation, which further is utilized as the input of the follow-up neural network. Despite both works~\cite{li2019deep, schneider2018high} utilizing the hidden layer to represent features, the actual neural network detecting anomalies can be different. One~\cite{li2019deep} takes sensor value and uses LSTM to generate prediction errors, while the other~\cite{schneider2018high} takes traffic data and uses autoencoder to generate reconstruction errors. Both methods are evaluated on data from testbeds. When expert knowledge is limited (\eg, face a new network protocol), this can be very useful.

\textcolor{black}{ \textit{Open problems}. Using deep learning methods to automatically learn features is known as one type of transfer learning~\cite{wang2018great}. When applied to anomaly detection task, DL-based feature representation faces two challenges. First, the rules to choose the number of features and learning parameters are not clear. Currently, it is quite subjective to decide the features and parameters to be selected. Second, not all features are appropriate for the detection task. It is possible that some features fail to capture the essential characteristics of data.}

\noindent \textbf{Capturing temporal and spatial relationships with different architectures}. The value of one sensor or actuator is one-dimension data (\eg, time-series), many LSTM-based DLAD methods are proposed to learn temporal behaviors of the data. However, there exist correlations among several different sensors and actuators, which reflect logical relations in the control system. In other words, there are interdependent relationships among sensors and actuators. Hence one challenge is to capture context (temporal, spatial, and logical) features in multi-dimensional (time-series of multiple sensors and actuators) data. To this end, CNN can extract features of multi-dimensional data jointly via convolution operations. Several approaches~\cite{canizo2019multi, wu2018weighted, kravchik2018detecting} adopt a convolutional layer as the first layer of the neural network to obtain correlations of multiple sensors in a sliding time window. Further, the extracted features are fed to subsequent layers to generate output scores. These methods can be employed to detect both attacks and faults. All methods take sensor and actuator value as input and generate prediction error or predicted labels. Meanwhile, Canizo \etal~\cite{canizo2019multi} and Wu \etal~\cite{wu2018weighted} utilized RNN to take the output of the CNN layer and form the prediction layer. Moreover, both methods use datasets from real industrial plants. Precision, recall,  $F_1$, and ROC are evaluation metrics. 

\textcolor{black}{ \textit{Open problems}. CNN models are used to capture correlations of sensor readings. However, the input to CNN models is still manually designed. A clear guideline to create the structure of input data is needed. On the other hand, for time-series data, anomaly detection action is conducted based on a time window. If the length of input time-series data is too short (shorter than a time window), a suitable padding mechanism is needed to expand input data.} 

\noindent \textbf{Exploration of GAN-based methods}. Li \etal~\cite{li2019mad} proposed a GAN-based framework to capture the spatial-temporal correlation in the multi-dimension data. Both the generator and discriminator are utilized to detect anomalies by reconstruction and discrimination errors. Also, LSTM models are used to build the generator and discriminator. The framework takes sensor and actuator values as input and aims to detect false control signals. Compared to a GAN-based anomaly detection method~\cite{zenati2018efficient} that is not focused on ICS, this method finds that capturing temporal correlation is the key to improve performance. The method outperforms baseline methods (\eg, Principal component analysis, One-Class SVM, K-Nearest Neighbour, Feature Bagging) in precision, recall, and $F_1$. This is an interesting attempt to utilize GAN-based models. Also, a well-tuned generator can be used to produce training data.

\textcolor{black}{ \textit{Current limitations}. It is an interesting early effort to explore GAN-based DLAD methods. However, two questions remain unsolved. First, false positives are high. The precision of the SWaT dataset is 70\% and the WADI dataset is 53.75\%. The cause of false positives is not fully investigated and discussed. Second, the DL models for the generator and discriminator are empirically selected. Other models can be examined to find the best choice.}

\noindent \textbf{Applying conventional and DLAD methods parallelly through ensemble learning}. We have introduced that conventional methods can be used as filters before applying DLAD methods. However, to increase the accuracy, these two kinds of methods can be placed parallelly to learn the characteristics of input data. Zohrevand \etal~\cite{zohrevand2017deep} proposed a framework named MBPF that ensembles two components: (1) a statistical method named TBATS (Trigonometric Box-Cox transform, ARMA errors, Trend, and Seasonal components)~\cite{de2011forecasting}, and (2) Multi-branch Deep Network Component. First, seasonality evaluation and outlier elimination are applied to remove noise. Then, pre-processed data is fed to TBATS and deep learning models simultaneously to capture linear and sequential relations. Finally, a Multi-Layer Perceptron (MLP) takes the output of TBATS and deep learning models, which will vote between the two methods and predict the next value. The MBPF framework can analyze any time-series data. The Mean Absolute Error (MAE) and Root Mean Square (RMSE) are utilized to measure prediction errors. Evaluated on a real-world SCADA water supply system, the method outperforms baseline methods (\eg, Multilayer Perceptron, Stacked LSTM, Regularized LSTM) when measured by MAE, RMSE, Absolute deviation (AbsDev) and Relative Errors (ReErr).

\textcolor{black}{ \textit{Open problems}. The ensemble of parametric methods and DL-based methods makes DLAD methods more reliable and less sensitive to stochastic data. However, there are two open questions that MBPF can address in future work. First, MBPF prefers capturing seasonal patterns, which often exist in systems such as water management systems. For systems that without seasonality, the method may not work. Second, complex pre-processing techniques and multiple models may add computational costs to CPSs. Thus it will limit the application of MBPF.}



\subsection{\textsc{DLAD methods in smart grid}}

\begin{figure}[!h]
	\centering
	\includegraphics[width=0.75\textwidth]{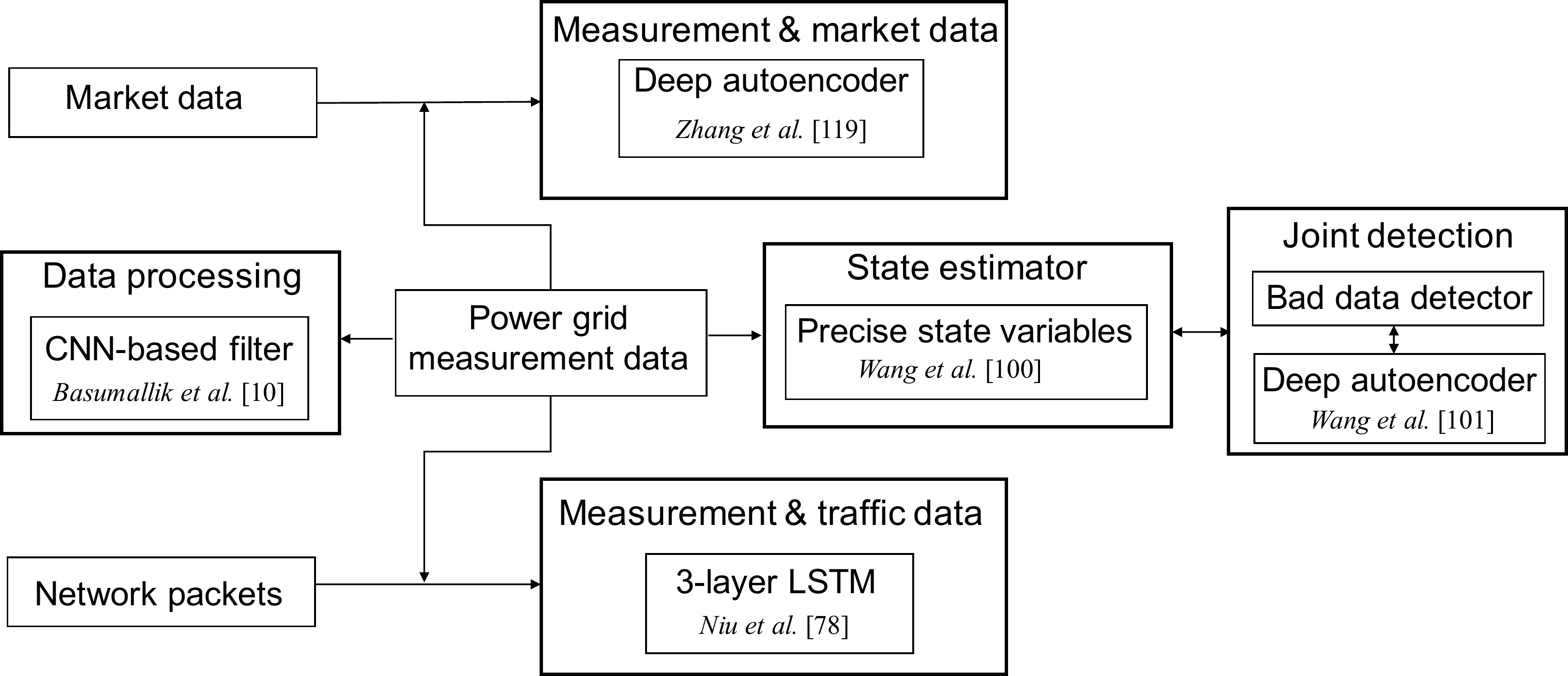} %
	\vspace{-3pt}
	\caption{An illustration of representative new techniques in smart grid.}
	\label{smartgrid} 
\end{figure}

\textbf{Characteristics of DLAD methods in smart grid}. False data injection (FDI) attacks~\cite{liu2011false} usually inject malicious packets (\eg, traffic of $C_0$, $C_1$) to create small measurement errors (\eg, alter $S_2$, $D_1$) to compromise the state estimation component of a smart grid. FDI attacks are stealthy and difficult to detect, which have attracted most of the research efforts~\cite{wang2018distributed, deng2018false, niu2019dynamic, wang2019deep, basumallik2019packet, wang2019wide}. Meanwhile, few studies detect faults~\cite{fan2018analytical} and injected anomalies~\cite{tasfi2017deep}. Although FDI attacks are accomplished via network packet injection, the majority of current work focuses on analyzing sensor data (\eg, voltage magnitude, power flow, electricity consumption). We find one work~\cite{niu2019dynamic} to analyze network packet data. We did not find work protecting control systems and datasets about system logs or traces in the smart grid published by DLAD methods. This is may partly because real control systems are hard to obtain. Autoencoders and RNNs (and their variations) are almost equally adopted architectures, which have been proven effective by existing works. So reconstruction and prediction errors are both used to detect anomalies. Simulations are mainly utilized to evaluate the performance of methods. Specifically, the IEEE X-bus~\cite{athay1979practical} (\eg, 9-bus, 14-bus) power test system is employed to simulate attacks and collect data. There are various evaluation metrics, \eg, precision, recall, $F_1$, and accuracy. As shown in Figure~\ref{smartgrid}, we present representative new techniques in smart grid.

\subsubsection{Representative new techniques}
\vspace{3pt}

\noindent \textbf{Deep learning aided state estimator}. In the smart grid, a state estimator is utilized to monitor the running state of the grid~\cite{moslehi2010reliability}, which is a key component to protect the power system. The input data of one state estimator is usually collected from SCADA systems, which obtain measurements from sensors and field devices. A bad data detector or filter~\cite{kosut2011malicious} is connected to the state estimator to eliminate false or injected data, which usually utilizes normalized residuals of measurements~\cite{huang2013bad}. However, attacks such as false data injection (FDI) and PMU data manipulation attack (PDMA) can evade the detection of conventional state estimators. These attacks deliberately mimic legitimate state variables and thus evade the detection. To thwart these attacks, several deep learning-based methods are proposed to improve state estimator, which adopt three strategies:

\begin{enumerate}

\item \textit{Remove false data before bad data detectors}. Basumallik \etal~\cite{basumallik2019packet} added a filter, which is based on deep learning techniques, to eliminate false data, which then could transfer sanitized data to the bad data detector. This filter contains two convolutional layers and takes voltage values as input. The output is the probabilities of various attacks (\eg, FDI attack). If attacks are detected, the false data is removed to protect the state estimator.

\item \textit{Improve bad data detectors via joint detection}. Wang \etal~\cite{wang2018distributed} utilized a deep autoencoder with RBM layers to form a joint detection framework with the bad data detector. The input of the autoencoder is extracted 108 features from PMU measurements, \eg, the three-phase magnitude, angles, and voltages. If the reconstruction error is above a pre-defined threshold, then attacks are detected from the raw data. Only attacks that are identified by both the autoencoder and bad data detector will be flagged as alerts in the management system, which significantly reduces false positives of conventional bad data detector.

\item \textit{Improve state estimators via predicting precise state variables}. Wang \etal~\cite{wang2019deep} proposed a DBN network with ten hidden layers to take generator and market time-series information as input and predict electric load in real-time. The predicted electric load intervals are the normal range of state variables. The method pinpoints precise state variables, thus attacks that cause abnormal states are detected.

\end{enumerate}

\textcolor{black}{ \textit{Current limitations}. The major threat to the smart grid is false data injection (FDI) attacks. Improving bad data detectors can thwart FDI attacks. However, current DL-aided detectors have limitations. First, the sampling rate of PMU can be at a pretty high frequency. Using all the data as an input needs massive computing resources. How to keep essential historical data and filter out irrelevant data is a relevant challenge to DLAD methods. Second, current bad data detectors are mainly evaluated on synthetic data, which are usually generated by simulation. The impact of real-world FDI attacks is not studied.}

\noindent \textbf{Combining characteristics from sensors and network layers}. Most existing studies adopt the threat model that a limited number of data points (\ie, point anomalies) are manipulated by FDI attacks. Niu \etal~\cite{niu2019dynamic} indicated that sophisticated attackers can inject multivariate malicious data points in a period (\ie, collective and contextual anomalies). Since such FDI attacks are more stealthy, inspecting measurement data alone may fail to detect such stealthy attacks. They proposed a mixed neural network architecture that combines sensor measurements and network packets. First, the one-dimension convolutional layer is utilized to extract features from the source data. Originally, raw data of the two sources are in different dimensions, which is further transformed into the same dimension by the convolution operation. Then, the features of two sources from past values are fully connected and fed to a 3-layer LSTM network to predict the next data point. Data points that generate large prediction errors are classified as anomalies. The method is evaluated on an IEEE 39-bus system. Overall, the accuracy of the method is above 0.8.

\textcolor{black}{  \textit{Open problems}. Utilizing features from both sensors and network traffic can help to learn system states precisely. But two issues need further investigation. First, two types of features are now directly connected using a fully connected layer. Methods that not only integrate all features but also provide interpretability have not been explored yet. Second, contributions of different features to the detection performance are not compared. The performance of one type of feature (\eg, using only sensor features) has not been explored.}

\noindent \textbf{Detecting anomalies both in the market and physical system}. Most existing methods concentrate on ensuring the stability of the running status of physical systems in the smart grid. However, considering merely sensor measurement and traffic packets data may fail to secure the robustness. In modern transactive power systems, the market plays an important role in adjusting the state of the system. Specifically, the electricity price and consumption also impact the grid by affecting the workload. Indeed, FDI attacks have already targeted markets~\cite{wang2013cyber, xie2010false}. We believe that it is closer to reality to consider cyberattacks in the market utilities and networks. Zhang \etal~\cite{zhang2019cyber} studied measurements of both the electricity market and the physical system. In particular, price, voltage magnitude, and power consumption are monitored. The proposed framework utilizes a stacked autoencoder and generates reconstruction errors of the market and physical system separately. If anomalies are detected in the market, network traffic and server logs are checked to locate the error. The framework is evaluated on a 9 bus bulk system modeled in MATPOWER~\cite{matpower}. Results show that 79\% of outages and 96.9\% of attacks can be detected.

\textcolor{black}{  \textit{Current limitations}. Bid price and bid quantity (market information) are two important factors that distinguish the smart grid from traditional electricity systems. These factors are the context of electricity consumption. However, this also indicates that these two features can be impacted by economic booms and busts. DLAD methods that use market information need to update models frequently so as to capture the characteristics of changed market information. Another limitation is that current detection schemes focus on securing the status of the whole grid. The specific compromised component is not identified.}



\subsection{\textsc{DLAD methods in ITSs}}

\begin{figure}[!h]
	\centering
	\includegraphics[width=0.75\textwidth]{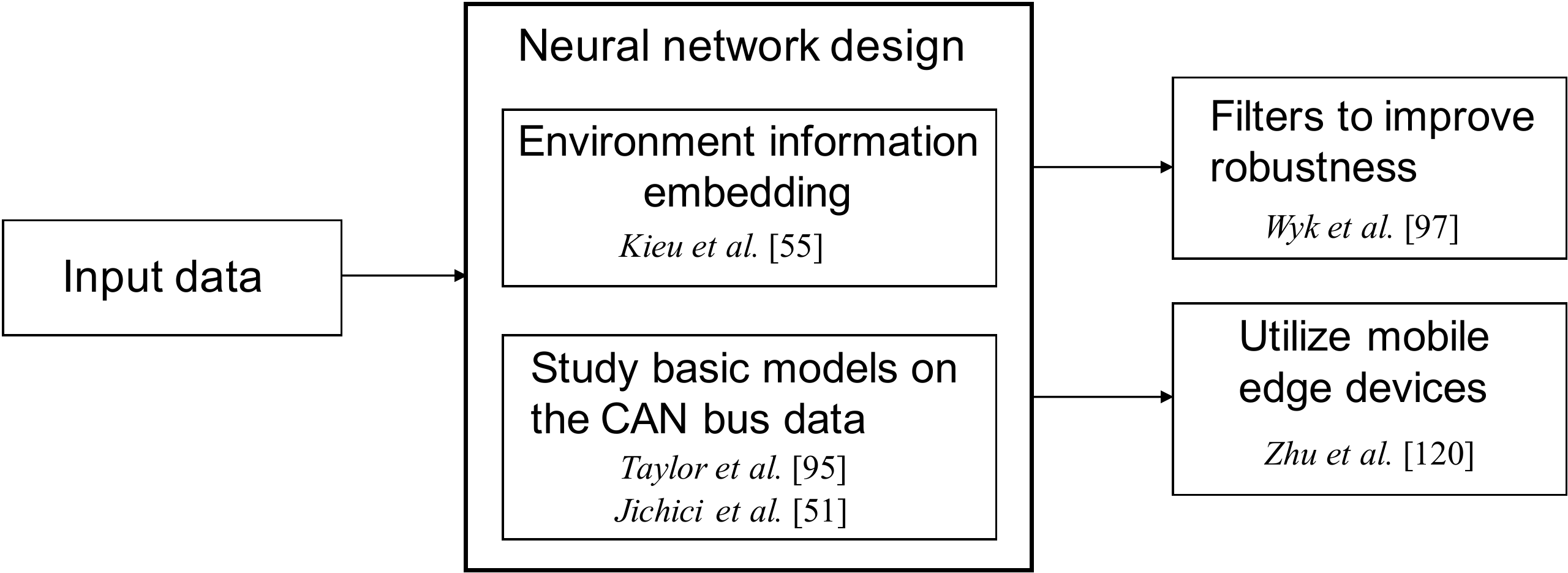} %
	\vspace{-3pt}
	\caption{An illustration of representative new techniques in ITS.}
	\label{itsfigure} 
\end{figure}

\textbf{Characteristics of DLAD methods in ITSs}. Most studies in ITS aim to detect attacks on the CAN bus~\cite{taylor2016anomaly, russo2018anomaly, zhu2019mobile, jichici2018examining}, which is responsible for the communication between devices (\eg, airbags) and Electronic Control Units (ECUs)~\cite{cho2016fingerprinting}. Khanapuri \etal~\cite{khanapuri2019learning} targeted vehicle platoons to avoid collisions among a sequence of cars. Kieu \etal~\cite{kieu2018outlier} studied aggressive manners of drivers while Wyk \etal~\cite{van2019real} also considered anomalies caused by faulty sensor readings. Attacks on the CAN bus include traffic drop, traffic sequence in reverse order, competing commands from two sources, false packet injection, and traffic replay attack, \etc. Given that most research efforts analyze CAN bus network data, sensor data from LIDAR, RADAR, GPS speed, acceleration sensor, \etc, are also utilized. Few works directly analyze control systems. For network architectures, there are no obvious dominant neural networks. Typically, LSTM models are used to capture temporal relations and CNN models are utilized to learn context respectively. Most methods generate prediction errors to detect anomalies while this work~\cite{kieu2018outlier} uses the reconstruction error. Most CAN-bus datasets are obtained from real-world vehicles. Precision, recall, accuracy, false positives, $F_1$, ROC are typically used to measure the performance. We present representative new techniques in ITS as summarized in Figure~\ref{itsfigure}.

\subsubsection{Representative new techniques}
\vspace{3pt}

\noindent \textbf{The embedding of contextual information}. Smart vehicles interact with the surrounding environment constantly. Cameras, radars, speed sensors are utilized to obtain the position, velocity, status of on-going vehicles. Existing studies use data from the above sensors and devices to ensure that vehicles perform in normal behaviors. However, the influence of environments is not captured if DLAD methods merely detect the condition of vehicles. Indeed, the environment information (\ie, context) is also important to decide the status of vehicles. For example, the same physical status can be classified as normal or anomalous depending on different weather, road, and traffic information. Kieu \etal~\cite{kieu2018outlier} utilized an embedding method to encode context information into matrixes. Further, context embedding matrixes are concatenated with feature-enriched time-series matrixes. Such enriched features contribute to higher precision and recall. This work aims to detect anomalies in time-series data and validates it on a driving behavior dataset. Thus, it can be used to identify reckless driving. The concatenated matrixes are fed to 2D CNN and LSTM autoencoders, which produce reconstruction errors to recognize outliers. The method outperforms two baseline methods (\ie, Local Outlier Factor~\cite{breunig2000lof}, One-Class
Support Vector Machines~\cite{manevitz2001one}) in precision, recall, and $F_1$ score.

\textcolor{black}{  \textit{Current limitations}. We compare the performance of the advanced model (with contextual information) and the basic model (without contextual information). We find that the advanced model improves the recall but the precision is unchanged. This shows that contextual information contributes to reducing false negatives. The advanced model can identify more anomalies than the basic model. However, if the model incorrectly classifies normal data as anomalies, adding contextual information cannot improve the model.}

\noindent \textbf{Utilizing mobile edge devices to boost computing}. Control commands are sent from ECUs to physical devices and mechanical parts of vehicles. With all these traffic transmitted on the CAN bus, a short delay of messages could cause severe casualties when users respond to sudden incidents. Meanwhile, DLAD methods typically consume a large number of computing resources. Restricted computing power on vehicles could add delay to send out benign commands when conducting the anomaly detection process. To this end, Zhu \etal~\cite{zhu2019mobile} proposed the multi-dimension LSTM framework to allow the parallel computing of certain LSTM layers, which can speed up the computing process. Also, part of the computing is delegated to mobile edge devices. In particular, two hidden layers are adopted to adjust the dimensions of input data, which are located at on-board computers. Further, data-based and time-based features are fed separately and simultaneously to two LSTM layers on edge devices. This work targets spoof, replay, flood, drop and tamper attacks to CAN bus messages. The cross entropy of the predicted message and the next message is calculated to detect the anomaly. With the accuracy reaching 90\%, the detection only takes about 0.61 milliseconds. 

\textcolor{black}{  \textit{Current limitations}. Mobile edge devices and parallel computing significantly reduce the computing time, which is 100 times faster than OBU as illustrated in this work. However, the communication costs between edge devices and control systems are not measured. DLAD methods have to balance computing and communication costs. Another issue is that parallel computing and its communication channel may also be vulnerable. Attackers may attack the communication network among edge devices instead of ITS.}

\noindent \textbf{Applying filters after DLAD methods to improve robustness}. DLAD methods are used to remove anomalous measurements so that control systems can generate correct responses to environmental changes. Thus, DLAD methods on ITS systems must be robust and work in real-time. To achieve robustness, Wyk \etal~\cite{van2019real} adopted a mixed framework. This work applies a three-layer CNN-based model first to eliminate false sensor readings. Then, scrutinized data is fed to Kalman filters (KF) to further remove anomalies that are undetected by the CNN model. The authors find that the CNN-KF model surpasses the KF-CNN model in general. Also, they observe that deploying a Kalman filter as the last layer makes the detecting process more reliable~\cite{schmidhuber2015deep}. This work aims to detect and remove false sensor readings caused by both false injection attacks and failures. The sensors include speed, acceleration, and GPS speed sensors. The CNN model consists of three CNN layers and two fully connected layers. Benign sensor readings are transferred to the control system from the CNN-KF model. The method is validated on a two-year real-world dataset obtained from the Safety Pilot Model Deployment (SPMD) program~\cite{bezzina2014safety}. Accuracy, precision, and $F_1$ are used to measure the performance, which outperforms two baseline (\ie, KF, CNN) methods.

\textcolor{black}{ \textit{Current limitations}. Overall, the $F_1$ score of the CNN-KF model is about 2\% higher than the CNN model. Hence if the computing resources are sufficient, the CNN-KF model can be adopted. If computing resources on the vehicle are limited, the CNN model may be used. This method does not distinguish between attacks and faults. Thus the root cause may not be identified. Also, since the experiments are conducted on synthetic datasets, the performance in real-world sensor networks are not measured.}

\noindent \textbf{Studying the performance of basic models on the CAN bus data}. As an important part of the communication system, the CAN bus has attracted most of the research efforts as we have shown in this section. With various heterogeneous neural network models introduced, the performance of basic neural networks is not clear. Serving as building blocks of sophisticated models, these basic architectures of neural networks have to be fully explored to better build and tune complex models. To this end, Taylor \etal~\cite{taylor2016anomaly} investigated the performance of a two-layer LSTM (with two hidden layers) model on different types of anomalies. Five types of anomalies (\eg, packet drop) are adopted to simulate attacks. Fifty million of traffic packets are captured from real-world vehicles as training and test dataset. The area under curve (AUC) is measured under different loss functions (\eg, maximum bit loss). Meanwhile, Jichici \etal~\cite{jichici2018examining} evaluated the performance of a three-layer DNN with different settings of training, validation, and testing proportion of datasets. The parameters of the gradient, epochs and Mean Squared Error (MSE) are reported. The replay of traffic frames and the injection of data attacks are used to simulate the anomaly. True negatives, false positives, true positives, and false negatives are calculated on a real-world dataset. Results show that basic models can achieve high true positives and low false positives. 

\textcolor{black}{ \textit{Current limitations}. CAN bus is the core of the control system in a smart car. Despite impressive detection performance, basic DL models also have problems. First, the false positives of basic models are relatively high. The authors reported that the false positive rate can be between 2\% to 10\%. It is not suitable to deploy such DLAD methods in practical smart car systems. Second, current detection is based on individual ECU components. The correlations among ECUs have not been studied, which can be used to reduce false positives.} 


\subsection{\textsc{DLAD methods in aerial systems}}

\begin{figure}[!h]
	\centering
	\includegraphics[width=0.8\textwidth]{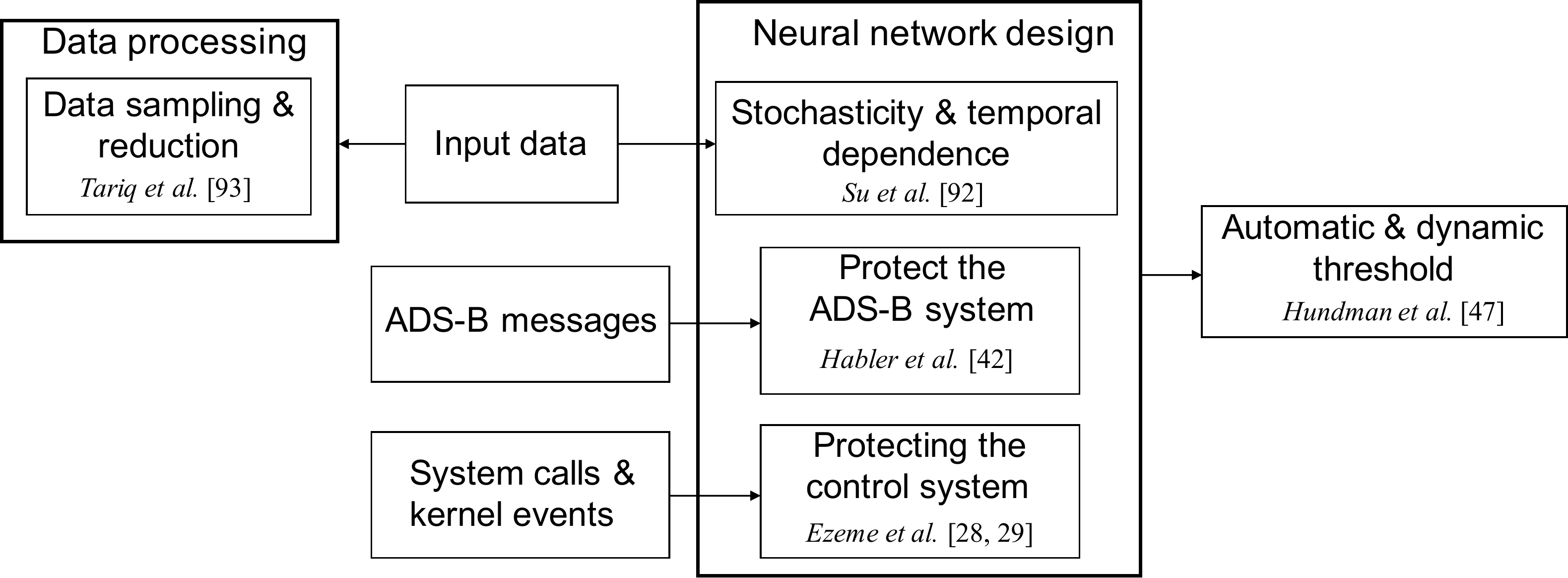} %
	\vspace{-3pt}
	\caption{An illustration of representative new techniques in aerial systems.}
	\label{aerialfigure} 
\end{figure}

\textbf{Characteristics of DLAD methods in aerial systems}. There are methods studying faults in aircraft~\cite{nanduri2016anomaly} and spacecraft~\cite{hundman2018detecting, tariq2019detecting, gunn2018anomaly}. The faults consist of point and contextual anomalies in sensor and communication data. Some research efforts are on attacks in unmanned aerial vehicles (UAVs)~\cite{ezeme2019dream, ezeme2019deepanom} and aircraft~\cite{habler2018using}. The attacks include malicious code in control systems, eavesdropping, and spoofing in the communication network, \etc. With network and sensor data as conventional input data, two studies~\cite{ezeme2019dream, ezeme2019deepanom} investigate attacks to control systems and utilize kernel events and logs as input. Most approaches use LSTMs and variants to generate prediction errors. Most aircraft and spacecraft data are collected from real airplanes and satellites. Although running data is obtained from real UAVs, attacks are simulated and injected into normal traces. It is hard to find a commonly used platform in aerial systems. Precision, recall, $F_1$, true positives and false positives are calculated to measure the performance. As shown in Figure~\ref{aerialfigure}, we present the details of representative techniques in aerial systems as follows. We argue that these methods can be used in other domains as well.

\subsubsection{Representative new techniques}
\vspace{3pt}

\noindent \textbf{Automatic and dynamic threshold}. For all DLAD methods, whether to generate prediction or reconstruction errors, a threshold is expected to decide if a value is normal or anomalous. Typically, this threshold is determined empirically via trying different values by an expert. To automate this process, an unsupervised yet accurate method is needed to produce a threshold without the expert knowledge. Hundman \etal~\cite{hundman2018detecting} proposed a dynamic and automatic method to calculate the threshold. Firstly, smoothed prediction errors are generated. An exponentially-weighted average (EWMA) is adopted to smooth a sequence of past prediction errors, which usually contain spikes when there are sharp changes in raw values. Secondly, a formula composed of the mean and standard deviations is utilized to dynamically adjust the threshold. The key observation is that the filtration of max smoothed errors is used to eliminate false positives. The unsupervised thresholding method outperforms Gaussian tail-based methods and can be used in other DLAD methods as well. This work utilizes the LSTM model to detect faults in the telemetry data of the spacecraft. Precision, recall, $F_{0.5}$ scores are computed to measure the performance.

\textcolor{black}{ \textit{Current limitations}. Automatic threshold tuning makes DLAD methods more practical in a real-world deployment. However, EWMA and smoothed errors based dynamic thresholds can be improved. Firstly, a large time window is needed to calculate errors. In a real-time system, attacks or faults may already cause catastrophic disasters. Thus the time window may cause a delay in detecting anomalies and prevent losses. Secondly, false positives are still high. Given the large volume of telemetry data, even a small false positive rate will cost massive time and effort of users to investigate these false alarms.}

\noindent \textbf{Input data sampling and reduction}. Spacecrafts generate a large quantity of telemetry data when operating at space. The size and noise of the data could reduce the efficiency and accuracy of DLAD models. Conventional average sampling methods adopt a time window to compress a sequence of data into a data point. But the disadvantage is that the anomalous data is also shifted into the normal range. Tariq \etal~\cite{tariq2019detecting} proposed an archive sampling method to reduce the data amount while maintaining the characteristics of raw data. To this end, a list is used to record each telemetry in one component. For each row data in the original dataset, different values are saved in the new database. In other words, rows with the same value will not be saved. With archive sampling, the characteristics have not been changed and remain the same with raw telemetry data. The method utilizes ConvLSTM and Mixtures of Probabilistic PCA (MPPCA) jointly to detect anomalies, where a higher error will be accepted as the final error score. The model is evaluated on a real-world satellite dataset with 22 million telemetry data points. The precision and $F_1$ score of the method outperforms four baseline methods (\eg, One-Class Support Vector Machines, Isolation Forest) to a large extent. But the recall is at a similar level with baseline methods.

\textcolor{black}{ \textit{Current limitations}. Archive sampling significantly reduces the amount of data and keeps the original characteristics. But two issues remain unsolved. First, due to the different sampling rates of telemetry channels, the first rows of the dataset may not be stored. This may cause a loss of data. Second, collective anomalies may not be detected. Because the temporal dependency among different telemetry channels may be altered by archive sampling (although numeric values have not changed).}

\noindent \textbf{Protecting the control system}. Attacks targeting control systems are covert and devastating, which do not necessarily change the values of sensors and network traffic. Detection methods rely on sensor measurements and communication patterns may fail to identify such attacks. Typically, malicious code that injected into control systems intentionally changes the running logic of controllers (\eg, PLCs), hence it can potentially cause physical damage to CPSs. However, conventional methods may fail to identify elaborate attacks that generate similar sequences of events to that of normal code blocks. Ezeme \etal~\cite{ezeme2019dream, ezeme2019deepanom} utilize system calls and kernel events to ensure the running status of control systems. Concretely, through log preprocessing, features (\eg, events) are extracted from raw log traces. Further, an LSTM model with an attention layer is adopted to predict subsequent event sequences. The prediction error is measured to identify the anomaly. Four scenarios (\ie, full-while, ffo-ls,
hilRF-InFin, and sporadic) are used to simulate the status of a UAV, where the data is retrieved. The method outperforms three approaches~\cite{ezeme2017imputation, ezeme2018hierarchical, salem2016anomaly} by evaluating true positives and false positives.

\textcolor{black}{ \textit{Open problems}. The correlation of system calls and kernel events is captured through attention-based LSTM neural networks. However, the number of system calls and kernel events in UAV control systems are small compared to other CPSs (\eg, ICSs). Whether this method can effectively handle enormous system calls needs to be measured. Moreover, the computational performance is measured on a desktop computer. A real embedded device can be used to evaluate the running performance.}

\noindent \textbf{Capture the stochasticity and temporal dependence}. The multivariate time series data is produced widely in CPSs (\eg, spacecraft, ICS), which contains both stochasticity and temporal dependence. To better learn the patterns of normal data, capturing both characteristics can improve the accuracy of the detection. To this end, Su \etal~\cite{su2019robust} adopts a deep Bayesian model (named VAE)~\cite{kingma2013auto} to map input data into stochastic variables. Further, to learn temporal dependence, these variables are connected to hidden Gated recurrent units (GRUs) representations. Finally, planar NF~\cite{rezende2015variational} is used to learn non-Gaussian distributions of input data from hidden variables of the previous step, the output of which is fed to consecutive layers to reconstruct the original input data point. Reconstruction errors are utilized to detect anomalies in time-series data. The method outperforms three baseline methods (LSTM-NDT\cite{hundman2018detecting}, DAGMM\cite{zong2018deep}, LSTM-VAE\cite{park2018multimodal}) in $F_1$, precision, and recall when evaluated on three datasets.

\textcolor{black}{ \textit{Open problems}. Applied to three different datasets, the recall of this work is much higher than the precision (about 5\%-20\%). And the precision is similar to the three baseline methods. This may be caused by the stochasticity of CPS data (this work intends to capture such stochasticity). If the DL model is sensitive to random noise, small changes of noise (\eg, relatively large but normal engine vibrations) could cause the model to generate a lot of false positives. This issue can be further studied and experimentally investigated.} 

\noindent \textbf{Detecting anomalies in the ADS-B system}. As a key component of the air traffic control management, the Automatic dependent surveillance–broadcast (ADS–B) system is utilized to notify the position of an airplane to ground stations and other aircraft. However, attackers could eavesdrop messages to learn activities and position of aircraft or spoof messages to disturb the air traffic. Also, DoS attacks can cause airplanes to fail to report and receive information. Existing countermeasures require additional sensors to send signals or modification of the ADS-B protocol to provide authentication and encryption, which may not be possible due to the strict regulation. To detect the above attacks, Habler \etal~\cite{habler2018using} used ADS-B messages as the data source to detect anomalies. They utilized an LSTM autoencoder to reconstruct features of a window of messages. The input features include speed, latitude, longitude, altitude and distance delta. Reconstruction errors are used to detect the anomaly. The method is evaluated on a large-scale flight tracking dataset from Flightradar24~\cite{flightradar24}, which outperforms five baseline methods (\eg, Hidden Markov model with Gaussian mixture emissions (GMM-HMM)~\cite{haider2017detecting}, one-class SVM, Isolation Forest, DBSTREAM~\cite{bar2014large}) when measured by true positives and false positives.

\textcolor{black}{  \textit{Open problems}. The LSTM-based autoencoder is effective to detect injected anomalies as illustrated in this work. However, two issues can be improved. First, the anomalies are manually simulated and the types are simple. For example, random noise and sensor value drift are injected into the origin data. However, sophisticated attacks such as replay attacks are not studied. Second, contextual information could be extracted as features. For example, the flying states (parameters) and current geolocations of airplanes can be combined as the context of a flight.}


\noindent \textcolor{black}{ \textbf{Summary of major categories of limitations}. We briefly summarize five major categories of limitations as discussed in this section. 1) A clear guideline is needed to create the input structure of neural models. 2) The cause of false positives is not fully investigated. 3) The evaluation is on synthetic anomalies. 4) Feature representation process requires a methodology. 5) Computational performance is usually not measured.}

\section{Exploration of Deep Learning-based Anomaly Detection Models}
\textcolor{black}{In this section, we use and customize DL models in the existing work~\cite{schneider2018high, kravchik2018detecting, zhang2019deep} to illustrate the process to detect anomalies in an industrial control system. Note that the purpose of experiments is not to provide sophisticated DLAD methods. Instead, with these experiments, we aim to show the usage of DL models, the typical workflow of DLAD methods, and the running performance of DL models. We hope these efforts can provide readers with some insights to develop DLAD methods that are fit for their own research problems. The source code of these experiments has been open-sourced\footnote{https://github.com/leonnewton/DLAD-Survey/tree/main/DLADexperiments}. Specifically, we conduct two series of experiments.}

\begin{itemize}
\item \textcolor{black}{We explore LSTM-based DLAD methods to capture the temporal dependency of time-series data~\cite{hundman2018detecting, tariq2019detecting, feng2017multi}. Prediction errors are used as anomaly scores}.

\item \textcolor{black}{We explore CNN-based autoencoders to capture the correlation of different sensors~\cite{canizo2019multi, van2019real, zhang2019deep}. Reconstruction errors are used as anomaly scores}. 
\end{itemize}

\noindent \textbf{\textcolor{black}{Testbed \& Dataset}}. \textcolor{black}{We utilize a scale-down and fully functional ICS testbed named  the Secure Water Treatment (SWaT)~\cite{itrust} to conduct experiments. The testbed is a water treatment plant with six stages. Each stage is responsible for a specific treatment process (\eg, filtration). The testbed consists of sensors, actuators, communication networks, and control systems. The testbed provides a dataset, which consists of 7 days of normal data and 4 days of attack data. The normal data contains 496,800 data points. The attack data contains 36 types of attacks and 449,917 data points. The details of the testbed can be found at~\cite{mathur2016swat}. This testbed is widely used in the anomaly detection research community in CPS.}

\noindent \textbf{\textcolor{black}{Implementation}}. \textcolor{black}{We conduct all experiments on a desktop computer (OS Linux x86\_64, 3.7GHz Intel i7 8700K CPU, 32 GB memory). The GPU is NVIDIA GeForce GTX 1080 Ti (12 GB memory). All neural models are implemented using the Keras~\cite{team_keras} development platform. The calculation operations are performed with NumPy~\cite{numpy}.}

\subsection{LSTM-based models}\label{lstmmodelexp}
\textcolor{black}{Time-series data are pervasive in CPS. Sensor values (usually with a sampling rate) are usually utilized to record the physical properties of CPS. For example, the water level sensor can report a water level value every second. Sensor values of different periods usually have dependencies. For instance, the water level at present will impact the value in the future. Also, whether increasing or decreasing, the changes of the water level should be continuous. To capture such dependencies and correlations, LSTM-based neural models are used to build DLAD methods. Indeed, LSTMs have been successfully applied to sequence learning tasks, \eg, time series prediction, speech recognition. The key enabler of LSTM is its special design of internal vectors. A cell state vector is utilized to obtain long-term "memories". Meanwhile, a forget vector named forget gate is used to selectively ignore information that is kept in the cell state. The hidden state vector stores the information to be fed to the next time step. Memorizing information of extensive time steps is useful to capture the temporal dependency of time-series data.}

\begin{figure}[h]
\centering
\begin{subfigure}[t]{.5\textwidth}
  \centering
  \includegraphics[width=.9\textwidth]{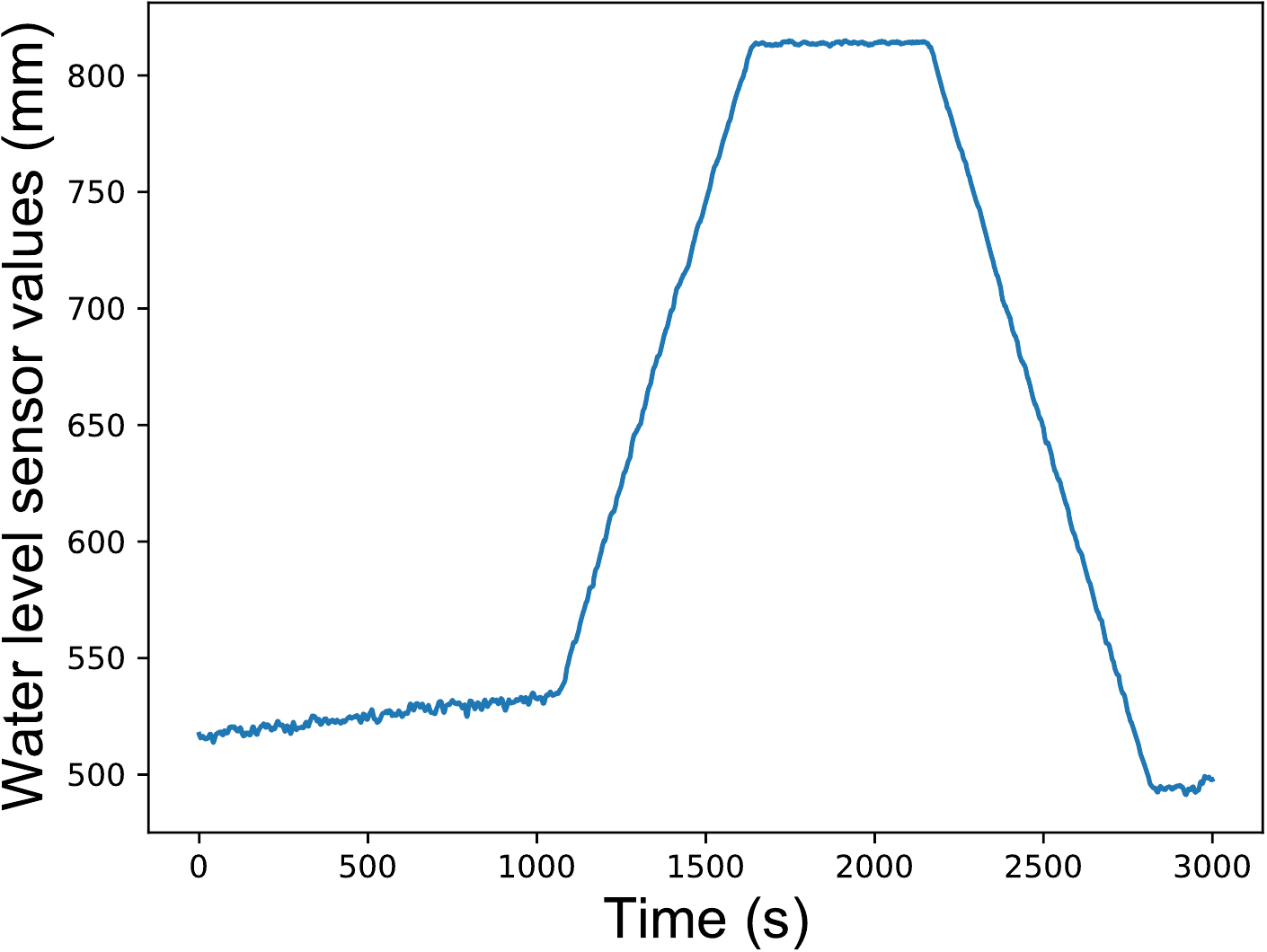}
  \caption{\textcolor{black}{Water level values in a normal period, which increase and decrease periodically as the system treats the water.}}
  \label{normalperiod}
\end{subfigure}%
\begin{subfigure}[t]{.5\textwidth}
  \centering
  \includegraphics[width=.9\textwidth]{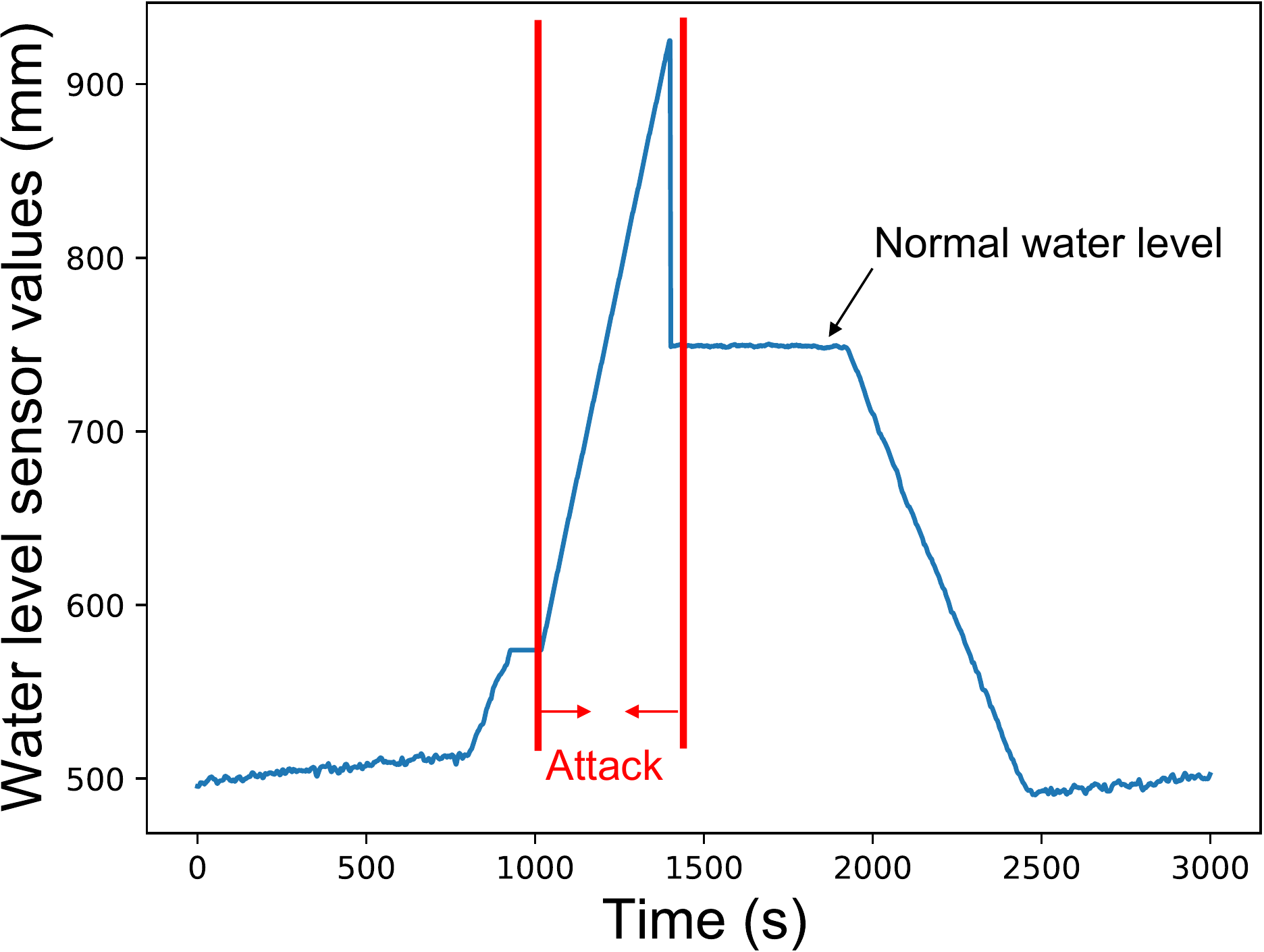}
  \caption{\textcolor{black}{Water level values in an anomalous period. A data injection attack has changed the normal increasing process of water level.}}
  \label{attackperiod}
\end{subfigure}
\caption{\textcolor{black}{Water level values of a normal period and an anomalous period.}}
\label{waterattacklstm}
\end{figure}

\noindent \textbf{\textcolor{black}{The attack used in the experiments}}. \textcolor{black}{We present a false data injection attack~\cite{mo2010false, mo2012integrity} (which is very common in attacks against CPS) in the SWaT dataset. As illustrated in Figure~\ref{waterattacklstm}, Figure~\ref{normalperiod} is the water level (sensor LIT-101) of a water tank in a normal period. The water level changes since the system is treating water. However, as presented in Figure~\ref{attackperiod}, the false data injection attack has manipulated the water level readings that far exceed the measurement range of sensors. Thus the temporal dependency of water level has been altered. The attack can be conducted through spoofing attacks or MITM attacks.}

\noindent \textbf{\textcolor{black}{Design of LSTM-based models}}. \textcolor{black}{We elaborate the design from three aspects: (1) Input data \& preprocessing. (2) Neural model architecture. (3) Anomaly scores.}

\textit{\textcolor{black}{Input data \& preprocessing}}. \textcolor{black}{The input data is unidimensional sensor time-series data. Data preprocessing approaches are usually leveraged to transform input data into the format that can be applied to neural networks. For example, we first apply a Min-Max~\cite{al2006normalization} scaler to map raw data into range 0 to 1.}

\begin{figure}[h]
\centering
\begin{minipage}{.5\textwidth}
  \centering
  \includegraphics[width=\textwidth]{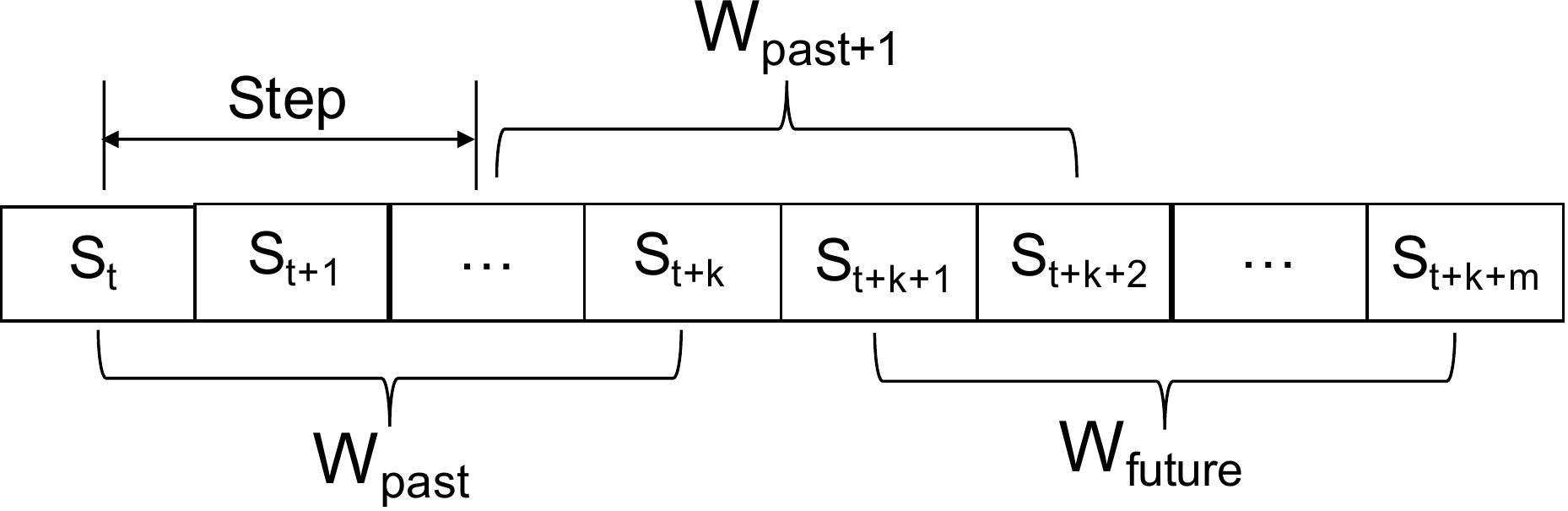}
  \captionof{figure}{\textcolor{black}{The input to the LSTM model. We use a $W_{past}$ time window to predict a $W_{future}$ time window. The distance between two $W_{past}$ time windows is the size of \textit{step}.}}
  \label{lstminputwindow}
\end{minipage}%
\hfill
\begin{minipage}{.4\textwidth}
  \centering
  \includegraphics[width=0.5\textwidth]{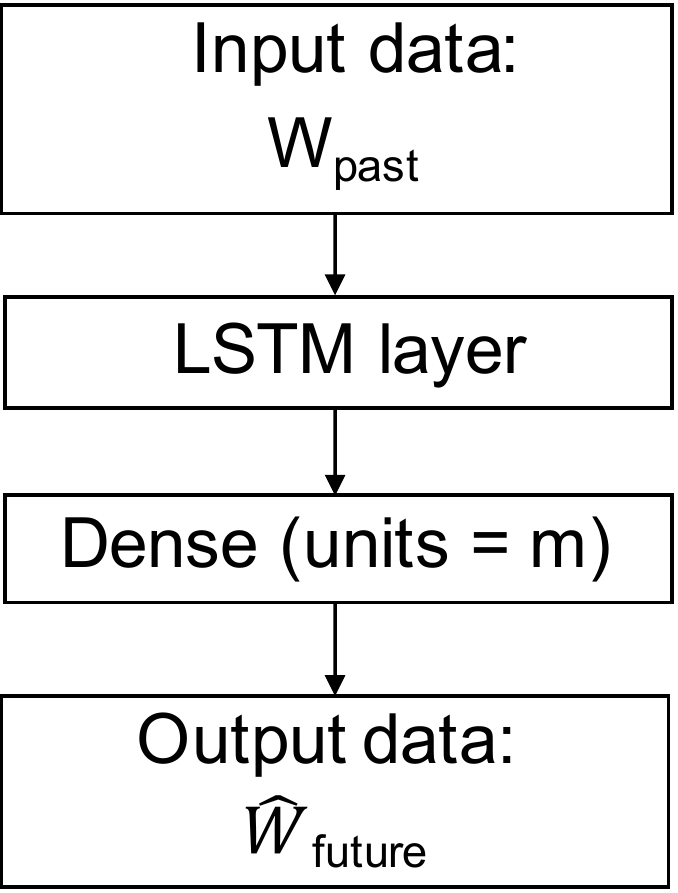}
  \captionof{figure}{\textcolor{black}{The architecture of the LSTM-based model. }}
  \label{lstmarchitecture}
\end{minipage}
\end{figure}

\textit{\textcolor{black}{Neural model architecture}}. \textcolor{black}{We utilize LSTM models~\cite{schneider2018high, kravchik2018detecting} to predict future values from past values. As illustrated in Figure~\ref{lstminputwindow}, a time window of past values can be represented as $W_{past}$ = ($S_{t}$, $S_{t+1}$, $\cdots$, $S_{t+k}$), where $S_{t}$ is a sensor value at time $t$ and the size of time window is $k$. Then, this sequence can be used to predict future time-window sequence $W_{future}$ = ($S_{t+k+1}$, $S_{t+k+2}$, $\cdots$, $S_{t+k+m}$), where the size of time window is $m$. In this experiment, we predict a single future value. Namely, $W_{future}$ = ($S_{t+k+1}$), $m = 1$. $k$ is set to be 60. A \textit{step} (distance) can be set between $W_{past}$ and $W_{past+1}$. If the start of $W_{past}$ is $S_{t}$, the start of $W_{past+1}$ is $S_{t+step}$. We set \textit{step} to be 1. As illustrated in Figure~\ref{lstmarchitecture}, for the neural model, we use one LSTM layer and a dense layer to build a basic model. The units of the LSTM layer can be tuned to get optimized values. The units of the dense layer are set to the size of $W_{future}$ (namely $m$). The model will predict values in the future time window $\hat{W}_{future}$.}

\textit{\textcolor{black}{Anomaly scores}}. \textcolor{black}{We adopt prediction errors as anomaly scores. Namely, the model will calculate and minimize the Mean Squared Error (MSE) between $W_{future}$ and $\hat{W}_{future}$. At the training phase, the model learns the characteristics of normal data. At the testing phase, if prediction errors are above a pre-defined threshold, an anomaly is detected. We use 80\% of the dataset to train the model and 20\% to validate the model.}

\begin{figure}[h]
\centering
\begin{minipage}{.48\textwidth}
  \centering
  \includegraphics[width=0.85\textwidth]{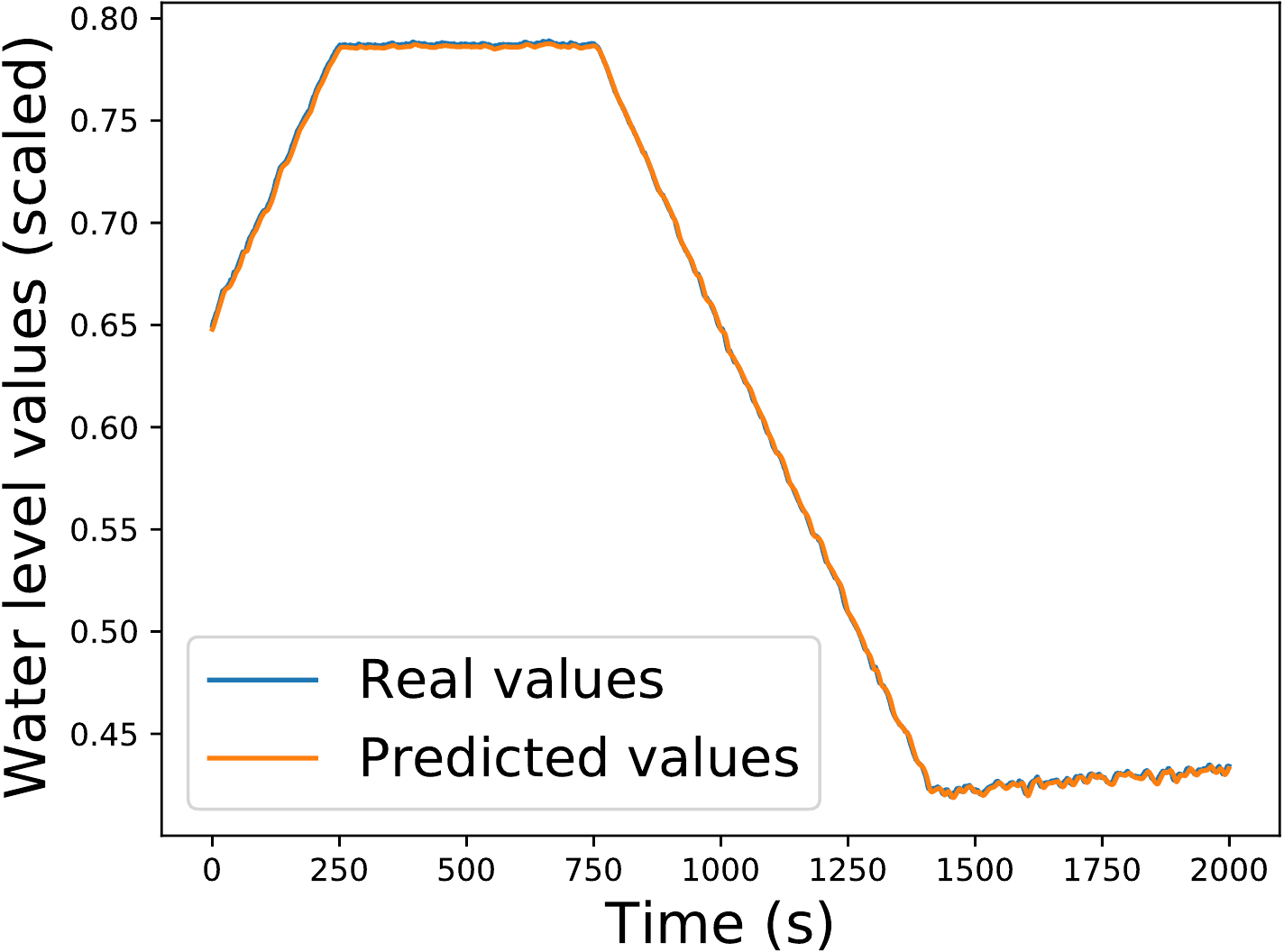}
  \captionof{figure}{\textcolor{black}{Real sensor values and predicted sensor values in a normal period.}}
  \label{originandpredict}
\end{minipage}%
\hfill
\begin{minipage}{.48\textwidth}
  \centering
  \includegraphics[width=0.85\textwidth]{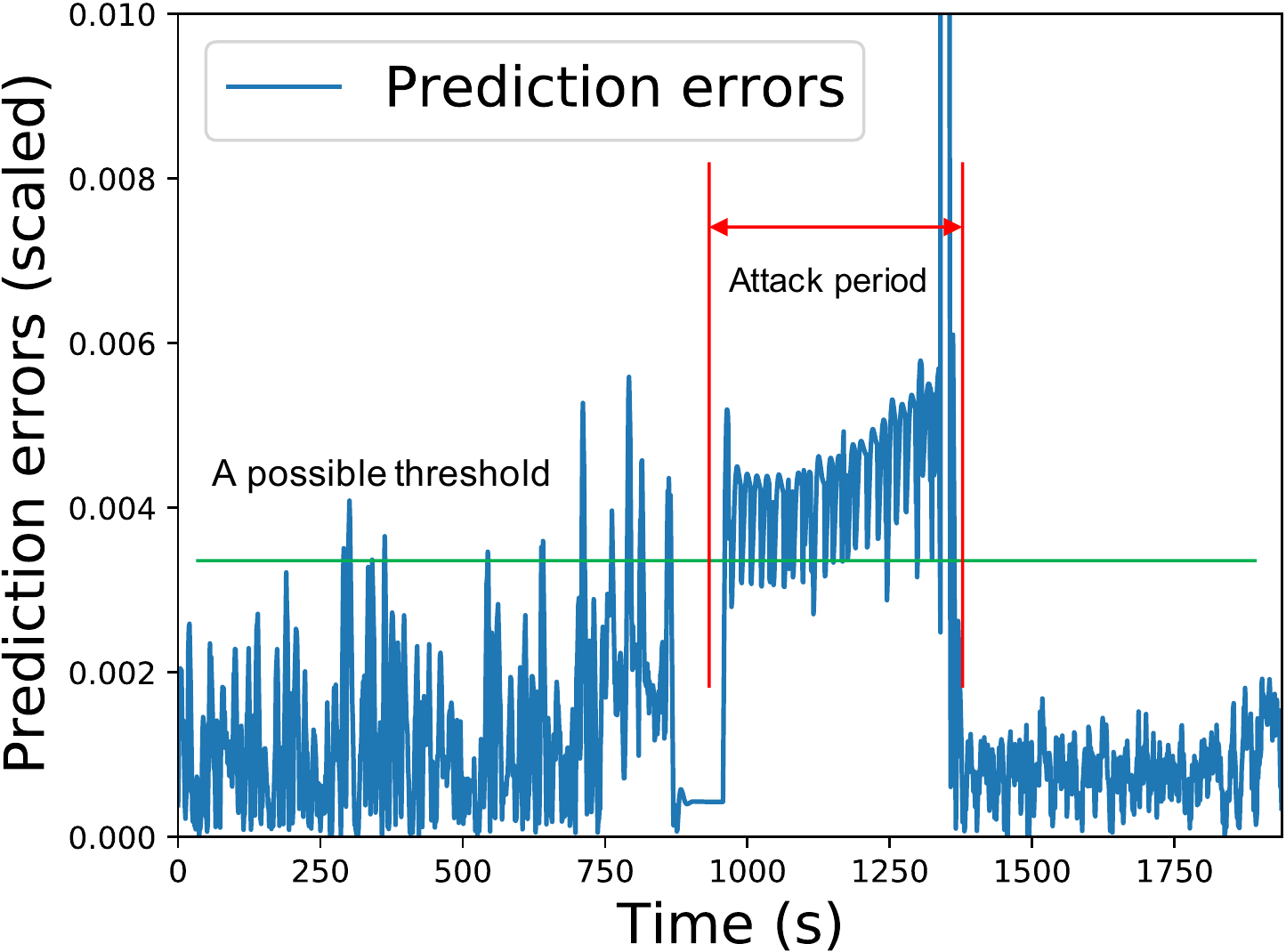}
  \captionof{figure}{\textcolor{black}{The prediction errors of the false data injection attack period. Prediction errors are larger than those in a normal period.}}
  \label{predictionerrors}
\end{minipage}
\end{figure}

\noindent \textbf{\textcolor{black}{Detection process}}. \textcolor{black}{The hyperparameters of the neural model include the number of layers, batch size, epochs, learning rate, the units of each layer (LSTM, Dense layer), \etc. It depends on the learning task and the characteristics of the data. Typically, researchers need to empirically obtain optimized combinations of hyperparameters. We set the batch size to be 100, the learning rate to be 0.001, and the units of LSTM to be 30. We illustrate the real sensor values and predicted sensor values in Figure~\ref{originandpredict}. We can see that prediction errors are rather small in the normal period (real values and predicted values are close). However, we present the prediction errors (the difference between real values and predicted values) in Figure~\ref{predictionerrors} when there is a false data injection attack (as illustrated in Figure~\ref{attackperiod}). We find that prediction errors are larger than those in a normal period. Specifically, a threshold can be used to decide whether data are normal or anomalous. This threshold is a hyperparameter and can be adjusted in terms of different purposes. A possible threshold is marked in the green line in Figure~\ref{predictionerrors}. Generally, we aim to achieve a balance between false positives and false negatives.}

\begin{figure}[ht!]
\centering
\begin{minipage}{.48\textwidth}
  \centering
  \includegraphics[width=0.85\textwidth]{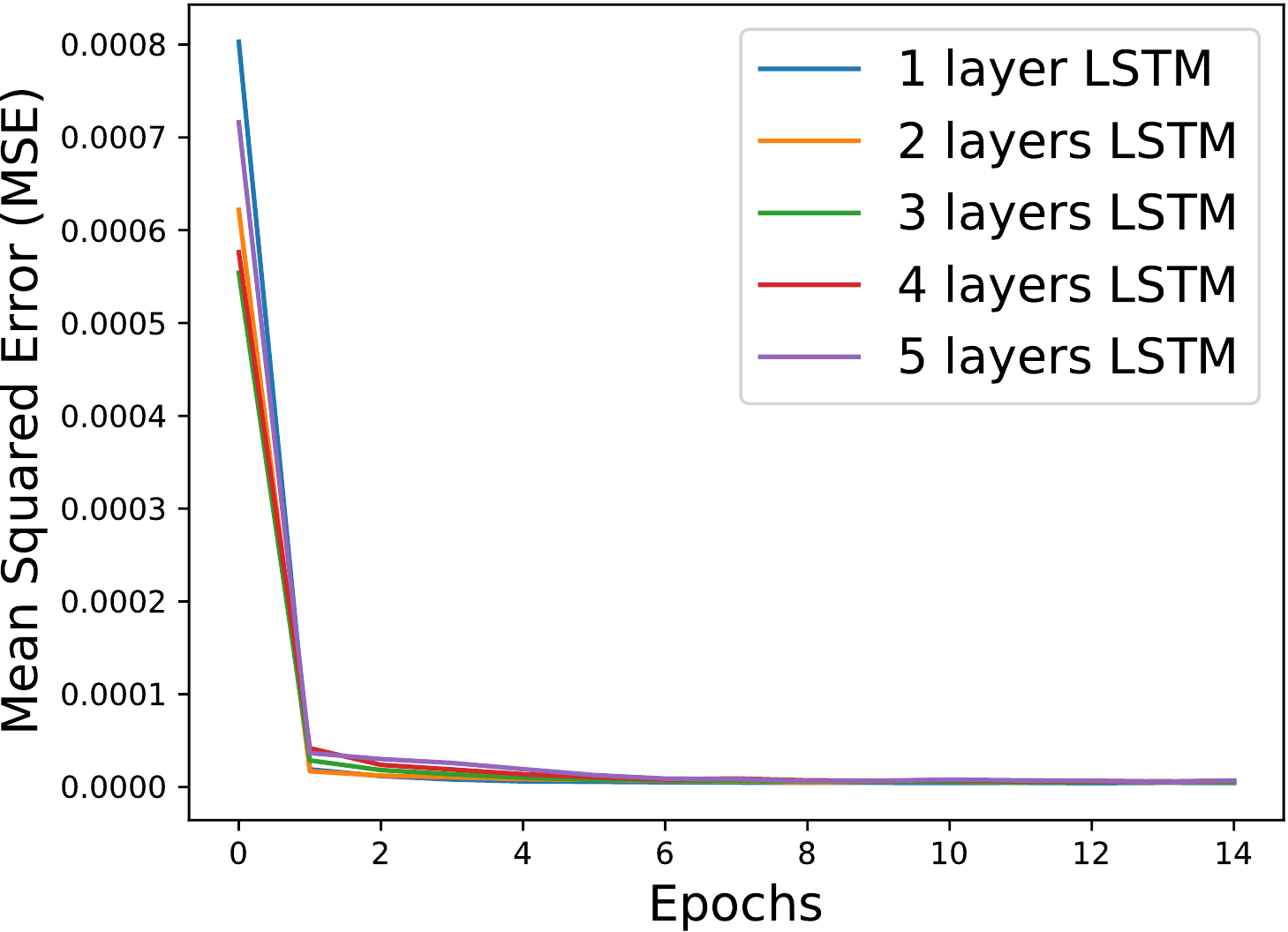}
  \captionof{figure}{\textcolor{black}{The training losses of LSTM-based models with different LSTM layers.}}
  \label{traininglosses}
\end{minipage}%
\hfill
\begin{minipage}{.48\textwidth}
  \centering
  \includegraphics[width=0.85\textwidth]{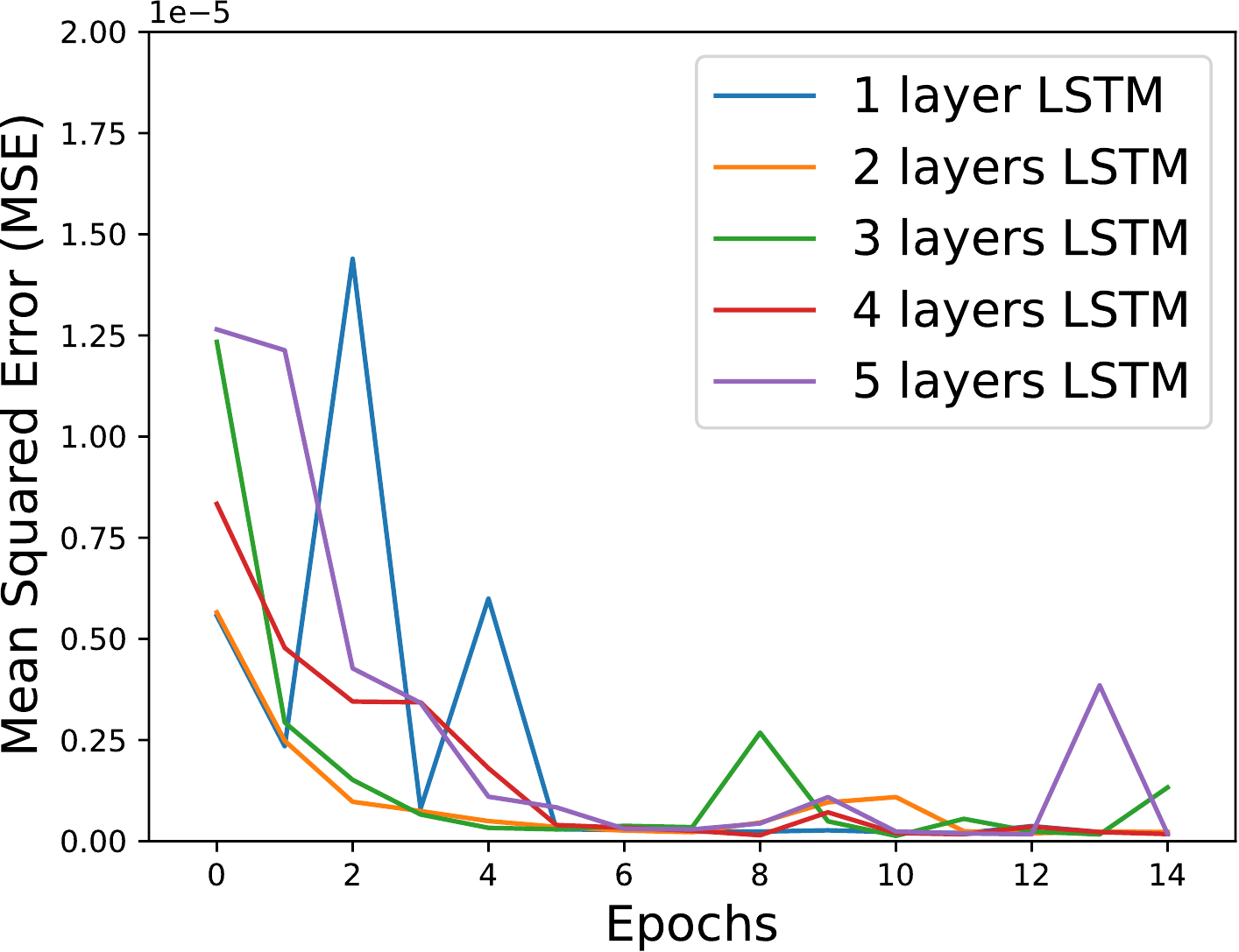}
  \captionof{figure}{\textcolor{black}{The validation losses of LSTM-based models with different LSTM layers. }}
  \label{validationlosses}
\end{minipage}
\end{figure}

\begin{figure}[ht!]
\centering
\begin{minipage}{.48\textwidth}
  \centering
  \includegraphics[width=0.85\textwidth]{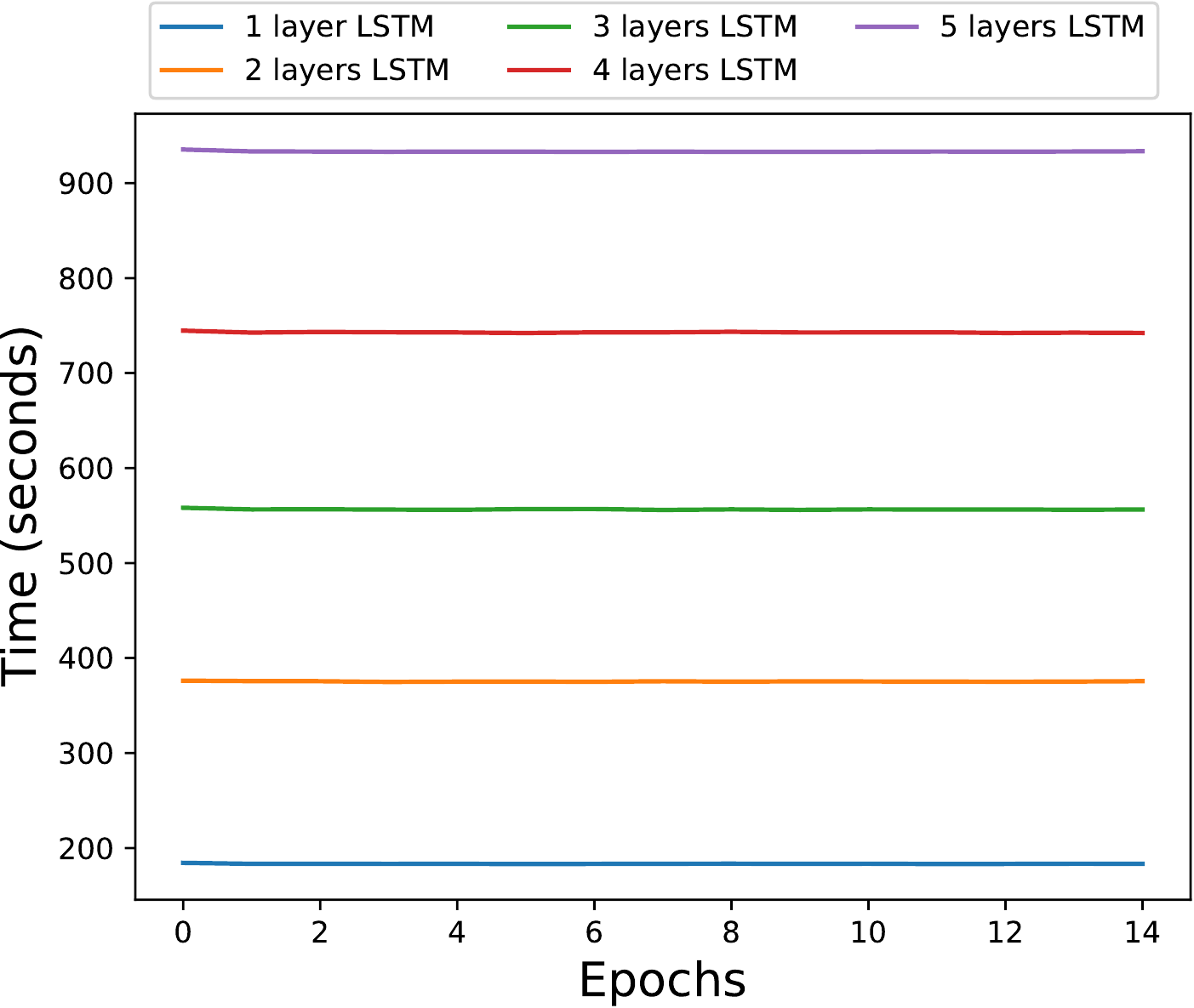}
  \captionof{figure}{\textcolor{black}{The training time for each epoch.}}
  \label{trainingtime}
\end{minipage}%
\hfill
\begin{minipage}{.48\textwidth}
  \centering
  \includegraphics[width=0.85\textwidth]{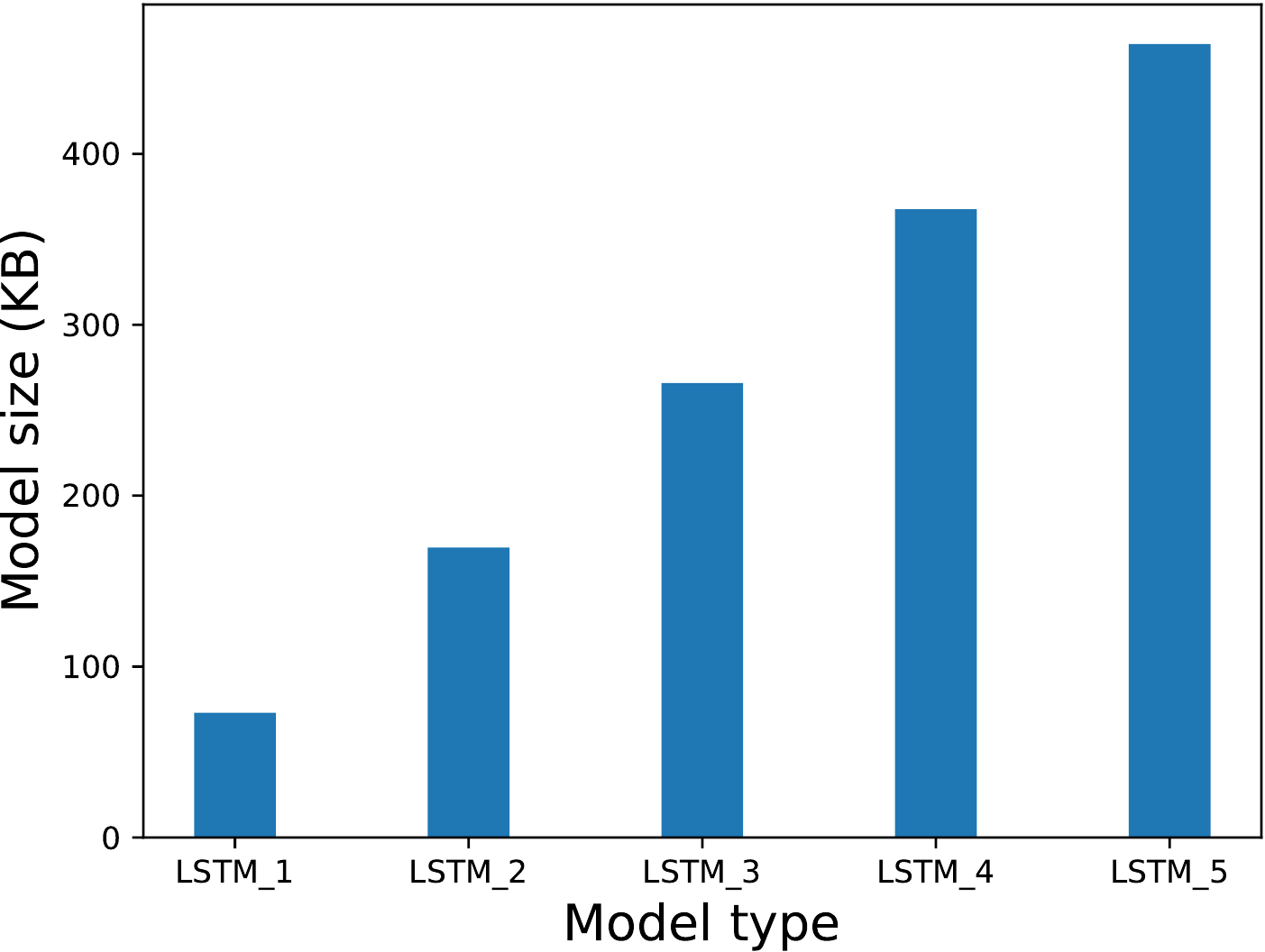}
  \captionof{figure}{\textcolor{black}{Model size}}
  \label{modelsize}
\end{minipage}
\end{figure}

\noindent \textbf{\textcolor{black}{Performance evaluation}}. \textcolor{black}{We evaluate and report the running performance for different layers of LSTM models. As illustrated in Figure~\ref{traininglosses}, with more training epochs, the training losses decrease to a certain level. Also, adding more layers slightly reduces the losses. As illustrated in Figure~\ref{validationlosses}, the 2-layer model seems to have the best performance on unseen data. This indicates that simply adding more layers may not necessarily improve the performance. Also, the performance of basic models is not stable. As illustrated in Figure~\ref{trainingtime}, more layers incur longer training time to optimize the model. Also, the model size increases steadily with more layers as illustrated in Figure~\ref{modelsize}. In CPS, given limited computing power, training time and model size need to be considered before the deployment of DLAD methods.}

\subsection{CNN-based Autoencoders}
\textcolor{black}{Physical devices in CPS not only contain temporal dependencies but also logical dependencies. For example, the water level of a tank (water level sensor) will impact the states of a pump (thus affect the water flow sensor). Namely, there are correlations among different sensors. The running status of a physical device has its context. To capture such context, CNN-based neural models are explored to develop DLAD methods. Convolutional operations can represent correlations of multivariate data~\cite{li2020detecting}. We build CNN-based models in the form of autoencoders. Autoencoders can learn essential features automatically and reduce the dimension of feature space~\cite{borghesi2019anomaly, gong2019memorizing}.}

\begin{figure}[!h]
	\centering
	\includegraphics[width=0.45\textwidth]{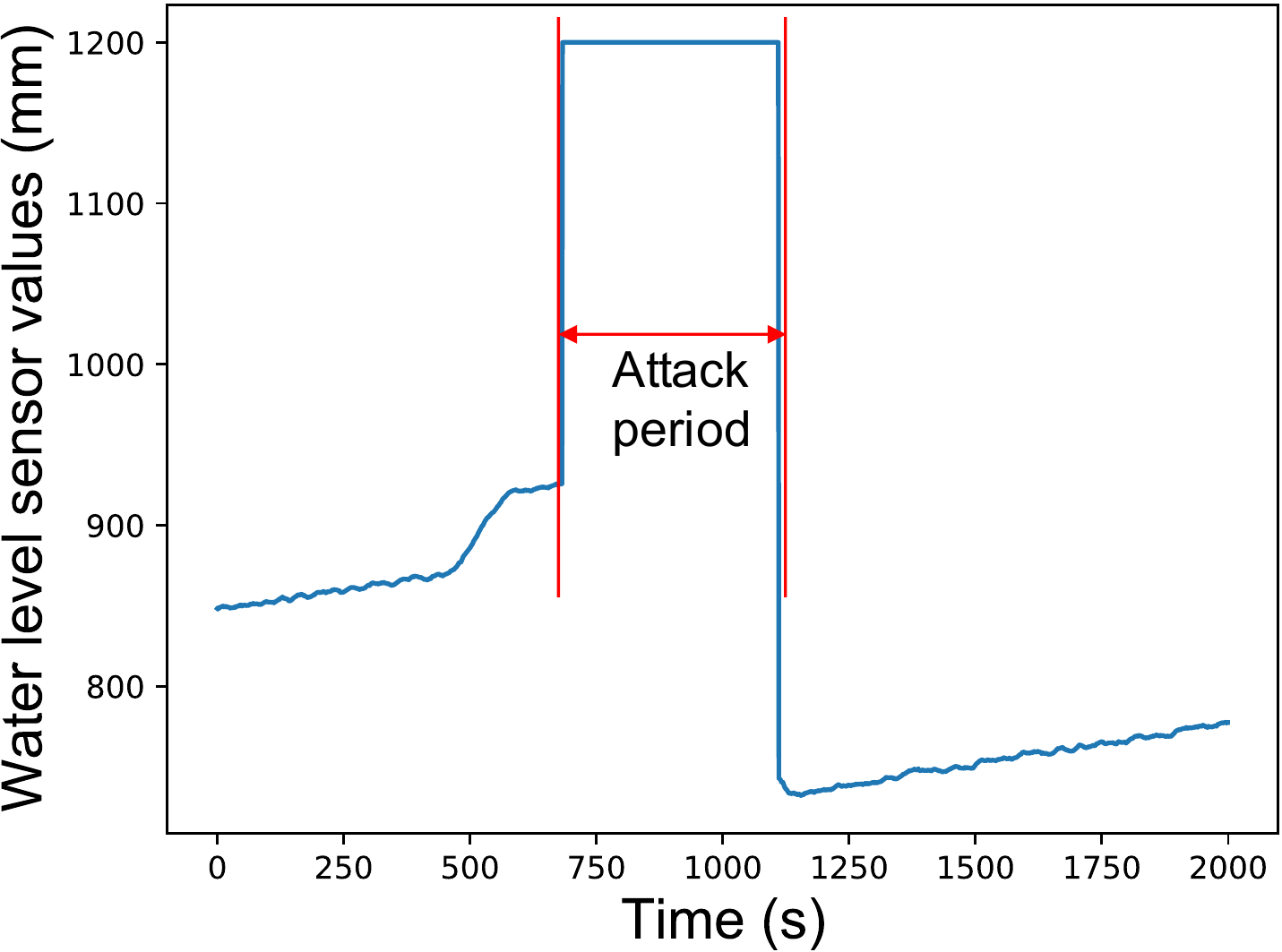} %
	\vspace{-3pt}
	\caption{\textcolor{black}{A false data injection attack with a longer duration.}}
	\label{lit301attack} 
\end{figure}

\noindent \textbf{\textcolor{black}{The attack used in the experiments}}. \textcolor{black}{As illustrated in Figure~\ref{lit301attack}, we present another false data injection attack used in the experiments. This attack causes a longer duration of anomalous high water level compared to the attack in Figure~\ref{attackperiod}. Thus, it impacts several sensors such as inflow, outflow, water level sensors.}

\noindent \textbf{\textcolor{black}{Design of CNN-based Autoencoders}}. \textcolor{black}{The architecture of the CNN model in our experiments is inspired by the work of Zhang~\etal~\cite{zhang2019deep}.  We elaborate the design from three aspects: (1) Input data \& preprocessing. (2) Neural model architecture. (3) Anomaly scores.}

\textit{\textcolor{black}{Input data \& preprocessing}}. \textcolor{black}{We first scale the input data to 0 to 1 with a Min-Max scaler. We design a matrix $M$ to represent correlations of sensors. First, we define two time windows $W_i = (S_{i}^{t},S_{i}^{t+1}, \cdots, S_{i}^{t+k}) $ and $W_j = (S_{j}^{t},S_{j}^{t+1}, \cdots, S_{j}^{t+k})$, which stands for a time window of size $k$ for sensor $i$ and $j$ respectively. $S_{i}^{t}$ denotes a value of sensor $i$ at time $t$. An element $m$ of matrix $M$ can be defined as:}
\begin{equation}
\begin{gathered}
m_{i j}^{t+k}=\frac{W_i * W_j}{k},  W_i * W_j = \sum_{n=0}^{k} S_{i}^{t+n} S_{j}^{t+n}
\end{gathered}
\end{equation}
\textcolor{black}{Thus we get a matrix $M \in \mathbb{R}^{l \times l}$ at time $t+k$, where $l$ is the number of sensors. This matrix is the input to the CNN neural model. We also set a time step $T$ between two matrixes.}

\begin{figure}[!h]
	\centering
	\includegraphics[width=0.5\textwidth]{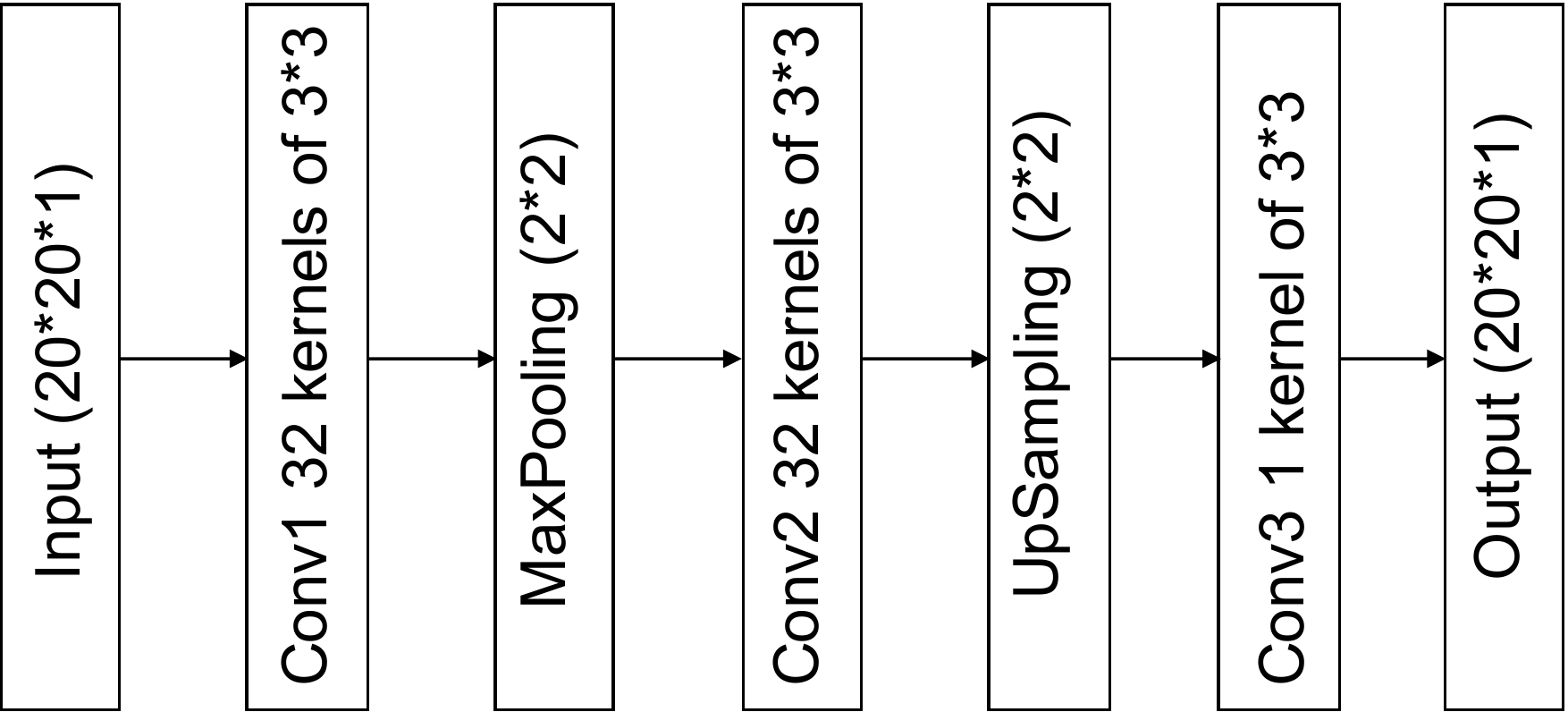} %
	\vspace{-3pt}
	\caption{\textcolor{black}{The architecture of the CNN-based model.}}
	\label{cnnarchitecture} 
\end{figure}

\textit{\textcolor{black}{Neural model architecture}}. \textcolor{black}{The CNN-based model takes a matrix $M$ as input and tries to reconstruct the matrix. As illustrated in Figure~\ref{cnnarchitecture}, the dimension of input is $20*20*1$, since the number of sensors is 20. Then, we use three convolutional layers to learn logical dependencies. MaxPooling is used to reduce the spatial dimension and UpSampling is used to recover the dimension to the size of the input.}

\textit{\textcolor{black}{Anomaly scores}}. \textcolor{black}{Reconstruction errors are utilized as anomaly scores. The  Mean Squared Error is calculated between matrix $M$ and the reconstructed matrix $\hat{M}$. At the testing phase if reconstruction errors are above the threshold, anomalies are detected. We also use 80\% of the dataset to train and 20\% to validate the model.}

\begin{figure}[h]
\centering
\begin{minipage}{.48\textwidth}
  \centering
  \includegraphics[width=\textwidth]{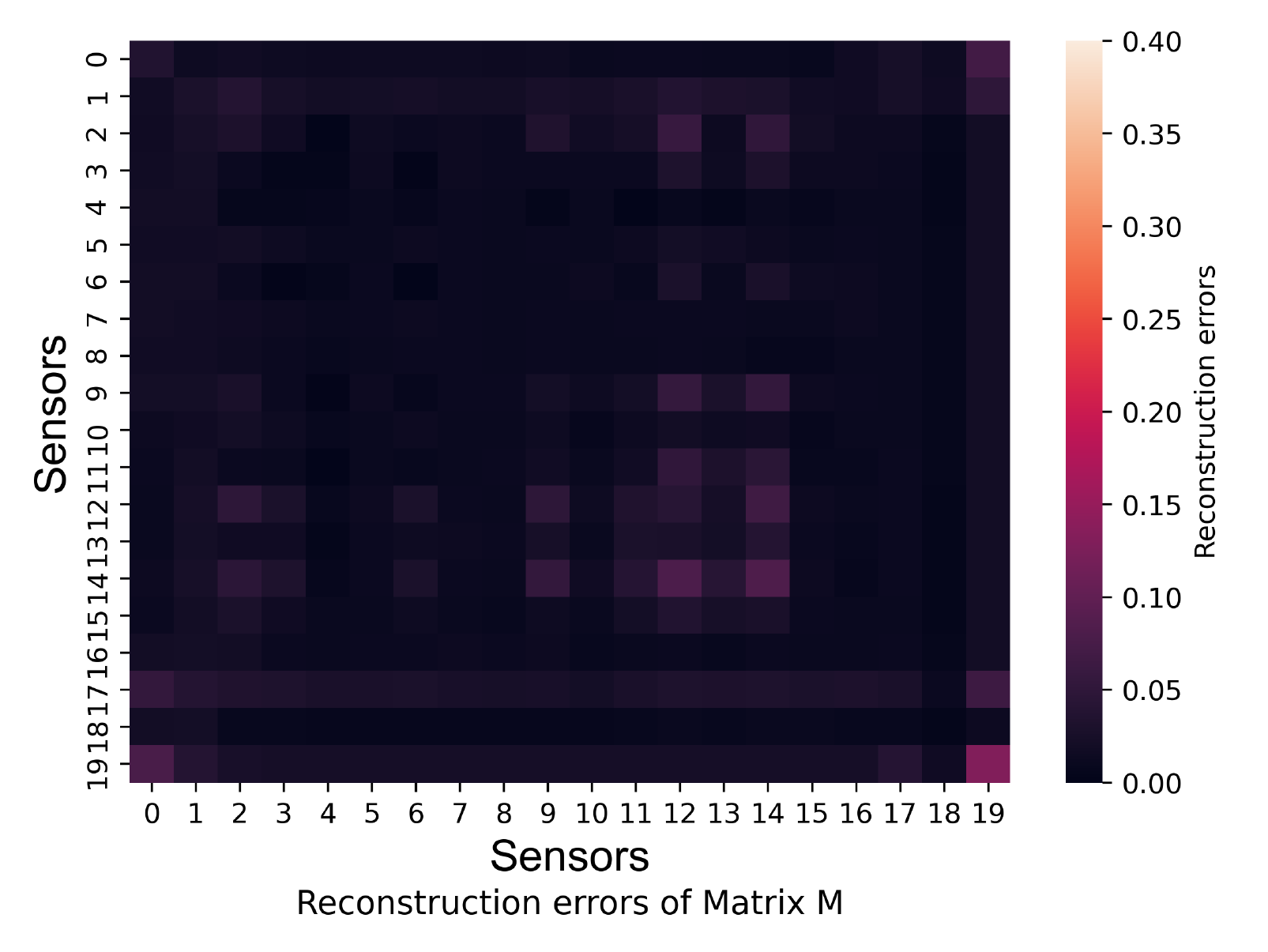}
  \captionof{figure}{\textcolor{black}{Reconstruction errors of Matrix $M$ in a normal period.}}
  \label{normalmatrix}
\end{minipage}%
\hfill
\begin{minipage}{.48\textwidth}
  \centering
  \includegraphics[width=\textwidth]{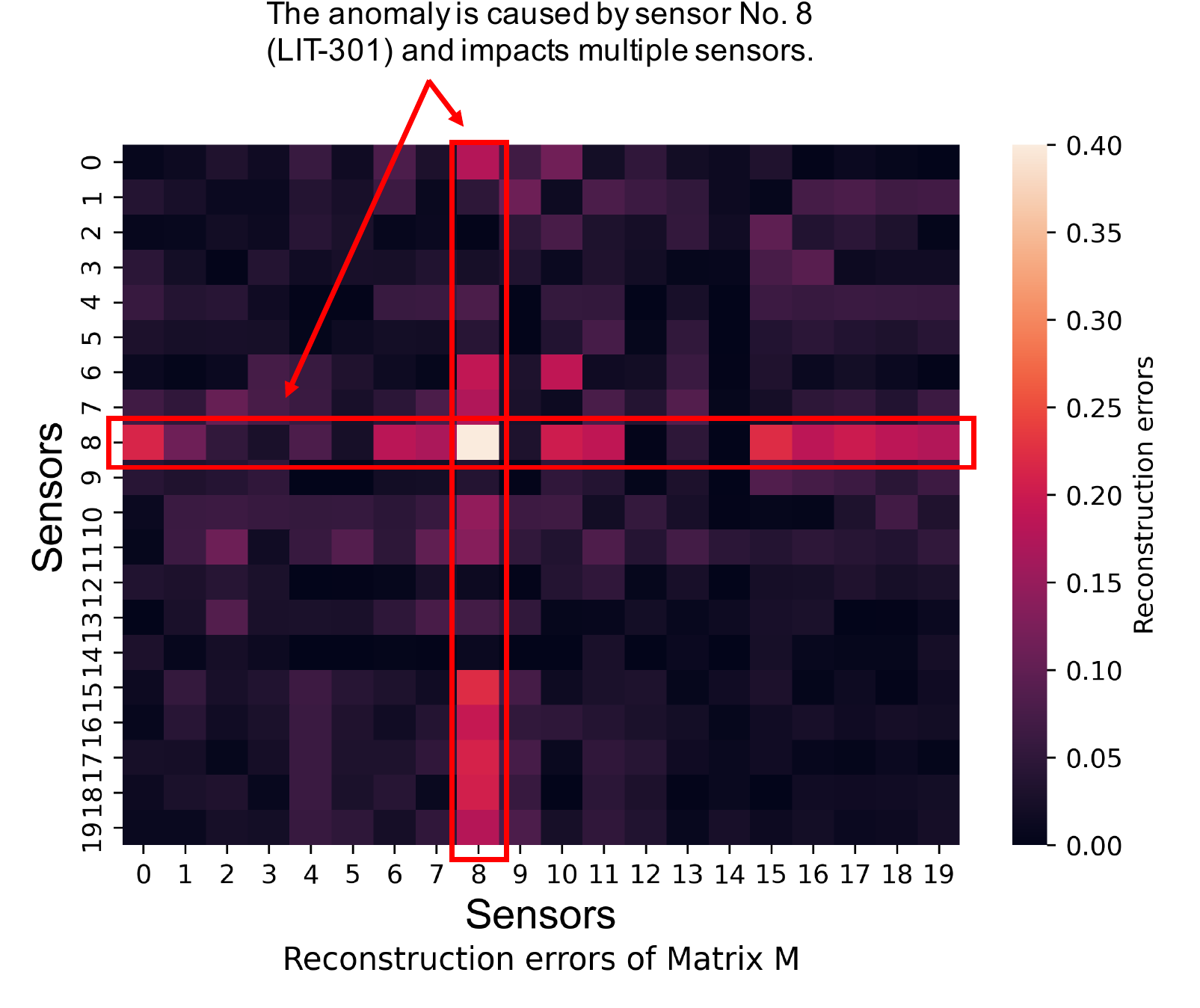}
  \captionof{figure}{\textcolor{black}{Reconstruction errors of Matrix $M$ in a false data injection attack period.}}
  \label{attackmatrix}
\end{minipage}
\end{figure}

\noindent \textbf{\textcolor{black}{Detection process}}. \textcolor{black}{We use reconstruction errors of matrix $M$ to detect anomalies. As illustrated in Figure~\ref{normalmatrix}, reconstruction errors are small in a normal period. However, we find that the false data injection attack causes large errors as shown in Figure~\ref{attackmatrix}. Also, we can see the correlations of sensors through the errors. For example, errors between sensor 8 and 11 are large, but errors between sensor 8 and 14 are small.}

\begin{figure}[h]
\centering
\begin{minipage}{.48\textwidth}
  \centering
  \includegraphics[width=\textwidth]{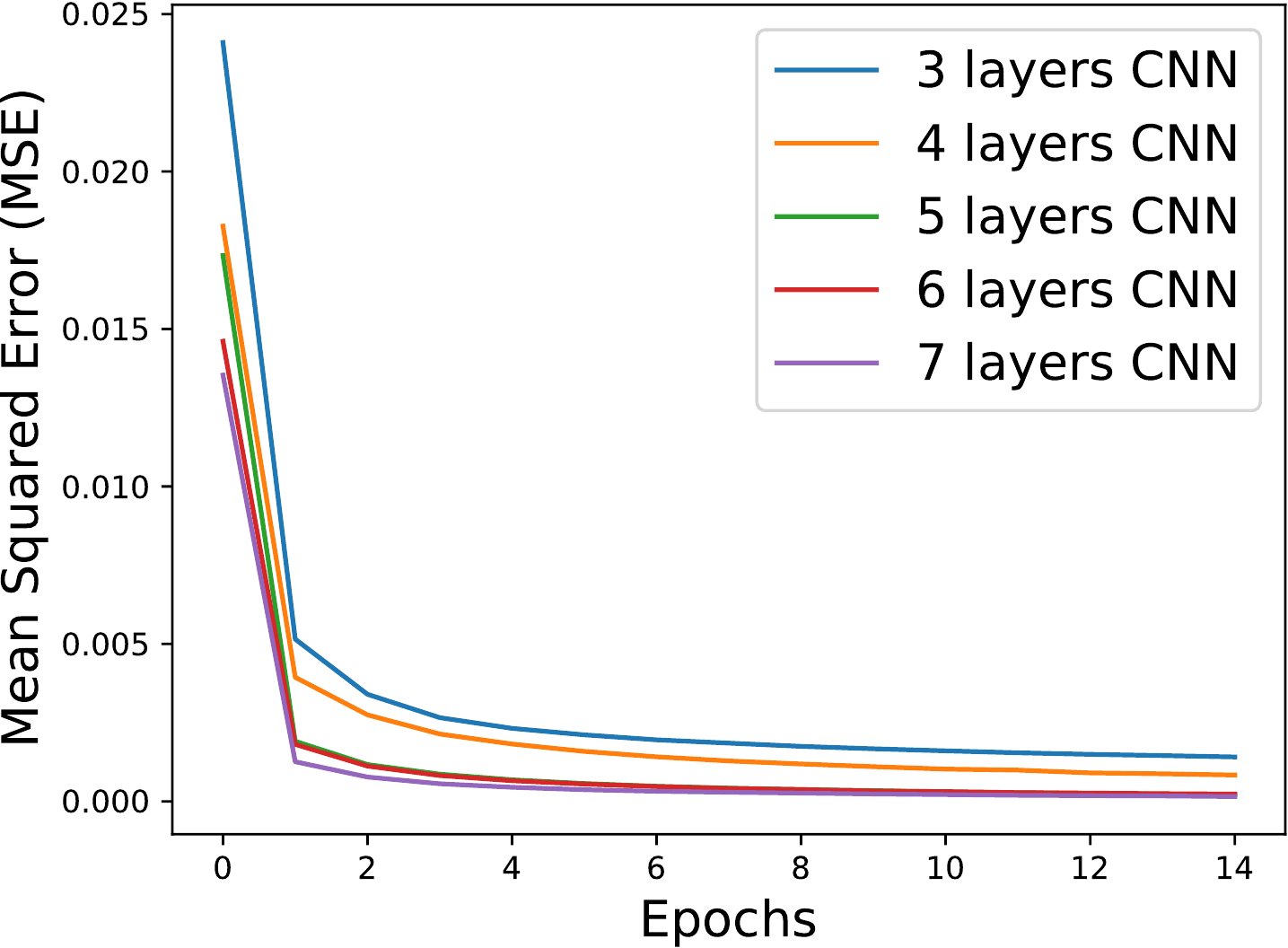}
  \captionof{figure}{\textcolor{black}{The training losses of CNN-based models with different CNN layers.}}
  \label{cnntrainlosses}
\end{minipage}%
\hfill
\begin{minipage}{.48\textwidth}
  \centering
  \includegraphics[width=\textwidth]{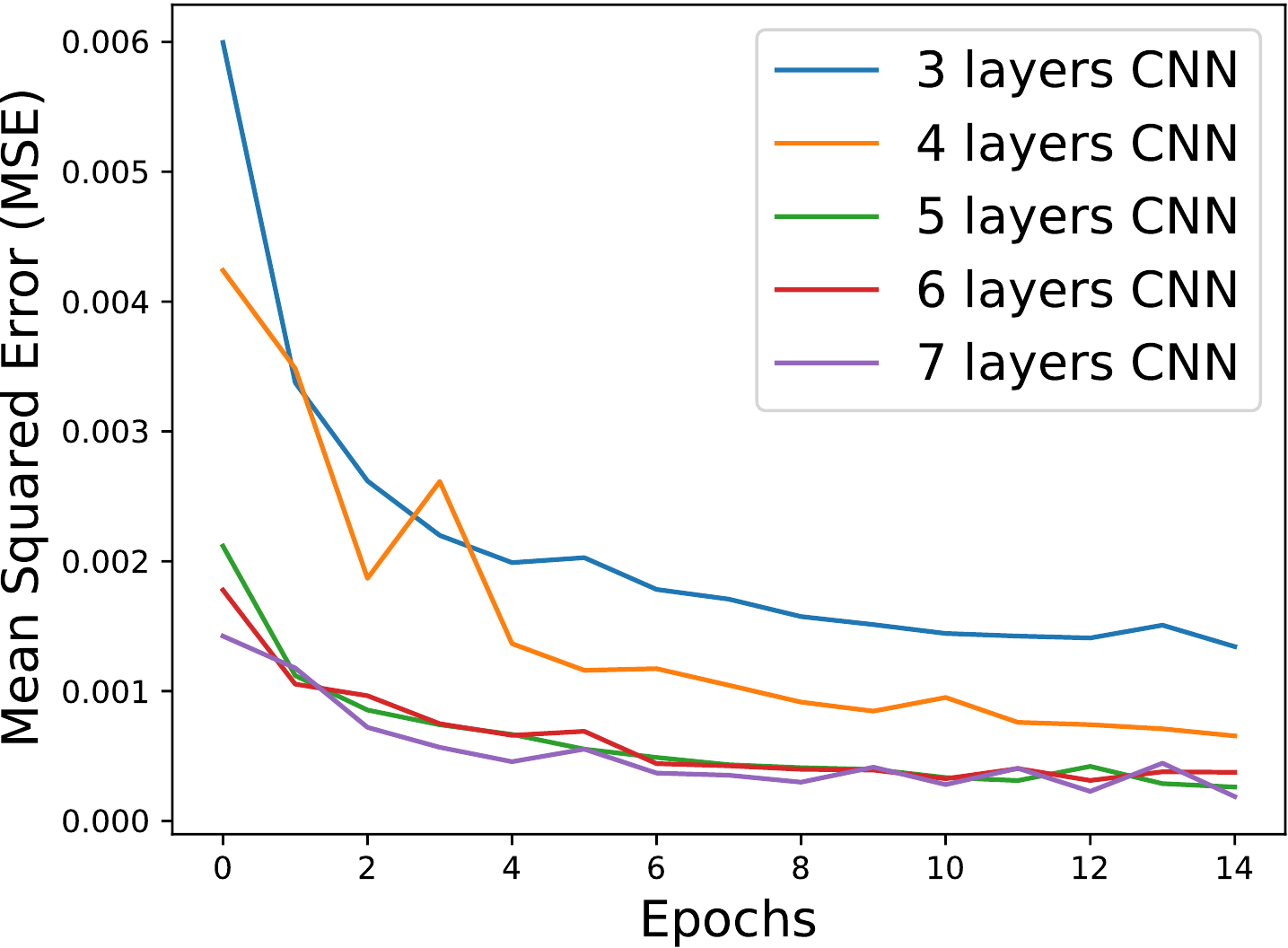}
  \captionof{figure}{\textcolor{black}{The validation losses of CNN-based models with different CNN layers. }}
  \label{cnnvaliloss}
\end{minipage}
\end{figure}

\begin{figure}[h]
\centering
\begin{minipage}{.48\textwidth}
  \centering
  \includegraphics[width=\textwidth]{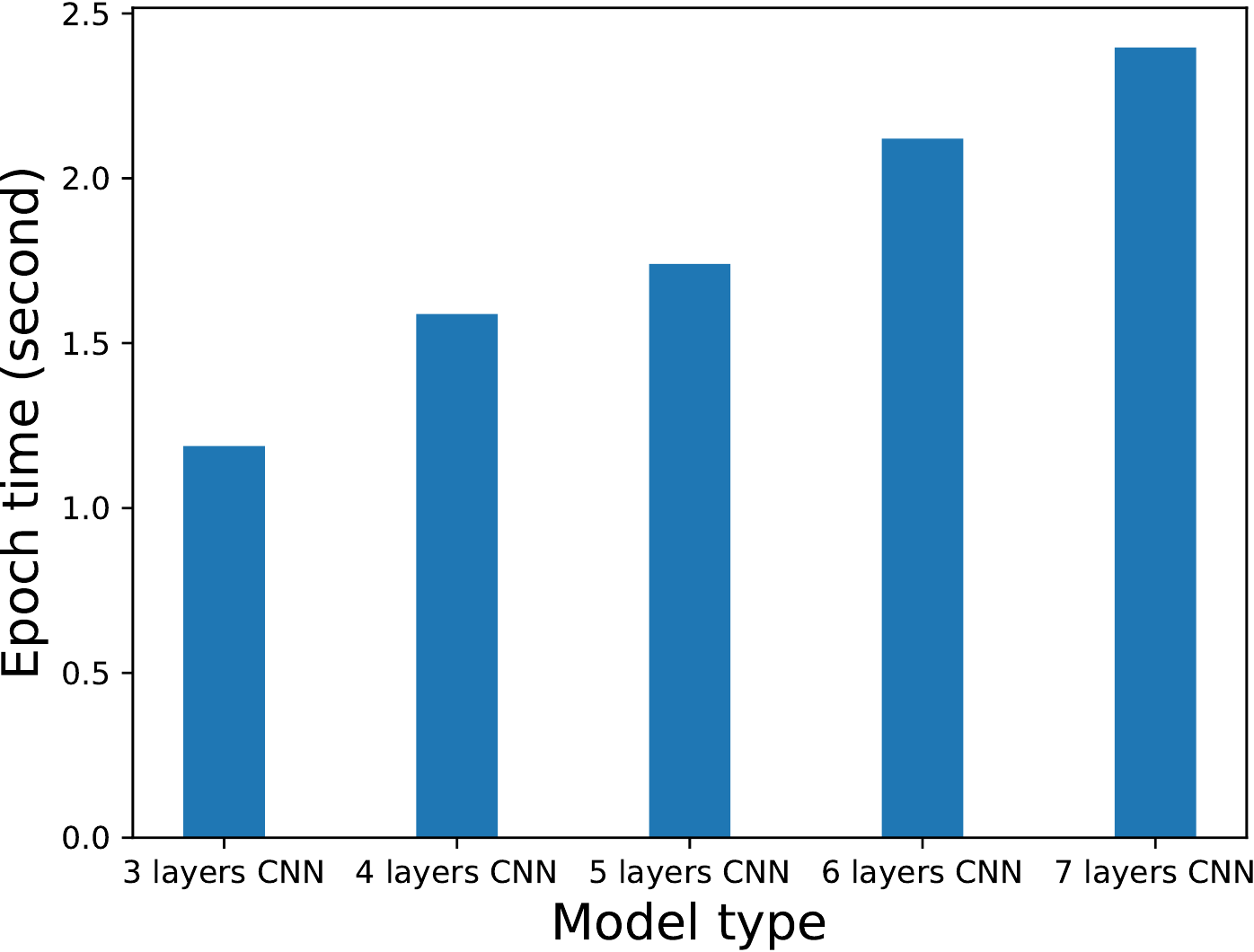}
  \captionof{figure}{\textcolor{black}{The training time for one epoch (mean).}}
  \label{cnntimeused}
\end{minipage}%
\hfill
\begin{minipage}{.48\textwidth}
  \centering
  \includegraphics[width=\textwidth]{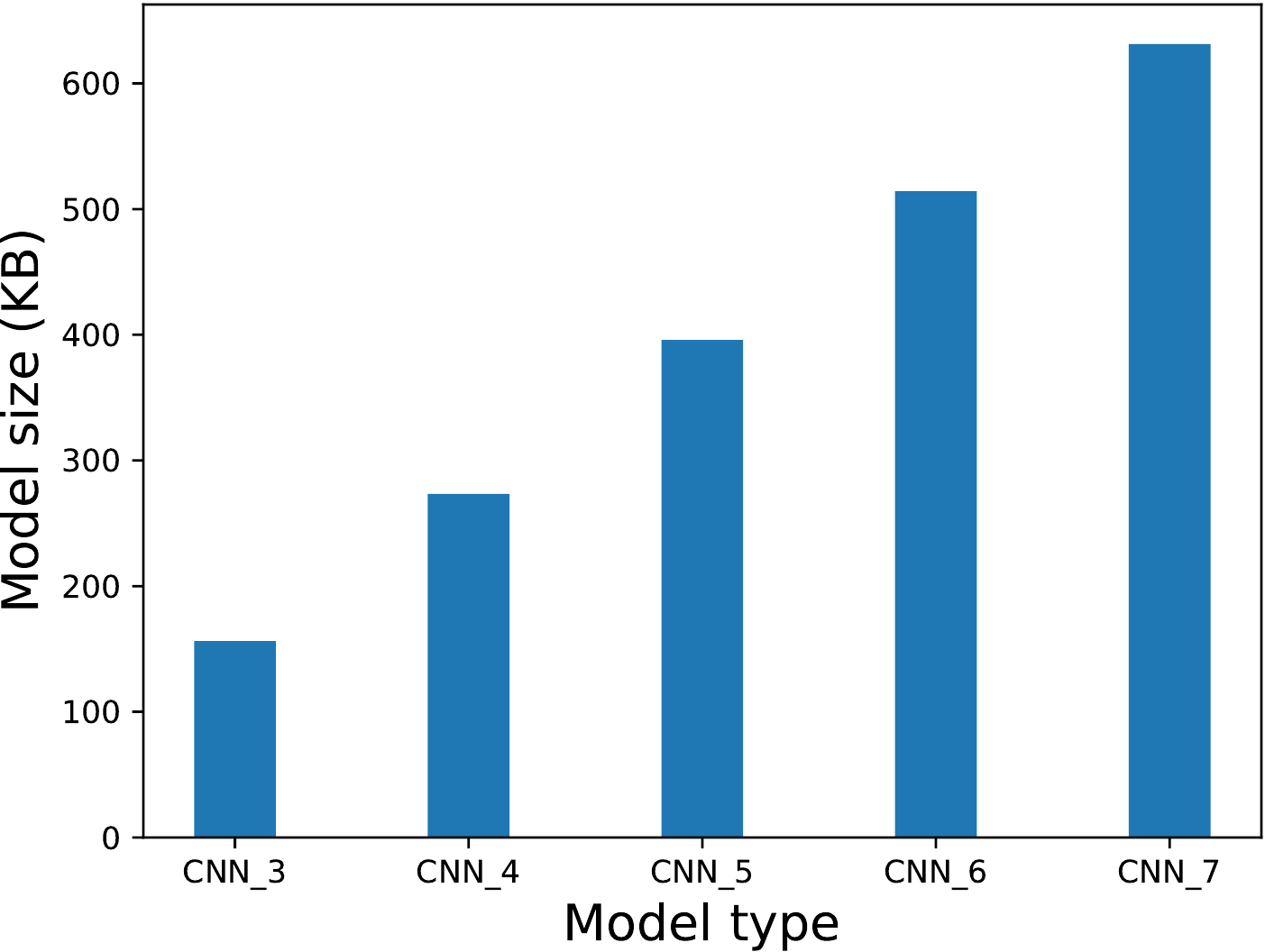}
  \captionof{figure}{\textcolor{black}{Model size}}
  \label{cnnsizebar}
\end{minipage}
\end{figure}

\noindent \textbf{\textcolor{black}{Performance evaluation}}. \textcolor{black}{We show the training and validation losses in Figures~\ref{cnntrainlosses} and ~\ref{cnnvaliloss}. Generally, using more CNN layers can reduce the validation errors of the model. In Figure~\ref{cnntimeused}, we report the mean training epoch time of each CNN-based model. We find that CNN-based models can be trained much faster than LSTM-based models. As illustrated in Figure~\ref{cnnsizebar}, adding more layers will also increase the size of CNN-based models.}

\textcolor{black}{
\noindent \textbf{Summary of key observations}. We summarize key observations from our experiments as follows.
\begin{itemize}
\item The typical workflow and essential procedure of a DLAD method include: (1) identify anomalies that intend to detect, (2) design the input layer, neural model, and anomaly scores, and (3) define evaluation metrics to measure the performance. In fact, we present our taxonomy in light of this workflow.
\item Typically, LSTM-based neural networks are capable of modeling the temporal dependency of time-series sensor data. CNN-based neural networks can capture correlations of multivariate sensor data. 
\item A threshold is usually selected as the boundary of normal and anomalous data.
\item Adding more neural layers may not necessarily improve the performance. Namely, only making models deep may not help.
\item Generally, adding more layers increases training time and model size.
\item Generally, CNN-based models can be trained faster than LSTM-based models. 
\end{itemize}
}

\section{Limitations of Deep Learning-based Anomaly Detection Methods}

\textcolor{black}{
Despite the fast and exciting research advances on deep learning-based anomaly detection, there are still many opportunities to improve the state-of-the-art solutions. In this section, we summarize the current technical limitations and discuss the circumstances where DLAD methods may fail in practice.
}

\noindent \textcolor{black}{ \textbf{The lack of interpretability}. We describe the interpretability of anomaly detection methods as the cause of a detection result that users can understand~\cite{ismail2020benchmarking}. Typically, conventional machine learning methods have rather good interpretability. For example, we can check the decision path of decision tree-based methods and examine the subset that the input sample belongs to. Current DLAD methods focus on improving detection performance (\eg, precision and recall). Thus the interpretability of the model is less discussed and studied~\cite{zhang2020interpretable}. However, the lack of interpretability may impact the usage of DLAD from the following aspects. 
}

\begin{itemize}

\item  \textcolor{black}{\textit{Users may not trust the detection result}. For tasks such as object detection in the computer vision domain, the output of deep learning models is intuitive. For example, it is easy for a human to check whether a cat in an image has been correctly identified. And users can know whether the model works as expected. However, for an anomaly reported by DLAD methods, users do not know the decision process of the model and \textit{what} the anomaly represents is not clear~\cite{giurgiu2019additive}. Namely, users do not have the "whole picture" as they have in the object detection task. At least, users have to inspect system status (\eg, check system logs) to find out what happened (\eg, unknown attacks or false positives). Thus, given a detection result, it is hard for users to decide under which circumstances it can be trusted and adopted~\cite{amarasinghe2018improving}.}

\item  \textcolor{black}{\textit{The root cause of the anomaly is not identified}. Users utilize DLAD methods to detect anomalies and protect CPSs. More importantly, users are more concerned with why and where (the root cause) the result is generated~\cite{du2017deeplog}. However, CPSs consist of vast and heterogeneous components such as communication networks and physical devices. For example, given an anomalous chemical component level in a water treatment plant, the root cause could be anomalous water inflow, anomalous water outflow, or a false chemical dosing process. In particular, root cause analysis helps users to choose appropriate mitigation actions.}
\end{itemize}

\noindent \textcolor{black}{ \textbf{Building and maintaining costs are high}. The performance of DLAD methods comes at a price, which is mainly from two aspects. First, the computing resources of devices in CPS usually cannot directly run DL models. Dedicated computing devices (\eg, GPUs) are expensive. Second, developers spend plenty of time on designing and maintaining neural network models. At the designing stage, researchers have to create corresponding models based on certain CPS architectures and anomalies. At the training stage, extensive efforts are needed to tune hyperparameters. At the maintaining stage, the transmission of parameters adds a burden to the communication network of CPSs. We elaborate the details as follows.
}

\begin{itemize}
\item \textcolor{black}{\textit{Computational costs}. Challenges of computing costs are from several aspects. First, the computing resources of devices in CPS are usually limited. Dedicated devices such as GPUs are needed to run DL models. Another solution is known as edge computing~\cite{wang2019edge}, which is used to support DL models with resource-constrained devices. Second, CPSs nowadays tend to generate a large quantity of data (or big data)~\cite{biron2020big}. This will require huge computational resources. Third, the scale of DL models also increases~\cite{almeida2019embench}. The growing depth and width of models requests increasing computing power. Finally, an effective DLAD method regularly updates models~\cite{salem2020updates}. This process demands computations constantly (not a one-time effort).}

\item \textcolor{black}{\textit{Hyperparameter tuning costs}. Hyperparameters include learning rate, batch size, epochs, \etc. Finding appropriate hyperparameters requires not only experience but also an in-depth understanding of DL models~\cite{sivaprasad2020optimizer}. With current large-scale models, it can be time-consuming to determine the right hyperparameters.}

\item \textcolor{black}{\textit{Maintaining costs}. Since sensor and traffic data keeps being generated, DLAD models need to be updated regularly. Maintaining costs come from two aspects. First, users have to set up a maintaining team (experts in DLAD) to update DLAD models, \eg, retrain and redeploy models. Second, CPSs are distributed systems, and thus the deployment of models will impact the communication networks of CPSs.}
\end{itemize}

\noindent \textcolor{black}{ \textbf{The lack of high-quality data}. As a data-driven technique, DLAD methods require a large volume and high-quality data. We summarize three challenges in terms of training data in CPS. First, CPS environments are changing. Physical components are added or removed, and new attacks may emerge. Second, the type of data (\eg, normal or attack types) is usually not labeled. Finally, anomalous cases are manually created. The details are as follows.}

\begin{itemize}
\item \textcolor{black}{ \textit{CPS environments keep changing}~\cite{ras2017bridging}. Physical devices in CPS can be added or removed after the initial deployment. In this case, DLAD models trained on the old dataset may not learn the characteristics of the new CPS system dynamics. Also, new attacks are developed to avoid the detection of DLAD methods. Hence DLAD methods need to be trained on new attack cases to identify them. }

\item \textcolor{black}{ \textit{The lack of labeled data}. Most datasets in CPS are data in a normal period~\cite{mujeed2020challenges}. Challenges to obtain labeled data include: (1) Labor costs are needed. Developers have to be trained before they can tag data. For example, they need to reach a consensus on the definition of a certain type of attack. (2) Lack of anomalies. Attack cases are less than normal data. Also, it is hard to extract and replay real-world attack cases, \eg, attacks to a smart car system may cause devastating losses. (3) Data in CPS is of a wide variety. It means normal data exist in multiple forms, \eg, time series, system logs, and network traffic. Also, it indicates that there are plenty of attack types. It is difficult to identify and collect all types of attacks. }

\item \textcolor{black}{ \textit{The lack of real-world anomalies or attack cases}~\cite{narayanan2020abate}. As we can see from Table~\ref{summarytable}, a large portion of existing studies evaluate their methods on manually created anomalies. To date, three strategies are utilized to generate anomalous samples. (1) Implementing attack or fault cases and scenarios. (2) Changing simulation model parameters. (3) Injecting noising measurements (\eg, Gaussian distributed noise). However, we argue that these synthetic anomalies either may not happen in the real world or obey a certain statistical distribution, which, unfortunately, may not represent the characteristics of threats in the real world. Hence, a well-designed DLAD method may not detect attacks or faults well when deployed on real CPSs. Moreover, based on different anomaly-creation methods, it is difficult to compare the performance of different methods even they tend to solve the same problem. }
\end{itemize}

\begin{figure}[h]
\centering
	\includegraphics[width=0.5\textwidth]{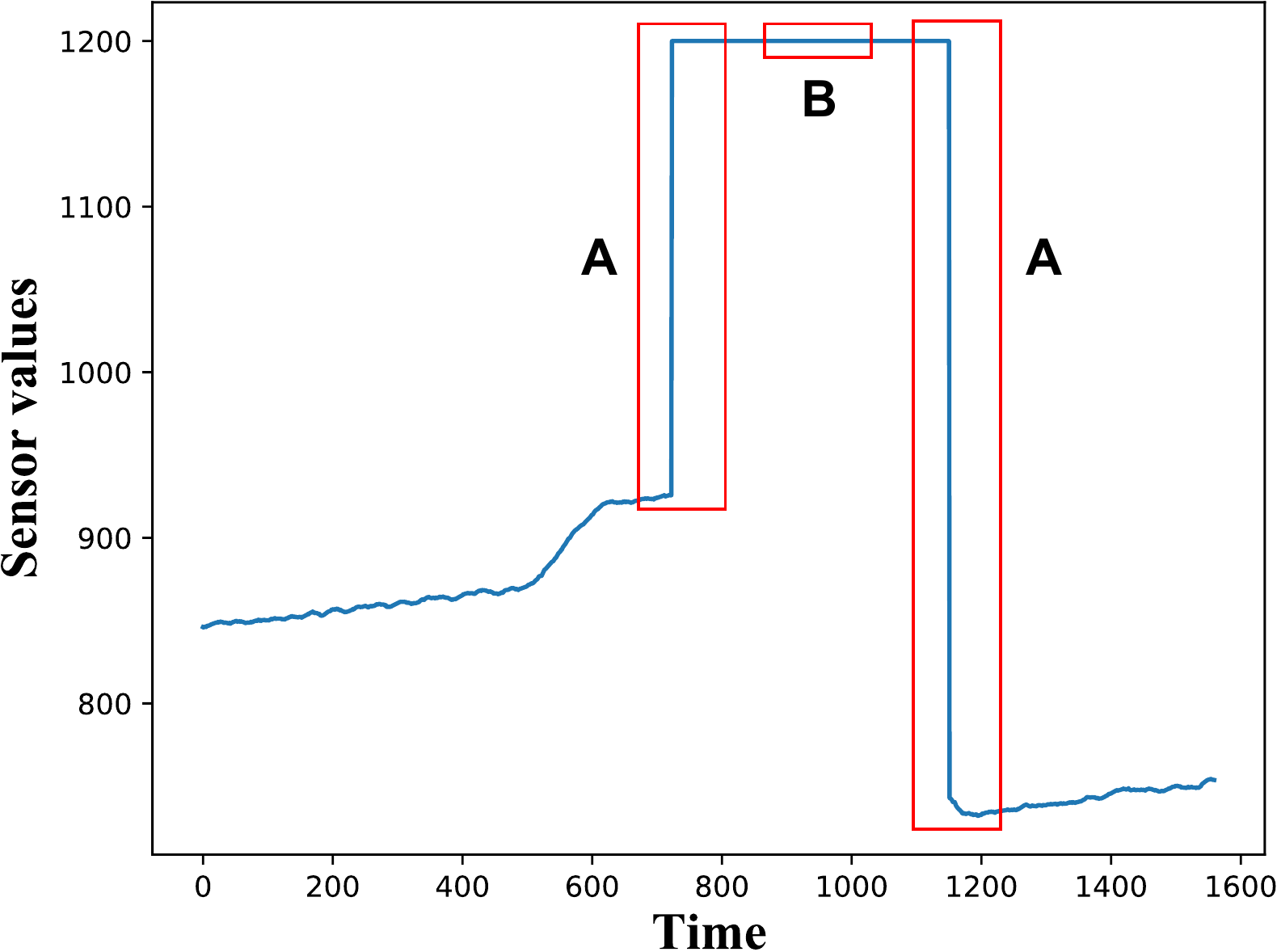} %
	\caption{\textcolor{black}{ Different periods of a long-time attack. Period \textbf{A} denotes sudden changes to the original values. Period \textbf{B} denotes anomalies with no or small changes. }}
	\label{fig:discussion1} %
\end{figure}

\noindent  \textcolor{black}{ \textbf{Fail to detect anomalies that exist for a long-time period}. We find that DLAD methods are sensitive to point and contextual anomalies (\eg, sudden changes). But they may fail when there are collective anomalies (\eg, anomalies exist for a long time). As illustrated in Figure~\ref{fig:discussion1}, a false data injection attack manipulated sensor values, which consist of period A and period B. For period A, DLAD methods can detect these sudden changes. However, for period B, DLAD methods may not raise prediction or reconstruction errors. Because anomalies exist for a long time, DLAD methods (especially LSTM-based methods) may treat data before and after period B as normal data since there is no sudden change. We argue that this is a challenge that future studies need to address. }

\noindent  \textcolor{black}{ \textbf{DLAD methods can be compromised}. DLAD methods need data from CPSs to update neural models. Attackers could inject false data to contaminate DL models or avoid detection. Specifically, there are two types of attacks: (1) Poisoning attack~\cite{xu2019data}. Attackers pollute training data to lead models to produce wrong detection results. (2) Adversarial examples~\cite{meng2017magnet}. With subtle perturbations to normal data, tampered data can avoid the detection of DLAD methods and still damage CPSs. These attacks can either take black-box or white-box assumptions~\cite{dlsecuritytse}. The details are as follows.}

\begin{itemize}
\item \textcolor{black}{ \textit{Poisoning attack}. Instead of directly attacking CPS, attackers could inject manipulated sensor or traffic data to attack DL models. Typically, after being deployed, DLAD methods require data from CPSs to update models. During this process, attackers could deliberately inject false training samples. So DLAD methods cannot learn the true characteristics of CPS and deviate from the decision boundary. Thus DLAD models cannot identify anomalies. }

\item \textcolor{black}{ \textit{Adversarial examples}. Different from poisoning attacks, adversarial examples do not compromise the training process of DLAD models. Instead, adversarial examples add subtle perturbations to normal data. The purpose is to deceive DLAD models to classify anomalous data (\eg, manipulated sensor data or control commands) as normal data. Or if attackers want to disable CPSs, they could trick DLAD methods to generate tremendous false positives (let DL models classify normal data as anomalies). }

\end{itemize}


\noindent  \textbf{No running performance evaluation}. We observe that almost all studies have not evaluated the running performance of DLAD models. To avoid catastrophic events, CPSs such as smart vehicles and aerial systems operate in real-time and need to respond to attacks or faults immediately~\cite{peri2020towards}. In this case, response time is an important factor to measure. For example, DLAD methods startup time and prediction time can be calculated. Furthermore, the computing power of certain CPSs is limited. Or, the computing resources that left to the DLAD methods are constrained at least. Typically, RAM on a commodity UAV is about 2 GB. Hence running costs like RAM usage and CPU overhead can be assessed.

\noindent  \textbf{No updating or online learning mechanism}. Existing research efforts mainly focus on developing new models to improve detecting performance (\eg, reducing false positives and false negatives). However, the deployment of these methods has not been studied yet. Specifically, there is no updating mechanism of trained models to thwart new attacks. Meanwhile, time-series streaming data keeps being generated all the time, which can be utilized to enhance the model constantly. When design one DLAD method, we can consider how to update the model (\eg, updating frequency and time) and keep learning from new data.

\noindent  \textbf{The threshold is empirically selected}. As a key part of DLAD methods, the number of layers and the sliding window size are hyperparameters that researchers have to decide. Such parameters can be empirically selected to design the network. However, once the network architecture (\eg, layers) is determined, the anomaly threshold needs to be resolved. Since it is the boundary of anomalies, the threshold plays an important role in the performance of DLAD models. Currently, the threshold is empirically set or selected in a brute force way~\cite{ghafouri2016optimal}. The value may not be optimized due to various reasons (\eg,  weak validation process, the lack of experience), which could be time-consuming and error-prone. Also, the threshold is fixed and not adaptive, which may not suitable for new data.

\section{Takeaways \& Conclusion}
In this section, we highlight several future research directions to improve deep anomaly detection methods. Based on our findings, these opportunities are proposed to solve the limitations of current DLAD methods.

\subsection{\textsc{Our findings}}\label{ourfinding}

\begin{enumerate}

\item Most studies do not explicitly present a clear threat model. Although these methods usually claim to target either attacks or faults, they do not provide types and details of specific threats that they tend to detect. Also, in different CPSs, prevalent anomalies are usually different. For example, most studies in the smart grid aim to detect the false data injection attack.

\item Sensor measurements in time-series form are the most adopted training and testing data source. First, almost all CPSs contain sensors, hence sensor readings can be easily obtained. Furthermore, sensor values reflect the working status of CPSs reasonably well. Last, sensor values can be accumulated in large quantities, which makes them perfect for deep learning-based methods. Meanwhile, network traffic is the second utilized data source. 

\item RNNs (especially LSTMs) and autoencoders are the most commonly adopted architectures in DLAD methods (and their variants). RNNs are leveraged to capture temporal relation contained in univariate and even multivariate data. Autoencoders are employed to conduct unsupervised learning, which overcomes the absence of labeled data. A mixture of RNN and autoencoder is also adopted to exploit both advantages. In particular, RNN plus CNN combined networks are usually utilized to capture both temporal and context relations. 

\item Prediction and reconstruction errors are equally employed to construct the loss function. All autoencoder-based DLAD methods utilize reconstruction error to build loss functions, which computes the difference between values reconstructed by the model and origin values. Other architectures tend to use prediction error, which computes the difference between values predicted by the model and real values. Prediction labels are typically adopted when labeled data is sufficient. 

\item Depending on different CPSs, different implementation strategies are selected. For methods that work on industrial control systems, scale-down yet fully functional testbeds are often used to collect data. For example, SWaT is a popular water treatment testbed, which contains sensors, actuators, control PLCs, and network traffic. For the smart grid, simulation is most frequently used. In fact, the IEEE X-bus system is the de facto evaluation platform. Meanwhile, for intelligent transportation systems, real-world datasets are applied. Typically, CAN bus data is entirely obtained from real vehicles. In terms of aerial systems, real-world datasets are also preferred. Satellite, UAV, ADS-B data are all collected from real devices. 

\item Precision, recall, $F_1$ are the most used evaluation metrics. In some cases, baseline methods are also presented to emphasize improvement. Note that these metrics are also commonly used in conventional statistical and machine-learning based methods. In particular, false positives and false negatives are balanced through the $F_1$ score. However, there is no specialized metric to measure the performance of DLAD methods. For example, training time and updating frequency are not considered at present. The computing and storage overhead has not been adequately evaluated. 

\end{enumerate}

\subsection{\textsc{Improving deep anomaly detection methods}}

\textbf{Determine the anomaly threshold automatically and adaptively}. We argue that the threshold should be decided: (1) Automatically. The conventional threshold tuning process is not efficient and error-prone. To this end, Su \etal~\cite{su2019robust} utilize the Extreme Value Theory (EVT)~\cite{siffer2017anomaly} to learn the threshold automatically. The key idea is to use a generalized Pareto distribution (GPD) to fit extreme values. Prediction errors of training datasets are used to optimize the threshold. No data distribution assumption is needed. Another method is to test a series of threshold values at a fixed interval and check the performance. Intuitively, the value that produces the best result can be selected. (2) Adaptively. A threshold is decided and fixed when a model is trained on a known dataset. However, with the development of CPSs, the boundary of anomalies is changing. The threshold should evolve as new data comes. A naive strategy is to update the model regularly based on newly collected data. Then, a threshold is generated according to the data. Moreover, online learning could be adopted to let models learn from recent incremental data. Meanwhile, when each time the model is updated, a new threshold is calculated to replace the old one.            

\textbf{Benchmarks with sufficient labeled and real-world anomalous data}. To date, we have not found many benchmarks in CPSs that can be used to compare different  DLAD methods. Although there exist some frequently used datasets (\eg, SWaT), different DLAD methods tend to tailor the dataset and adopt the processed data on their own. We suppose that benchmarks in each CPS domain (\eg, aerial systems) can help to improve the evaluation process. Different methods may compare performances on the same benchmark. Specifically, we conclude several requirements for benchmarks. (1) Cover enough data types. Ideally,  sensor, actuator, network, and control system logs data can be provided. DLAD methods can choose any type of data based on their design goals. Also, some models tend to work better on specific data types (\eg, sensor time-series data), which could be produced separately. (2) Include labeled anomalous data. One challenge to evaluate DLAD methods is the lack of labeled anomalies. Researchers have to design and simulate attack or fault cases. Standard and rich attack data and cases can improve detection performance and reduce data-processing efforts. (3) Collect from the real world.  Although simulation is widely adopted in certain domains (\eg, smart grid) due to hardware constraints, real measurements and anomalies can represent the status of systems better. For example, the sequential order and interval of packets in CAN bus traffic in a smart vehicle can be utilized as factors to decide whether there is an anomaly. Simulation may not fully contain and represent these important factors.

\textbf{Enhance the running performance to a real-time level}. We observe that many studies~\cite{van2019real, canizo2019multi, russo2018anomaly} in the smart vehicle domain discussed the running performance of DLAD methods. This is because the response time is critical to avoid devastating accidents in smart cars. To make DLAD methods more practical, we argue that running performance is important in other CPS systems as well.  Concretely, the design can be improved from two aspects. (1) Accept real-time input measurements. Instead of using data from offline datasets, DLAD methods could obtain online real-time measurements and traffic from host systems. The data amount, sampling rate and format can be decided based on computing resources and network architectures. For example, DLAD methods that run on edge devices can achieve a high detection speed, which is owing to powerful computing ability. (2) Take real-time actions. While it is essential to detect anomalies, actions to prevent catastrophic losses can also be adopted. In some sense, actions should be taken into account when design and train DLAD models. For example, when designing the loss function, we could study how to choose appropriate actions in terms of different anomalies. 

\textbf{Locate the anomalous device or root cause}. The detection performance (\eg, true positives, precision)  is high in current DLAD methods. However, the location and the root cause of the anomaly is usually not identified. Users still do not know where an anomaly is from and how to handle the anomaly even DLAD methods detect anomalous status. Moreover, anomalies in different parts of CPSs present different impacts. We argue that DLAD methods could improve the detection granularity to component level. For example, Zhang \etal~\cite{zhang2019deep} adopt ConvLSTM to detect anomalies in each sensor or actuator. Once an anomaly is identified, the compromised device is also recognized. Then certain actions could be taken to prevent the loss. Further, this process can be automatically conducted without the intervene of users. 

\textbf{For different CPSs and problems, different compatible neural network architectures can be adopted}. We observe that there exist typical data types and anomalies in different CPSs. In ICS, sensor time-series measurement data is commonly collected. Gradual sensor and sudden actuator change anomalies will break time relations in the data. LSTM-based models and variants are utilized to capture such time relation. Meanwhile, FDI attacks are prevalent in the smart grid. We find that DLAD methods are used to aid conventional state estimator methods. LSTM and autoencoder can both be adopted. Moreover, attacks on the CAN bus system in ITS are mostly seen. Thus LSTM and CNN are used to capture both time relation and context information (\eg, packet order and content). In aerial systems, most anomalies are injected. LSTM-based methods are utilized to capture time relations. We suggest that researchers custom their models based on these findings. 

\subsection{\textsc{Conclusion}}
In this work, we systematically reviewed the current research efforts on deep learning-based anomaly detection methods in cyber-physical systems. To this end, we first proposed a taxonomy to characterize the key properties of DLAD methods. Further, we highlighted prevailing DLAD methods and research findings based on our taxonomy. We also discussed the limitations and open problems of current methods. Meanwhile, we conducted experiments to explore the characteristics of typical neural models, the workflow of DLAD methods, and the running performance of DL models. We presented deficiencies of DL approaches, our findings, and possible future directions to improve DLAD methods. We hope that this survey -- systematizing state-of-the-art deep learning-based CPS security solutions -- helps the community prioritize research efforts to address pressing deployment issues in CPS anomaly detection.


\appendix

\bibliographystyle{./ACM-Reference-Format}

\bibliography{review}

\end{document}